\documentclass[11pt]{article}
\usepackage{amsmath, amssymb,graphicx,psfrag}
\usepackage{setspace}

\def\N{{\mathcal N}}
\def\L{{\mathcal L}}

\def\O{{\mathcal O}}
\def\L{{\mathcal L}}

\def\a{\alpha}

\def\d{\delta}

\def\m{\mu}
\def\n{\nu}

\def\f{\phi}

\def\l{\lambda}

\def\p{\partial}
\def\rb{\right}
\def\lb{\left}
\def\axs{AdS_5\times S^5}
\def\tr{\mathrm{tr}}
\newcommand{\eq}[1]{\begin{equation} #1 \end{equation}}
\newcommand{\al}[1]{\begin{align} #1 \end{align}}
\newcommand{\ml}[1]{\begin{multline} #1 \end{multline}}

\DeclareSymbolFont{AMSb}{U}{msb}{m}{n}
\DeclareMathSymbol{\IN}{\mathbin}{AMSb}{"4E}
\DeclareMathSymbol{\IZ}{\mathbin}{AMSb}{"5A}
\DeclareMathSymbol{\IR}{\mathbin}{AMSb}{"52}
\DeclareMathSymbol{\Q}{\mathbin}{AMSb}{"51}
\DeclareMathSymbol{\II}{\mathbin}{AMSb}{"49}
\DeclareMathSymbol{\IC}{\mathbin}{AMSb}{"43}
\DeclareMathSymbol{\IP}{\mathbin}{AMSb}{"50}
\DeclareMathSymbol{\IH}{\mathbin}{AMSb}{"48}
\DeclareMathSymbol\IA{\mathalpha}{AMSb}{"41}
\DeclareMathSymbol\IS{\mathalpha}{AMSb}{"53}

\begin{document}



\thispagestyle{empty}
\renewcommand{\thefootnote}{\fnsymbol{footnote}}

\bigskip

\begin{center} \noindent \Large \bf
Magnetic Catalysis of Chiral Symmetry Breaking. A Holographic Prospective.

\end{center}

\bigskip\bigskip\bigskip

\centerline{ \normalsize \bf Veselin G. Filev${}^a$,  Radoslav C. Raskov${}^b$\footnote{also Dept of Physics, Sofia University, Bulgaria}} 

\bigskip\bigskip\bigskip

\centerline{${}^{a}${\it School of Theoretical Physics,}}
\centerline{\it  Dublin Institute For Advanced Studies,}
\centerline{\it 10 Burlington Road, Dublin 4, Ireland}
\centerline{\small \tt  vfilev@stp.dias.ie}
\bigskip

\bigskip

\centerline{ ${}^b$\it Institute for Theoretical Physics,} 
\centerline{ \it Vienna University of Technology,}
\centerline{ \it Wiedner Hauptstr. 8-10, 1040 Vienna, Austria}
\centerline{\small \tt  rash@hep.itp.tuwien.ac.at}
\bigskip

\bigskip

\bigskip

\bigskip

\bigskip\bigskip

\bigskip\bigskip

\renewcommand{\thefootnote}{\arabic{footnote}}

\centerline{\bf \small Abstract}

\medskip
{\small \noindent We review a recent investigation of the effect of magnetic catalysis of mass generation in holographic Yang-Mills theories. We aim at a self-contained and pedagogical form of the review. We provide a brief field theory background and review the basics of holographic flavordynamics. The main part of the review investigates the influence of external magnetic field on holographic gauge theories dual to the D3/D5-- and D3/D7-- brane intersections. Among the observed phenomena are the spontaneous breaking of a global internal symmetry, Zeeman splitting of the energy levels and the existence of pseudo Goldstone modes. An analytic derivation of the Gell-Mann--Oaks--Renner relation for the D3/D7 set up is reviewed. In the D3/D5 case the pseudo Goldstone modes satisfy non-relativistic dispersion relation. The studies reviewed confirm the universal nature of the magnetic catalysis of mass generation. }

\newpage


\tableofcontents

\newpage

\section{Introduction}
An important concept in our attempt to describe the structure of our physical reality dating back to Democritus is the atomic principle. Namely the idea that macroscopic bodies are build out of fundamental particles. In modern perspective we are interested in studying the basic interactions between the building blocks of matter. It is experimentally well established that there are four fundamental interactions: electromagnetic, strong and weak interactions as well as gravity. Despite the remarkable success of the Standard Model of particle physics unifying the first four interactions it still remains a challenge to come up with a consistent quantum theory of gravity. At present one of the most promising directions towards a unified theory of the fundamental interactions lies in the framework of string theory.

Historically, string theory emerged as an attempt to describe the strong interactions by what was called dual resonance models. However, shortly after its discovery Quantum Chromodynamics (QCD) which is a $SU(3)$ Yang--Mills gauge theory, superseded it. The matter degrees of QCD consist of quarks which are in the fundamental representation of the gauge group, while the interaction between the fundamental fields is being mediated by the gluons which are the gauge fields of the theory thus transforming in the adjoint representation of $SU(3)$. 

A remarkable property of QCD is the fact that it is asymptotically free, meaning that at large energy scales, or equivalently at short distances, it has a vanishing coupling constant. This makes QCD perturbatively accessible at ultraviolet. However, the low energy regime of the theory is quite different. At low energy QCD is strongly coupled, the interaction force between the quarks grows immensely and they are bound together, they form hadrons. This phenomenon is called confinement. Additional property of the low energy dynamics of QCD is the formation of a quark condensate which mixes the left and right degrees of the fundamental matter and leads to a breaking of their chiral symmetry. It is extremely hard to examine the properties of the strongly coupled low energy regime of QCD, since the usual perturbative techniques are not applicable. 

The AdS/CFT correspondence  \cite{Maldacena:1997re}, as we shall describe in details in Section 3.1 of this review, is a powerful analytic tool providing a non--perturbative dual description of non--abelian gauge theories, in terms of string theory defined on a proper gravitational background. An important extension of the correspondence making it relevant to the description of flavored Yang-Mills theories was the introduction of fundamental matter via the introduction of probe flavor branes \cite{Karch:2002sh}. The most understood case is in the limit  when the number of different flavors is much less than the number of colors. This corresponds to the quenched approximation on gauge theory side and the probe approximation on supergravity side of the correspondence. We will review more details about the way the AdS/CFT dictionary works in Section 3.2. 

Despite its great potential direct application of the AdS/CFT correspondence to realistic non-abelian gauge theories such as QCD remains a challenge. A major limitation is that realistic field theories do not seem to have simple holographic backgrounds. Furthermore there are indications that most realistic gauge theories do not even poses exact  holographically dual geometries. Nevertheless applications of the AdS/CFT correspondence are still possible. One plausible direction is the investigation of non-abelian gauge theories exhibiting universal behaviour. Particularly interesting is to analyze the phase structure of strongly coupled Yang-Mills theories. An example of such application of the holographic approach is the study of properties of strongly coupled quark-gluon plasmas \cite{Gubser:2009md}.

Another possible direction is the study of phenomena known to have a universal nature. An important example is the phenomenon of mass generation in an external magnetic field. This phenomenon has been extensively studied in the conventional field theory literature \cite{Gusynin:1994xp}-\cite{Klimenko:1992ch}. The effect was shown to be model independent and therefore insensitive to the microscopic physics underlying the low energy effective theory. The essence of this effect is the dimensional reduction D $\rightarrow$ D-2 (3+1 $\rightarrow$ 1+1) in the dynamics of fermion pairing in a magnetic field. Magnetic catalysis of mass generation has been demonstrated in various 1+2 and 1+3 dimensional field theories. Given the universal nature of this effect it is natural to explore this phenomenon in the context of holographic gauge theories. In this review we focus on such studies for holographic gauge theories dual to the Dp/Dq--brane intersection.

The structure of the review is as follows:

In Section 2 we provide a short field theory background. In the first  subsection we describe the properties of flavored $SU(N_c)$ Yang-Mills theory focusing on the global symmetries of the lagrangian. We remind the reader about some basics of the phenomenon of Chiral Symmetry breaking. We describe the effective field theory approach to the of  Chiral Symmetry breaking and provide a brief derivation of the famous Gell-Mann-Oaks-Renner relation \cite{GellMann:1968rz}. In the second subsection we review the mechanism of the chiral symmetry breaking due to the presence of an external magnetic field \cite{Gusynin:1994xp}. 

Section 3 of this review is dedicated to the AdS/CFT correspondence and its extension to include matter in the fundamental representation of the gauge group. In the first subsection we outline the main ideas that lead to the formulation of the Maldacena's conjecture \cite{Maldacena:1997re}. We discuss some qualitative and quantitative aspects of the correspondence and provide a brief description about the way the AdS/CFT dictionary operates. The second subsection focuses on the addition of flavor degrees of freedom to the correspondence. We review the approach of ref. \cite{Karch:2002sh} and provide some basic extracts from the AdS/CFT dictionary, which will be important for the studies presented in Section 4.

Section 4 is the main part of the review. We present the studies of the influence of external magnetic field on holographic gauge theories dual to the D3/D5 -- and the D3/D7 --branes intersection performed in refs. \cite{Filev:2007gb}, \cite{Filev:2007qu}, \cite{Filev:2009xp},\cite{Erdmenger:2007bn}. In the case of the D3/D7 system we review the general properties of the holographic set up and the way chiral symmetry breaking is realized as a separation of the color and flavor branes in the infrared. We review the properties of the light meson spectrum of the theory and uncover Zeeman splitting of the energy levels as well as the existence of Goldstone modes corresponding to the spontaneously broken Chiral Symmetry. In the limit of small bare masses review the an analytic derivation of the Gell-Mann-Oaks-Renner relation obtained in ref.~\cite{Filev:2009xp} from dimensional reduction of the eight dimensional effective action of the probe D7--brane. We also review the analogous studies of the D3/D5--system. Again there are mass generation, Zeeman splitting and Goldstone modes. Interestingly the broken Lorentz invariance in this case leads to non-relativistic dispersion relations.

We end with a brief summery of the presented material and a short discussion in the conclusion section of the review.


\section{Field Theory Preliminaries}
In this section we provide a basic field theory background. Our goal is to remind the reader about some of the properties of strongly coupled flavored Yang-Mills theories. In particular their global symmetries and the corresponding spontaneous symmetry breaking. We outline the effective field theory description of Chiral Symmetry breaking and provide a brief derivation of the Gell-Mann-Oaks-Renner relation \cite{GellMann:1968rz}. We also provide a short review of the effect of magnetic catalysis of mass generation. 
\subsection{Flavored Yang-Mills Theory and Chiral Symmetry Breaking}
\paragraph{Flavored Yang-Mills Theory.}
The lagrangian of four dimensional pure $SU(N_c)$ Yang-Mills theory coupled to $N_f$ flavors of fermionic fields is given by:
\begin{eqnarray}
&&{\cal L}=-\frac{1}{4g_{YM}^2}Tr[F_{\mu\nu} F^{\mu\nu}]+\frac{\theta}{32\pi^2}Tr[{F}_{\mu\nu}{\tilde F}^{\mu\nu}]+\sum_{i=1}^{N_f}\bar\psi^i(i{\gamma^{\mu}D_{\mu}}-m_i)\psi^i\label{YMlagr}\ ,\\
&&F_{\mu\nu}=\partial_{\mu}A_{\nu}-\partial_{\nu}A_{\mu}-i[A_{\mu},A_{\nu}];~~~D_{\mu}=\partial_{\mu}-iA_{\mu};~~~{\tilde F}^{\mu\nu}=\frac{1}{2}\varepsilon^{\mu\nu\rho\sigma}F_{\rho\sigma};\nonumber
\end{eqnarray}
The first term in equation (\ref{YMlagr}) is a dynamical term for the gauge field $A_{\mu}$. The second term (the so called $\theta$-term) is topological and is related to the Pontryagin index of $A_{\mu}$. The parameter $\theta$ is ranges from $0$ to $2\pi$ and parametrizes topologically distinct sectors of the theory. The last term in (\ref{YMlagr}) describes the matter (fundamental) fields of the theory. Let us look closely at the last term. If one defines the left and right fermionic fields $\psi_{L,R}=1/2(1\pm\gamma^5)\psi$ it can be written as:
\begin{equation}
{\cal L}_{f}=\sum_{i=1}^{N_f}\left(\bar\psi_L^ii{\gamma^{\mu}D_{\mu}}\psi_L^i+\bar\psi^i_Ri{\gamma^{\mu}D_{\mu}}\psi_R^i-m_i(\bar\psi_L^i\psi^i_R+\bar\psi_R^i\psi_L^i)\right)\ .\label{masstermYM}
\end{equation}
It is clear that the mass of the matter fields $m_i$ can be interpreted as a coupling between the left and right field $\psi_{L/R}$. Therefore at vanishing $m_i$ we have two distinct sets of $N_f$ fermionic fields and at classical level the theory has a global $U(N_f)_L\times U(N_f)_R$ symmetry. It is instructive to split the global symmetry to:
\begin{equation}
U(N_f)_L\times U(N_f)_R=SU(N_f)_V\times SU(N_f)_A\times U(1)_V\times U(1)_A\label{GlobalSYM}
\end{equation}
Let us focus first on the abelian symmetry. In infinitesimal form we have the transformations:
\begin{eqnarray}
&&\delta\psi_L=-i\alpha\psi_L;~~~\delta\psi_R=-i\alpha\psi_R;~~~{\rm{for}}~~U(1)_V;\\\label{VectorU1}
&&\delta\psi_L=-i\alpha\psi_L;~~~\delta\psi_R=+i\alpha\psi_R;~~~{\rm{for}}~~U(1)_A;\label{axialU1}
\end{eqnarray}
Transformation (\ref{VectorU1}) is just a rigid  $U(1)-$gauge transformation and correspond to some quantum number. We will not be interested in breaking gauge symmetries in these notes this is why we focus on the transformation (\ref{axialU1}).

\paragraph{Anomalous Chiral Symmetry.}

 In terms of the fields $\psi,\bar\psi$ transformations (\ref{axialU1}) can be written as:
\begin{equation}
\delta\psi=-i\alpha\gamma^5\psi;~~~\delta\bar\psi=-i\alpha\bar\psi\gamma^5; \label{AnomChiral}
\end{equation}
The corresponding Noether current is given by:
\begin{equation}
j^{\mu5}=\bar\psi\gamma^{\mu}\gamma^5\psi;\label{axialcurrent}
\end{equation}
and is conserved upon applying the equations of motion. Clearly a non-zero fermionic condensate $\langle\bar\psi\psi\rangle$ would break the Chiral transformation (\ref{AnomChiral}). Naively one would expect the existence of a corresponding Goldstone boson. This is the famous $\eta'$-meson in QCD. However it turns out that the measure of the path integral has a non-zero Jacobian under the transformation (\ref{AnomChiral}) and the axial current defined in equation (\ref{axialcurrent}) is anomalous. In fact one can show that  the anomaly is given by\footnote{look at ref.~\cite{Smilga:2001ck} pages 185-192 for a brief derivation}:
\begin{equation}
\partial_{\mu}j^{\mu5}=-\frac{1}{16\pi^2}Tr[F_{\mu\nu}\tilde F^{\mu\nu}]; 
\end{equation}
and the chiral transformation can be absorbed into a redefinition of the theta angle of the theory $\theta\rightarrow\theta-2\alpha$. this suggests that the mass of the $\eta'$-meson can be related to the topological susceptibility of the theory $\chi_{YM}=\partial^2{\cal E}_{\rm{vac}}/\partial^2\theta|_{N_f=0}$. For a canonically normalized $\eta'$-field one can obtain the Witten-Veneziano formula \cite{{Witten:1979vv},{Veneziano:1979ec}}:
\begin{equation}
m_{\eta'}^2=\frac{4N_f}{f_{\eta'}^2}\chi_{\rm{YM}}\propto \frac{N_f}{N_c};\ ,
\end{equation}
where we have used the $N_c$ dependence of  $f_{\eta'}$ for large $N_c$, $f_{\eta'}\propto \sqrt{N_c}$ \cite{Barbon:2004dq}.

The fact that $m_{\eta'}^2\propto N_f/N_c$ has an important consequences for the large $N_c$ limit of the theory. It suggests that if the number of flavors is $N_f\ll N_c$ (the so called quenched approximation) the mass of the $\eta'$-meson is effectively zero and the anomalous $U(1)_A$ axial symmetry is restored. This result is essential for the holographic studies that we will review in Section 4. In fact the holographic supersymmetric gauge theory that we consider has an anomalous $U(1)$ $R$-symmetry, which mimics the anomalous $U(1)_A$ symmetry (\ref{AnomChiral}). It is the spontaneous breaking of this symmetry under external magnetic field that has been explored in the holographic set up presented in Section 4.1.
 
\paragraph{Non Singlet Chiral Symmetry}

Let us now focus on the non-abelian part of the global symmetry (\ref{GlobalSYM}). We will be interested in the dynamical breaking of this symmetry by a non vanishing fundamental condensate $\langle\bar\psi\psi\rangle$. Clearly only the axial $SU(N_f)_A$ is broken by the fundamental condensate. Goldstone theorem suggests the existence of $N_f^2-1$ goldstone fields. In $N_f=3$ Quantum Chromodynamics these are the $\pi_{\pm}$, $\pi_0$, $K_{\pm}$, $K_0$, $\bar K_0$ and $\eta$ mesons.

An important extension of this discussion is the case when the mass of the fundamental fields is not vanishing but is still a small parameter with respect to some relevant energy scale. In this case the Chiral Symmetry is an approximate symmetry of the theory and the corresponding goldstone particles (mesons) acquire small masses. At leading order there is an important relation between the mass of the mesons, the bare mass of the fundamental fermions and the fundamental condensate, namely the Gell-Mann-Oaks Renner relation \cite{GellMann:1968rz}. Since we will be interested in verifying this relation via holographic techniques in section 4, let us provide a brief derivation using an effective field theory description.

\paragraph{Effective Chiral Lagrangian and the Gell-Mann-Oaks-Renner relation.}
In what follows we will treat the fundamental condensate of the theory as an order parameter of dynamically broken Chiral Symmetry along the lines of ref.~\cite{Smilga:2001ck}. Let us define a condensate matrix as:
\begin{equation}
\Sigma^{ij}_0=\langle\bar\psi_L^i\psi_R^j\rangle_0\ .
\end{equation}
Non-breaking of the vector symmetry implies that the matrix order parameter $\Sigma^{ij}$ can be brought into the form:
\begin{equation}
\Sigma^{ij}_0=\frac{1}{2}\delta^{ij}\Sigma_0\label{orderpar}
\end{equation}
by the group transformation (\ref{GlobalSYM}). Here $\Sigma_0$ is in general a complex scalar. Now fluctuations of the order parameter (\ref{orderpar}) will be described by a unitary matrix $U(x)\in SU(N_f)_A$:
\begin{equation}
\Sigma(x)^{ij}=\frac{1}{2}\Sigma_0 U(x)^{ij}\label{parordfluct}
\end{equation}
as well as a fluctuations of the phase of $\Sigma_0$ parametrized by elements of $U(1)_A$. Let us focus on the non-abelian case first. It is convenient to express $U(x)$ as an exponential:
\begin{equation}
U(x)=\exp\left\{\frac{2i\pi^a(x)t^a}{f_{\pi}^2}\right\} \ ,\label{pionspar}
\end{equation}
where $\pi^a$ are the physical meson fields and $f_{\pi}$ is a constant of dimension of mass (the pion's decay constant in Quantum Chromodynamics) and $t^a$ are the generators of $SU(N_f)_A$. To fix the exact form of the effective lagrangian note that there is a unique invariant structure involving two derivatives:
\begin{equation}
{\cal L}_{\rm{eff}}^{\rm{(kin)}}=\frac{f_{\pi}^2}{4}Tr[\partial_{\mu}U\partial^{\mu}U^{\dagger}]\ .
\end{equation}
To leading order in $\pi^a$ we obtain:
\begin{equation}
{\cal L}_{\rm{eff}}^{(2,\rm{kin})}=\frac{1}{2}\partial_{\mu}\pi^a\partial^{\mu}\pi^a+\dots
\end{equation}
and hence we have a canonically normalized bosonic field. In order to fix the mass term of our pseudo-Goldstone fields let us note that in the mass term in equation (\ref{masstermYM}) can be traded for:
\begin{equation}
{\cal L}^m_{\rm{eff}}=-Re[Tr\{{\cal M}\Sigma(x)\}]\ ,
\end{equation}
where we have defined the mass matrix ${\cal M}=||m_i\delta^{ij}||$. Now using equations (\ref{parordfluct}) and (\ref{pionspar}) one arrives at:
\begin{equation}
{\cal L}^m_{\rm{eff}}={\rm const}+\frac{Re[\Sigma_0]}{f_{\pi}^2}(\sum_{i=1}^{N_f}m_i)\pi^a\pi^a+O\left(\pi^3\right)\ ,
\end{equation} 
therefore to quadratic order our effective action is given by:
\begin{equation}
{\cal L}_{\rm eff}^{(2)}=\frac{1}{2}\partial_{\mu}\pi^a\partial^{\mu}\pi^a+\frac{Re[\Sigma_0]}{f_{\pi}^2}(\sum_{i=1}^{N_f}m_i)\pi^a\pi^a\ . \label{eff2}
\end{equation}
Equation (\ref{eff2}) implies the following expression for the mass of the meson fields $\pi^a$:
\begin{equation}
M^2_{\pi}=-\frac{2Re[\Sigma_0]}{f_{\pi}^2}(\sum_{i=1}^{N_f}m_i)=-\frac{2\langle\bar\psi\psi\rangle}{f_{\pi}^2}m\ ,\label{Gelint}
\end{equation}
where in the last equality we have used that $\langle\bar\psi\psi\rangle=N_fRe[\Sigma_0]$ and defined:
\begin{equation}
m=\frac{1}{N_f}\sum_{i=1}^{N_f}m_i; \ .
\end{equation}
Equation (\ref{Gelint}) is the famous Gell-Mann-Oaks-Renner relation \cite{GellMann:1968rz}. Using similar arguments one can obtain similar expression for the mass of the $\eta'$-meson corresponding to the spontaneous breaking of the $U(1)_A$ Chiral Symmetry (which is non-anomalous in the quenched approximation $N_f\ll N_c$). 

\subsection{Magnetic Catalysis of Mass Generation in Field Theory}

In this subsection we will review the mechanism of the chiral symmetry breaking due to the
presence of an external magnetic field. We will follow closely the outline provided in ref.~\cite{Gusynin:1994re}.

To start with, let us make a few comments on the general properties of the chiral fermions in a
constant magnetic field turned on in $x^3$ direction of the $4d$ spacetime. The lagrangian of a relativistic fermion is of the standard form
\eq{
\L=\frac{1}{2}\lb[\bar\Psi, (i\gamma^\m D_\m -m)\Psi\rb],
}
where the covariant derivative is given by
\eq{
D_\m=\p_\m-ie A_\m^{ext}, \quad A_\m^{ext}=-Bx_2\d_{\m,1}.
}
One of the most important characteristic of the system is its spectrum, or so called Landau levels, which can be easily obtained from the above lagrangian
\eq{
E_n(k_3)=\pm\sqrt{m^2+2|eB|n+k_3^2}, \quad n=0,1,2, \cdots.
}
First of all, one can immediately see the degeneracy of the Landau levels. 
The energy is parametrized by a continuous parameter $k_3$, the momentum along the magnetic field, and a discrete parameter $n$ related to the finite dynamics in the plane orthogonal to the magnetic field.
The number of states for the lowest Landau level is different from the others - Landau degeneracy factor for the lowest level is $\frac{|eB|}{2\pi}$ while for the other is $\frac{|eB|}{\pi}$. 
Our purpose will be to show that the dynamics of the lowest Landau level (LLL) is the one playing crucial role in the chiral symmetry breaking.

The dynamics of the chiral condensates in an external magnetic field has many interesting and important features. To make conclusions for those which we will use in the next sections, let us start with expressing the chiral condensate through the fermion propagator $S(x,y)$
\eq{
\langle 0|\bar\Psi \: \Psi |0\rangle=\lim_{x\rightarrow y} \tr\: S(x,y).
}
Thus, the problem we are going to study is encoded in the properties, or more concretely the pole structure of the fermion propagator $S(x,y)$.

The fermion propagator is well known for long time and is usually defined as the matrix element
\al{
S(x,y)& =\lb(i\gamma^\m D_\m^x+m\rb)\langle x|\frac{-i}{(\gamma^\m D_\m)^2+m^2}
|y\rangle \notag \\
& =\lb(i\gamma^\m D_\m^x+m\rb)\int\limits_{0}^\infty ds \langle x|
e^{-is[(\gamma^\m D_\m)^2+m^2]}|y\rangle.
\label{matr-elem-1}
}
To calculate the matrix element $\langle x|
\exp(-is[(\gamma^\m D_\m)^2+m^2])|y\rangle $ one can use the Schwinger's proper time approach, which gives
\eq{
\langle x| e^{-is[(\gamma^\m D_\m)^2+m^2]}|y\rangle=
\frac{e^{-i\frac{\pi}{4}}}{8(\pi s)^{\frac{3}{2}}}
e^{i[S_{cl}-sm^2]}\lb(eBs\:\cot(eBs)+\gamma^1\gamma^2 eBs\rb).
}
In the above expression $S_{cl}$ is defined as
\eq{
S_{cl}=e\int\limits_y^x A_\n^{ext}dx^\n -\frac{1}{4s}(x-y)_\n
\lb[g^{\n\m}+\frac{\lb((F^{ext})^2\rb)_{\n\m}}{B^2}\lb(1-eBs\cot(eBs)\rb)
\rb](x-y)_\m,
}
where the integration is along a straight line connecting the two points since the result is path independent. 

Separating the phase factor containing the integration, the propagator can be represented in the following convenient form
\eq{
S(x,y)=e^{ie \int\limits_y^x A_\n^{ext}dx^\n}\tilde S(x-y),
}
where
\ml{
\tilde S(x)=-i\int\limits_0^\infty \frac{ds}{16(\pi s)^2}
e^{-is m^2}e^{-\frac{i}{4s}\lb[(x^0)^2-x_A^2(eBs)\cot(eBs)-(x^3)^2\rb]}\\
.\lb(m+\frac{1}{2s}\lb[\gamma^0x^0-\gamma^Ax_A(eBs)\cot(eBs)-\gamma^3x^3\rb]
-\frac{eB}{2}\epsilon_{AB}\gamma^A x^B\rb)\\
.\lb((eBs)\cot(eBs)-\gamma^2\gamma^2(eBs)\rb),\quad A=1,2, \quad \epsilon_{12}=+1.
}
It is more convenient to consider the propagator in the Euclidean momentum space. Transforming to the Euclidean momentum space ($k^0\rightarrow ik_4, \: s\rightarrow -is$), we get
\al{
\langle 0|\bar\Psi \Psi|0\rangle & =\frac{-i}{(2\pi)^2}\tr\int d^4k \tilde S_E(k) \notag \\
&= \frac{4m}{(2\pi)^2}\int d^4 k\int\limits_{1/\Lambda}^\infty ds 
e^{-s\lb(m^2+k_4^2+k_3^2+k_A^2\frac{\tanh(eBs)}{eBs}\rb)} \notag \\
& = \frac{eBm}{(2\pi)^2}\int\limits_{1/\Lambda}^\infty\frac{ds}{s}e^{-sm^2}
\coth(eBs),
}
where $\Lambda$ is UV cutoff. It is instructive to look at the behavior of the condensate for the infinitesimal $m$. The last expression in this limit takes the form
\eq{
\langle 0|\bar\Psi \Psi|0\rangle \overset{m\rightarrow 0}{\longrightarrow}
-|eBs|\frac{m}{4\pi^2}\lb(\log \frac{\Lambda^2}{m^2}+\O(m^0)\rb).
}
It is clear that the logarithmic singularity is due to the contributions from large distances, i.e. for large proper time $s$. The conclusion one can draw about the role of the magnetic field is that it confines effectively the dynamics in only two dimensions, i.e. we arrive at $1+1$ dynamical problem. 
To uncover the nature of the logarithmic singularity let us take a closer look at the fermion propagator
in Euclidean signature 
\ml{
\hat S_E(k)=-i\int\limits_0^\infty ds\: e^{-(m^2+k_4^2+k_3^2)s}\:
e^{-\frac{k_\perp^2}{eB}\tanh(eBs)}\lb(1+\frac{1}{i}\gamma_1\gamma_2\tanh(eBs)\rb)\\
.\lb(-k_\m+m+\frac{1}{i}\gamma_Ak_B\epsilon^{AB}\tanh(eBs)\rb).
}
It is obvious that all the terms can be obtained by differentiating on parameters or integrating by parts of the expression
\eq{
I=\int\limits_0^\infty ds\: e^{-\tilde\rho s}\: e^{-\frac{\l}{2}\tanh(eBs)},
}
where $ \l=\frac{k_\perp^2}{eB}, \:\tilde\rho=m^2+k_3^2+k_4^2$. The second exponent can be expanded over Laguerre polynomials $L_n^\a$ using the generating function ($z=-\exp(-2eBs)$)
\eq{
e^{\frac{\l}{2}}\: e^{\frac{\l}{2}.\frac{z-1}{z+1}}=\sum\limits_{n=0}^\infty
c_n(\l)z^n, \quad c_n(\l)=L_n(\l)-L_{n-1}(\l), \quad |z|< 1.
}
The final expression one can obtain after lengthy but straightforward calculations is
\eq{
\hat S_E(k)=-ie^{-\frac{k_\perp^2}{eB}}\sum\limits_{n=0}^\infty
(-1)^n\frac{D_n(eB,k)}{k_4^2+k_3^2+m^2+2eBn},
\label{propagator-1}
}
where
\ml{
D_n(eB,k)=4(k_1\gamma_1+k_2\gamma_2) L_n^1(2\frac{k_\perp^2}{eB})\\
+(m-k_4\gamma_4-k_3\gamma_3)\lb[(1-i\gamma_1\gamma_2)L_n(2\frac{k_\perp^2}{eB})
-(1+i\gamma_1\gamma_2)L_{n-1}(2\frac{k_\perp^2}{eB})\rb].\notag
}
Thus, the poles of the propagator are located at the Landau levels!
From this result one can draw the following important conclusions. 
Analyzing the terms in the propagator one can see that
\textit{the logarithmic singularity in the condensate is due to the lowest Landau level}. 
The second conclusion
 is that the above expression explicitly shows the $1+1$ nature of the lowest Landau level dynamics. 
Thus, the dynamics of the fermion pairing in a magnetic field in 4d is $1+1$ dimensional phenomenon.

Summarizing, we stress on the conclusion that the presence of a magnetic field drives spontaneous chiral symmetry breaking even when the field strength is weak. The mechanism is fairly universal since it catalyzes the fermion pairing at the lowest Landau level. The pairing dynamics is essentially 1+1 dimensional in the infrared region. Concluding this section, we note that the generation of dynamical masses can be illustrated on the examples of concrete models described in the literature
(see for example \cite{Gusynin:1994re,Gusynin:1994xp}).

\section{Holographic Flavor Dynamics in a Nutshell}

The idea of gauge/string duality is one of the most profound in the realm of fundamental
interactions. It influenced a lot both sides of the correspondence: since
the first papers on the subject several new important ideas and results have emerged in string and
gauge theories.

A crucial milestone was the large N limit proposed by t'Hooft \cite{thooft-planar}. Instead of using
$SU(3)$ as a gauge group, 't Hooft proposed to consider $SU(N)$ Yang-Mills theory and
take the limit $N \gg 1$, while keeping the so{called 't Hooft coupling fixed $\lambda = Ng^2_{YM}$.
't Hooft proved that in this limit only planar diagrams contribute to the partition
function which makes the theory more tractable. On the other side, the expansion in
$1/N$ corrections of the QCD partition function and the genus expansion of the string
partition function exhibit the same qualitative behavior, suggesting that perhaps a dual
description of the large N limit of non-abelian gauge theories might be attainable in
the frame work of string theory.

In this section we will discuss some qualitative and quantitative aspects the holographic correspondence between strings and gauge theories. Once the correspondennce is argued, our primary interest will be focused on the introduction of favors and their dynamics.

\subsection{The AdS/CFT Correspondence  }

Let us make the above ideas more concrete introducing the basic ingredients of the
so called AdS/CFT correspondence. In our outline we will assume basic knowledge of the concepts of superstring theory and the notion of D--branes.\footnote{We refer the reader to ref.~\cite{Johnson:2003gi} for a comprehensive introduction to the subject.}

One can conceive of two basic types of strings.
The first are the so-called closed strings, which 
at any moment of time have the topology $S^1$.  It turns out that the closed
strings define a consistent perturbation theory in and of themselves, and
that it is this case that leads to the type II supergravities the equation of motion.
One might also consider so-called open strings which, 
at any instant of time, have the topology of an interval.  In order
for the dynamics of such strings to be well-defined, one must 
specify boundary conditions at the ends. One possibility is to impose Neumannn boundary conditions to describe the free ends. Another possibility is to impose Dirichlet boundary conditions requiring the end points of the string to remain fixed at some point of the spacetime. One can also
consider a mixture of Dirichlet and Neumann boundary conditions, 
insisting that the end of the string remain attached to some submanifold
of spacetime, but otherwise leaving it free to roam around the
surface.  Surfaces associated with such Dirichlet boundary conditions
are known as Dirichlet submanifolds; i.e. D-branes. To shorten the long discussion we just stress that  it turns out that the Dirichlet submanifolds
are sources of the Ramond-Ramond gauge fields and of the gravitational
field.  That is, they carry both stress-energy and Ramond-Ramond
charges. Summarizing, one can say that
in many ways, the discovery of D-branes was a breakthrough for string theory. D-branes
provide non-perturbative solutions to the theory. They also couple naturally to both open
strings, which have gauge fields in their spectrum; and to closed strings, which have gravitons
as vibration modes. This complementary nature of D-branes makes for a powerful framework for further study of the ideas of gauge/string duality.

The idea of the gauge/string duality passed though many controversal developments 
 over few decades. 
The early hints about a possible gauge/string duality came very close to reality
with the development of the concept of Dp-branes and their identification as the
sources of the well-known black p-brane solutions of type IIB supergravity.
 The key observation was that the low energy dynamics of a stack of N coincident Dp--branes
can be equally well described by a $SU(N)$ supersymmetric Yang-Mills theory in p+1
dimensions and an appropriate limit of a p-brane gravitational background. The first
gauge theory studied in this context is the large $N$ $\N = 4$ $SU(N)$ supersymmetric
Yang-Mills theory in 1 + 3 dimensions which is a maximally supersymmetric conformal
field theory. The corresponding gravitational background is, as proposed by
Maldacena \cite{Maldacena:1997re}, the near horizon limit of the extremal 3--brane solution of type IIB
supergravity. This was the original formulation of the standard (by now) AdS/CFT
correspondence.

An important step was the understanding
of the role of one of the additional coordinates (supplementing the four usual ones)
as a renormalization-group scale. The idea was further promoted 
 demonstrating the possibility of self-consistent account of the
back reaction of the gravity on the D--branes and vise versa. Further development of this idea lead to the formulation of the AdS/CFT correspondence. We refer the reader to the extensive review \cite{Aharony:1999ti} for a detailed historical overview and a detailed list of references and focus on the physical aspects of the correspondence.


\paragraph{Low energy dynamics of D3-branes and the decoupling limit}

When there are $N$ parallel Dp--branes, their
low energy dynamics is described by a reduction of the $\N = 1$ supersymmetric ten dimensional
Yang-Mills theory of the gauge group $U(N)$ to the p + 1 dimensional supersymmetric Yang-Mills 
higgsed theory. The physics of the supersymmetric Yang-Mills systems can be understood by
that of the D--brane dynamics and vice versa.

Let us consider a stack of $N$ coincident D3--branes. This system has two different kinds of perturbative type IIB string theory excitations, namely open strings that begin and end on the stack of branes and closed strings which are the excitations of empty space. Let us focus on the low energy massless sector of the theory.

The first type of excitations corresponds to zero length strings that begin and end on the D3--branes. The orientation of these strings is determined by the D3--brane that they start from and the D3--brane that they end on. Thus, the states describing the spectrum of such strings are labeled by 
$\lambda_{ij}$, where $i,j=1,\dots,N$. It can be shown that in the case of oriented strings \cite{Johnson:2003gi}  $\lambda_{ij}$ transform in the adjoint representation of $U(N)$. On the other side, the massless spectrum of the theory should form a ${\cal N}=4$ supermultiplet in $1+3$ dimensions. The possible form of the interacting theory (if we take into account only interactions among the open strings) is thus completely fixed by the large amount of supersymmetry that we have and is the ${\cal N}=4$ $U(N)$ supersymmetric Yang--Mills theory. Note that $U(N)$ came from the transformation properties of $\lambda_{ij}$. On the other side, $U(N)$ can be thought of as a direct product of $U(1)$ and $SU(N)$, geometrically the $U(1)$ corresponds to the collective coordinates of the stack of D3--branes. We will restrict ourselves to the case when those modes were not excited, we refer the reader to ref.~\cite{Aharony:1999ti} for further discussion on this point.

The second kind of excitations is that of type IIB closed strings in flat space. The low energy massless sector is thus a type IIB supergravity in $1+9$ dimensions. 

The complete action of the system is a sum of the actions of those two different sectors plus an additional interaction term. This term can be arrived at by covariantizing the brane action after introducing the background metric for the brane~\cite{Leigh:1989jq}. It can be shown \cite{Aharony:1999ti} that in the $\alpha' \rightarrow 0$ limit the interaction term vanishes and the two sectors of the theory decouple. 

To summarize: the low energy massless perturbative excitations of the stack of D3--branes are given by two decoupled sectors, namely ${\cal N}=4$ $SU(N)$ supersymmetric Yang--Mills theory and supergravity in flat $1+9$ space-time. Our next step is to consider an equivalent description of this system in terms of effective supergravity solution.

Let us consider the extremal black 3--brane solution of type IIB supergravity. The corresponding gravitational background is given by \cite{Johnson:2003gi}:
\begin{eqnarray}
ds^2&=&H_3^{-1/2}\eta_{\mu\nu}dx^{\mu}dx^{\nu}+H_3^{1/2}dx^idx^i \ ,\label{D3--branes}\\
e^{2\Phi}&=&g_s^2 \ , \nonumber\\
C_{(4)}&=&H_3^{-1}g_s^{-1}dx^0\wedge\dots\wedge dx^3\ , \nonumber
\end{eqnarray}
where $\mu=0,\dots, 3$, $i=4,\dots,9$, and the harmonic (in six dimensions) function $H_3$ is given by:
\begin{equation}
H_3=1+\frac{4\pi g_3 N\alpha'^2}{r^4}\ .
\end{equation}
The integer number $N$ quantizes the flux of the five-form field strength $dC_{(4)}
$. It can also be interpreted as the number of D3--branes sourcing the geometry.
We note that the elementary D3--brane solutions for small $y$ have 
a warp factor describing a ``throat'' geometry. For very large $y$ the throat opens into a flat $\mathbb R^{1,4}$ space. 
Taking the near horizon limit of the geometry corresponds to sending $\alpha'\rightarrow 0$ while keeping the quantity $u=r/\alpha'$ fixed. Such a limit serves two goals: first it enables one to zoom in the geometry near the extremal horizon and second it corresponds to a low energy limit in the string theory defined on this background. After leaving only the leading terms in $\alpha'$, one can obtain the following metric \cite{Maldacena:1997re}:
\begin{eqnarray}
ds^2&=&\frac{u^2}{R^2}(-dx_{0}^2+dx_1^2+dx_2^2+dx_3^2)+R^2\frac{du^2}{u^2}+R^2d\Omega_{5}^2\ ,\label{AdS-intro}\\
C_{(4)}&=&\frac{1}{g_s}\frac{u^4}{R^4}dx^0\wedge dx^1\wedge dx^2 \wedge dx^3\ ,\nonumber\\\
e^\Phi&=&g_s\ ,\nonumber\\
R^4&=&4\pi g_{s}N_{c}\alpha'^2\ ,\nonumber
\end{eqnarray}
The background in equation (\ref{AdS-intro}) is that of an AdS$_5\times S^5$ space-time of radius $R$. Note that from a point of view of an observer at $r\rightarrow\infty$, the type IIB string theory excitations living in the near horizon area, namely superstring theory on the background (\ref{AdS-intro}), will be redshifted by an infinite factor of $\sqrt{g_{tt}}=H_3^{-1/4}$. Therefore, we conclude that type IIB superstring theory on the background of AdS$_5\times S^5$ should contribute to the low energy massless spectrum of the theory seen by an observer at infinity. However, an observer at infinity has another type of low energy massless excitations of type IIB string theory, namely type IIB supergravity on flat $1+9$ space or gravitational waves. Those two different types of excitations can be shown to decouple form each other. To verify this one can consider the scattering amplitudes of incident gravitons of the core of the geometry (the near horizon area). It can be shown that at low energies ($\omega\ll 1/R$) the absorption cross-section of such a scattering $\sigma_{\rm{abs}}$ goes like \cite{Gubser:1997yh,Klebanov:1997kc} $\sigma_{\rm{abs}}\propto \omega^3R^8$. Therefore, one has that $\sigma_{\rm{abs}}\rightarrow 0$ and those two types of excitations decouple in the low energy limit $\omega\rightarrow 0$.

Let us see what the decoupling limit means for the string sigma model
in the $D3$-brane background. We will
concentrate here on the metric part, thereby ignoring the contributions from the five-form field
$F_5$. We denote the $D = 10$ coordinates by $x^M, M = 0, 1,\cdots, 9$, and the metric by 
$G_{MN}(x)$.
We choose the first 4 coordinates to coincide with $x^\m $ of the Poincar\'e invariant $D3$ worldvolume, while
the coordinates on the 5-sphere are $x^M$ for $M = 5,\cdots, 9$ and 
the coordinate $x^4 = u$. The full $D3$-brane
metric takes the form $ds^2 = G_{MN}dx^Mdx^N = R^2 \tilde G_{MN}(x; R)dx^Mdx^N$, where the rescaled metric $\tilde G_{MN}$ is given by
\eq{
\tilde G_{MN}(x; R)dx^Mdx^N = 
 \lb(1 +\frac{R^4}{u^4}\rb)^{\frac{1}{2}}(\frac{du^2}{u^2}+d\Omega^2_5)
\lb(1 +\frac{R^4}{u^4}\rb)^{-\frac{1}{2}}\frac{\eta_{ij}}{u^2}dx^idx^j
 \label{resc-metr}
 }
Substituting this metric back into the non-linear sigma model, we obtain
 \eq{
S_G =\frac{1}{4\pi\a'}\int\limits_{\Sigma} \sqrt{-\gamma}\gamma^{mn} 
G_{MN}(x)\p_m x^M\p_n x^N 
=\frac{R^2}{4\pi\a'}\int\limits_{\Sigma} \sqrt{-\gamma}\gamma^{mn} 
\tilde G_{MN}(x)\p_m x^M\p_n x^N
\label{resc-act}
} 
The overall coupling constant for the sigma model dynamics is given by
\eq{
\frac{R^2}{4\pi\a'}= \sqrt{\frac{\lambda}{4\pi}}, \quad
\lambda \equiv g_sN 
\label{param}
}
Keeping $g_s$ and $N$ fixed but letting $\a'\rightarrow 0$ implies that $R\rightarrow 0$. 
It is easy to see that under this limit the sigma
model action admits a smooth limit, given by
\eq{
S_G = \sqrt{\frac{\lambda}{4\pi}}\int\limits_{\Sigma}
\sqrt{-\gamma}\gamma^{mn} 
\tilde G_{MN}(x)\p_m x^M\p_n x^N
}
where the metric $\tilde G_{MN}(x; 0)$ is the metric on $\axs$,
\eq{
G_{MN}(x; R)dx^Mdx^N =
\frac{\eta_{ij}}{u^2}dx^idx^j+\frac{du^2}{u^2}+d\Omega^2_5,
}
rescaled to unit radius. 
More over, the coupling $1/\sqrt{\lambda}$ has taken over the role of $\a'$ as the
non-linear sigma model coupling constant and the radius $R$ has canceled out.


\paragraph{The AdS/CFT correspondence}

As we learned from the above, the massless sector of the low energy dynamics of $N$ coincident D3--branes allows two possible descriptions. Conjecturing that these descriptions are equivalent is the core of Maldacena's AdS/CFT correspondence \cite{Maldacena:1997re}. Notice that in both descriptions one part of the decoupled sectors is a type IIB supergravity in flat space. Thus, we are naturally led to the conclusion that 
{\it type IIB superstring theory on the background of AdS$_5\times S^5$ background is dual to  ${\cal N}=4$ $SU(N)$ supersymmetric Yang--Mills theory in $1+3$ dimensions.}
We have presented this statement in a diagrammatic way in Figure~\ref{fig:correspondence}.
\begin{figure}[h] 
   \centering
   \includegraphics[ width=13cm]{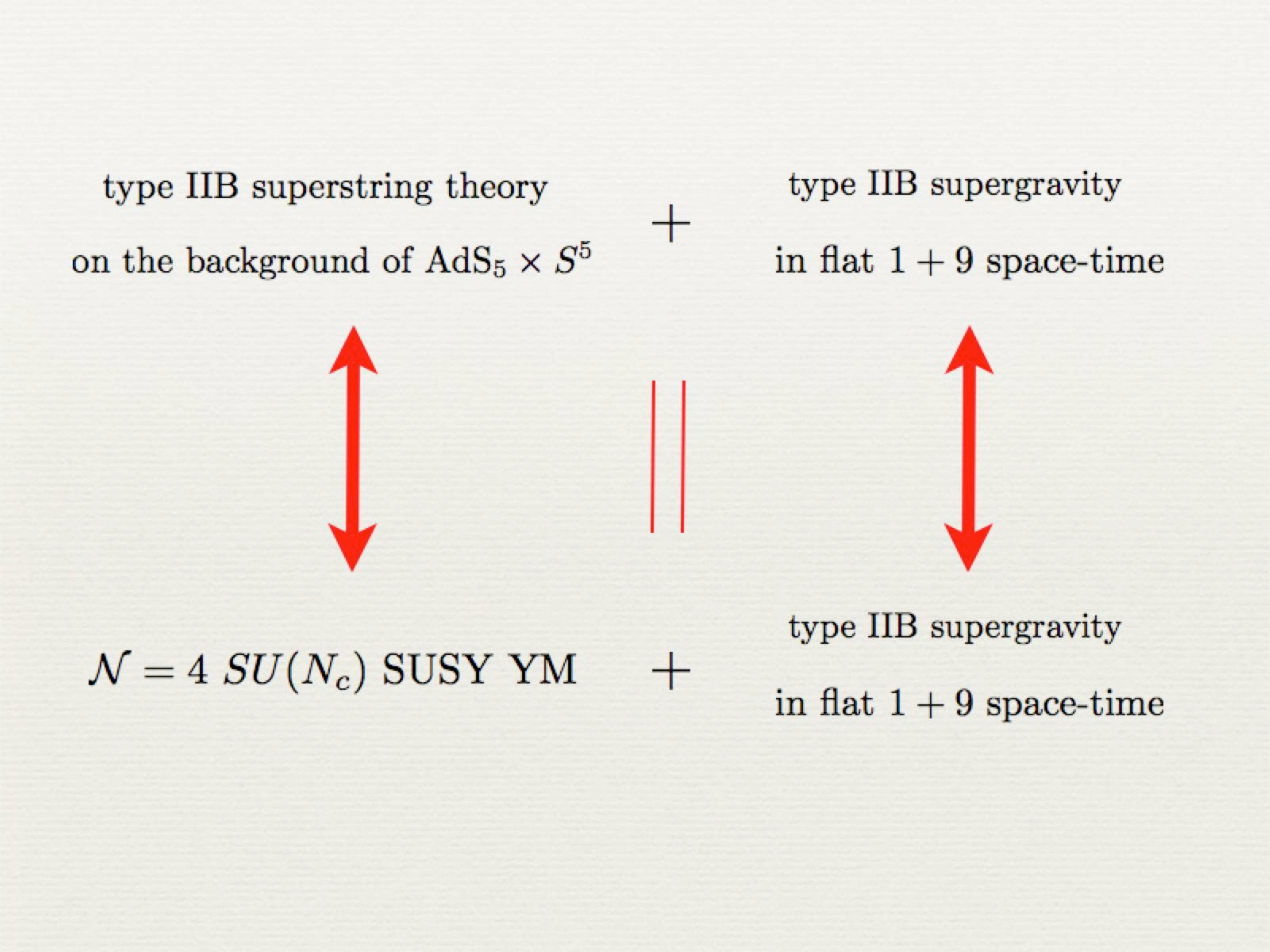}
   \caption{A diagrammatic statement of the AdS/CFT correspondence.}
   \label{fig:correspondence}
\end{figure}

A further hint supporting the AdS/CFT correspondence is that the global symmetries of the proposed dual theories match. Indeed, an AdS$_5$ space-time of radius $R$ can be embedded in a $\mathbb{R}^{2,6}$ flat space. It can then be naturally described as a hyperboloid of radius $R$ and thus has a group of isometry $SO(2,6)$ which is also the group of rotations in $2+4$ dimensions. On the other side, the $S^5$ part of the geometry has a group of isometry $SO(6)$ (the group of rotations in $6$ dimensions). This leads us to the conclusion that the total global symmetry of string theory on AdS$_5\times S^5$ gravitational background is $SO(2,4)\times SO(6)$. It is satisfying that the corresponding gauge theory has  the same global symmetry. Indeed, it is a well--known fact that the ${\mathcal N}=4$ supersymmetric Yang--Mills theory in $1+3$ dimensions is a conformal field theory. As such it should has the global symmetry of the conformal group in $1+3$ dimensions, but this is precisely $SO(2,4)$. Actually, since the theory is supersymmetric,  the full global symmetry group is the superconformal group which includes an $SU(4)$ global R--symmetry. In particular this group rotates the gauginos of the super Yang--Mills theory. However, it is well known that $SU(4)\cong SO(6)$, and therefore the global symmetry of the gauge theory is indeed $SO(2,4)\times SO(6)$!

An important aspect of the correspondence is the regime of the validity of the dual description. Depending on the precise way in which we are taking the $\alpha' \rightarrow 0$ limit, there are two basic forms of the AdS/CFT correspondence. The strongest form is that the string/gauge correspondence holds for any $N$. Unfortunately, this strong form of the conjecture cannot be tested directly since it is not known how to quantize superstring theory on a curved background in the presence of Ramond-Ramond fluxes \cite{Erdmenger:2007cm}. The second form of the conjecture holds in the t'Hooft limit when $N\rightarrow\infty$ and the t'Hooft coupling $\lambda\propto g_sN$ is kept fixed. In this way on gauge side of the correspondence only planar diagrams contribute to the partition function while on string side the required $g_s\rightarrow 0$ limit suggests semiclassical limit of the superstring theory on AdS$_5\times S^5$.  An important observation is that large $\lambda \gg1 $ suggests large AdS radius $R\propto \lambda^{1/4}$ and hence small curvature of the AdS background. This implies that the supergravity description is perturbative and thus provides an analytic tool for perturbative studies of non--perturbative field theory phenomena. On the other side, if we are at weak t'Hooft coupling ($\lambda \ll 1$) we can perform perturbative studies on gauge side of the correspondence and transfer the result to the perturbatively inaccessible regime of the supergravity description. Therefore, we conclude that the AdS/CFT correspondence is a strong/weak duality. In this work we will concentrate solely on the study of strongly coupled $\lambda \gg 1$ Yang--Mills theories. Hence, we will perform the analysis on the supergravity side of the AdS/CFT correspondence.

\paragraph{The AdS/CFT dictionary}

It did not take long until an explicit formulation of Maldacena’s conjecture was found.
Gubser, Klebanov and Polyakov \cite{Gubser:1998bc} and Witten \cite{Witten:1998qj} independently proposed to identify
the classical on-shell supergravity action, expressed in terms of given boundary values,
with the effective action of super Yang-Mills theory, where the supergravity boundary 
values play the roles of generating currents. Moreover, Witten suggested that via this identification
any field theory action on (d+1)-dimensional anti-de Sitter space gives rise to
an effective action of a field theory on the d-dimensional boundary of Anti-de Sitter space.
Most importantly, this field theory on the boundary must be a conformal field theory, because
the AdS symmetries act as conformal symmetries on the asymptotic boundary.
This duality has since been called the \textit{AdS/CFT correspondence}.

The general correspondence formula is \cite{Witten:1998qj}
\eq{
\int_{\Psi_0}D\Psi e^{-I_{ADS}[\Psi]}=\left\langle \exp\int d^dx\,\O(x)\Psi_0(x)\right\rangle
\label{sugra-1}
}
where the functional integral on the left hand side is over all fields $\Psi$ whose asymptotic
boundary values are $\Psi_0$, and $\O$ denotes the conformal operators of the boundary conformal
field theory.

In the classical limit, which will be considered exclusively throughout this Chapter, the
functional integral on the left hand side of equation \eqref{sugra-1} becomes redundant, and the
correspondence formula can be given in the simple form \cite{Gubser:1998bc,Witten:1998qj}
\eq{
I_{ADS}[\Psi_0]=W_{CFT}[\Psi_0]
\label{sugra-2}
}
where $I_{AdS}$ is the classical on-shell action of an AdS field theory, expressed in terms of the
field boundary values $\Psi_0$, and $W_{CFT}$ is the CFT effective action with generating currents,
given by minus the logarithm of the right hand side of equation \eqref{sugra-1}. However, one must
expect $I_{AdS}$ to be divergent as it stands, because of the divergence of the AdS metric  on
the AdS horizon, $x_0=0$. Thus, in order to extract the physically relevant information, the
on-shell action has to be renormalized by adding counter terms,which cancel the infinities.
After defining the renormalized, finite action by
\eq{
I_{ADS,fin}=I_{ADS}-I_{div}
\label{sugra-3}
}
where $I_{div}$ stands for the local counter terms, one identifies $I_{AdS,fin}$ with the CFT effective
action. Thus, the meaningful correspondence formula is
\eq{
I_{ADS,fin}\equiv W_{CFT}
\label{sugra-4}
}

Given a field theory action on AdS space and a suitable regularization method, it is
straightforward to calculate the renormalized on-shell action $I_{AdS,fin}$. On the other hand,
the CFT effective action
\eq{
W_{CFT}[\Psi_0]=-\ln\left\langle \exp \int d^dx\,\O(x)\Psi_0(x)\right\rangle
\label{sugra-5}
}
contains all information about the conformal field theory living on the AdS horizon, in that
all correlation functions of its operators can be obtained in a standard fashion. Thus, the
AdS/CFT correspondence formula \eqref{sugra-4} provides for the most amazing fact that the properties
of certain conformal field theories can be obtained from seemingly unrelated theories,
namely field theories on AdS spaces. Moreover, any field theory on AdS space, which
includes gravity, has a corresponding counter part CFT, whose action might not even be
known. Thus, the AdS/CFT correspondence might be an invaluable tool for formulating
non-trivial CFT's in various dimensions, although studies of this aspect.

 Let us focus on the precise way that the AdS/CFT correspondence is implemented on the example of a
 scalar field $\f$. After a closer 
look at the geometry of the AdS$_5\times S^5$ background, we conclude that it has five non--compact directions. Four of them are parallel to the D3--branes world volume and correspond to the $1+3$ directions of the dual gauge theory. The fifth non--compact direction is the radial direction $u$ (radial in the transverse, to the D3--branes, $\mathbb{R}^6$ space) and its interpretation in the dual gauge theory is not obvious. To shed more light on it, let us consider the action of a free massless scalar field in $1+3$ dimensions~\cite{Erdmenger:2007cm}:
\begin{equation} 
S=\int d^4x(\partial\phi)^2\ .
\label{free}
\end{equation}
The corresponding field theory is conformal and thus has a global symmetry $SO(2,4)$ which is the conformal group in $1+3$ dimensions. Therefore, we can consider the transformation properties of the scalar field $\phi$ under the action of the dilatation operator. One can verify that the transformation:
\begin{equation} 
x\rightarrow e^{\alpha}x;~~~\phi\rightarrow e^{-\alpha}\phi;
\end{equation}
leaves the action (\ref{free}) invariant. Furthermore, we learn that the scalar field $\phi$ has a scaling dimension one. On the other side, the $SO(2,4)$ group is the group of isometry of AdS$_5$ and one can verify from equation (\ref{AdS-intro}) that the transformation $x\rightarrow e^{\alpha}x$, suggests:
\begin{equation} 
u\rightarrow e^{-\alpha}u\ .
\end{equation}
Therefore, we learn that the coordinate $u$ scales as energy under dilations and thus has a natural interpretation as an energy scale of the dual gauge theory. 

The further development of the AdS/CFT correspondence resulted in a map between gauge invariant operators in ${\mathcal N}=4$ super Yang -Mills in a particular irreducible representation of the $SU(4)$ R-symmetry group and supergravity fields in the isomorphic representation of the $SO(6)$ global symmetry. These representations are obtained after Kaluza-Klein reduction of the supergravity fields on the internal $S^5$ sphere. Let us consider for simplicity the case of a scalar field of mass $m$, propagating on the AdS$_{d+1}$ background.
The relevant action is \cite{Erdmenger:2007cm}:
\begin{equation}
S=\int d^dxdu\sqrt{-g}(g^{ab}\partial_a\phi\partial_b\phi-m^2\phi^2)\ .
\end{equation}
The solution of the corresponding equation of motion have the following asymptotic behavior at large $u$:
\begin{equation}
\phi(u,x)=\left(\frac{1}{u}\right)^{4-\Delta}\phi_0(x)+\left(\frac{1}{u}\right)^{\Delta}\langle{\mathcal O}(x)\rangle\ ,
\label{asdict}
\end{equation}
where
\begin{equation}
\Delta=\frac{d}{2}+\sqrt{\frac{d^2}{4}+R^2m^2}\ .
\end{equation}
Note that the supergravity field $\phi(u)$ is a scalar field and is thus invariant under the action of the dilatation operator because the latter is one of the generators of the global symmetry $SO(2,4)$. Therefore, we conclude that $\phi_0$ and $\langle{\mathcal O}(x)\rangle\ $ carry scaling dimensions $4-\Delta$ and $\Delta$, respectively. In \cite{Gubser:1998bc} it was suggested that $\phi_0$ and $\langle{\mathcal O}(x)\rangle$ correspond to the source and the vacuum expectation value of the gauge invariant operator ${\mathcal O}(x)$. It is also worth noting that the expression:
\begin{equation}
\int d^dx\phi_0(x)\langle{\cal O}(x)\rangle=\mathrm{inv.}
\end{equation}
is invariant under the $SO(2,4)$ global symmetry. It was suggested that the exact form of the map is given by the relation \cite{Gubser:1998bc,Witten:1998qj}:
\begin{equation}
\langle e^{\int d^dx\phi_0(x)\langle{\cal O}(x)\rangle}\rangle_{\mathrm{CFT}}={\cal Z}_{\mathrm{Sugra}}[\phi_0(x)]\ ,
\label{dictwit}
\end{equation}
where
\begin{equation}
{\cal Z}_{\mathrm{Sugra}}[\phi_0(x)]=\lim_{\epsilon\rightarrow 0}{\cal Z}_{\mathrm{Sugra}}[{\phi_0(1/\epsilon,x)=\epsilon^{d-\Delta}\phi_0(x)}]\ .\label{regeps}
\end{equation}
{\it i.e.} the generating functional of the conformal field theory coincide with the generating functional for tree level diagrams in supergravity. We refer the reader to the extensive review ref.~\cite{Aharony:1999ti} for more subtleties on the precise way of taking the $\epsilon\rightarrow 0$  limit in equation (\ref{regeps}). 

Formula (\ref{dictwit}) has been tested by comparing correlation functions of the ${\mathcal N}=4$ quantum field theory with classical correlation functions in AdS$_d$. Note that the tree level approximation on supergravity side is valid only at strong t'Hooft coupling and therefore the corresponding conformal field theory is strongly coupled. This is why the correspondence was tested in this way only for correlation functions which satisfy non--renormalization theorems and hence are independent on the coupling \cite{Erdmenger:2007cm}. In particular it applies for the two- and three- point functions of $1/2$ BPS operators \cite{Freedman:1998tz, Lee:1998bxa}.

Further checks of the correspondence beyond the $1/2$ BPS sector was started with the so--called  plane--wave string/gauge theory duality, where one takes appropriate plane--wave limit of the AdS$_5\times S^5$ background \cite{Berenstein:2002jq}. Key point of this limit is that superstring theory on this background can be exactly quantized. Recently a significant progress towards quantizing superstring theory on AdS$_5\times S^5$ has been achieved using integrable spin chain models. We refer the reader to refs.~\cite{{Beisert:2004ry},{Plefka:2003nb}} for more details on these vast subjects.

\subsection{Adding Flavors to the Correspondence}

Direct consequence of the confining property of QCD is the fact that the low energy dynamics of the theory is governed by color singlets, such as mesons, baryons and glueballs. Mesons and baryons are bound states of quarks, the latter transform in the fundamental representation of $SU(3)$. The fact that at low energy QCD is strongly coupled suggests that it is not accessible for perturbative studies. This is why it is important to come up with an alternative non--perturbative techniques describing the strongly coupled regime of Yang--Mills theories and in particular Yang--Mills theories containing matter in the fundamental representation of the gauge group, such as QCD.
 
Further need of alternative non--perturbative techniques applicable to the properties of the fundamental fields in the strongly coupled regime of non--abelian gauge theories is required by the very recent discoveries of the properties of matter obtained in heavy ion collision experiments. More precisely the fact that the quark--gluon plasma which is the phase of matter of the fireballs obtained in such experiments, is not the expected weakly coupled quark--gluon plasma predicted by the standard perturbative QCD but is classified as a strongly coupled quark-gluon plasma. A novel phase of matter that provides challenge for the society of theoretical physicists.

One of the purpose of the study of the AdS/CFT correspondence is to develop the above mentioned analytic tools for the study of strongly coupled Yang--Mills theories. The original form of the conjecture that we described in the previous section, focuses on a gauge theory with a huge amount of symmetry, namely the ${\cal N}=4$ super Yang--Mills theory. On way to make the correspondence more applicable to realistic gauge theories, such as QCD, is to reduce the amount of the supersymmetry of the theory by introducing additional gauge invariant operators. This approach though fruitful still has the weakness that the matter content of the dual gauge theory, more precisely the fermionic degrees of freedom, transform in the adjoint representation of the gauge group. In other words there are no fundamental fields in the theory. The reason is that both ends of the strings, producing the field content of the gauge theory, are attached to the same stack of D3--branes and the corresponding states transform in the adjoined representation of the gauge group.  In order to introduce fundamental matter, one needs to consider separate stack of D--branes. 

The easiest way to introduce fundamental fields in the context of the AdS/CFT correspondence is to consider an additional stack of $N_f$ D7--branes \cite{Karch:2002sh}. (From now on we will use $N_c$ as a notation for the number of the D3--branes sourcing the gravitational background.) Since the D7--branes' world volume is higher dimensional and non--compact in the transverse to the D3--branes dimensions, the D7--branes have infinite ``internal" volume and thus their gauge coupling vanish making their gauge symmetry group a global symmetry. In this way we introduce family of fundamental matter with global flavor symmetry $SU(N_f)$. To be more precise let us consider two stacks of parallel $N_{c}$ D3--branes and $N_f$ D7--branes embedded in the following way: 
      \begin{table}[h]
\begin{center}
\begin{tabular}{|c|c|c|c|c|c|c|c|c|c|c|}
\hline
 &0&1&2&3&4&5&6&7&8&9\\\hline
 D3&-&-&-&-&$\cdot$&$\cdot$&$\cdot$&$\cdot$&$\cdot$&$\cdot$\\\hline
 D7&-&-&-&-&-&-&-&-&$\cdot$&$\cdot$\\\hline
 
\end{tabular}
\end{center}
\caption{ Embedding of the flavor D7--branes. }
\label{default}
\end{table}%
 
 The low energy spectrum of the strings stretched between the D3-- and D7--branes directions give rise to the ${\cal N}=2$ hypermultiplet containing two Weyl fermions of opposite chirality coming from the light-cone modes of strings stretched along the NN and DD directions (2,3,8,9) and two complex scalars coming form strings stretched along the ND directions, namely 4,5,6,7. Now if we consider $N_f \ll N_c$ and take the large $N_c$ limit. We can substitute the stack of D3--branes with a $AdS_5\times S^5$ space and study the $N_f$ D--branes in the probe limit using their Dirac--Born--Infeld action. On gauge side this corresponds to working in the quenched approximation ($N_f \ll N_c$) and taking the large $N_c$ t'Hooft limit.
 If the D3-- and D7--branes are separated in their transverse (8,9)-plane, then the strings stretched between them has a final length and hence final energy. It can be shown that \cite{Johnson:2003gi} the mass of the hypermultiplet is given by the energy of the string or the distance of separation $L$ multiplied by the string tension $m_q=L/{(2\pi\alpha')}$.
 
 Let us study closer the symmetry of the set up. If the D3-- and the D7--branes overlap the hypermultiplet is massless ($m_q=0$). In this case the $SO(6)$ rotational symmetry of the transverse $\IR^6$ space is broken to the product $SO(4)\times SO(2)$, corresponding to rotations along the ND directions (4,5,6,7) and the DD directions (8,9), respectively. This is equivalent to a $SU(2)_L\times SU(2)_R\times U(1)_R$ global symmetry, and suggests that the gauge theory has a R--symmetry group $SO(2)_R\times U(1)_R$ \cite{Kruczenski:2003be}, which is indeed the case, when the hypermultiplet is massless. If the D3-- and D7--branes are separated it is known that the R--symmetry is just $SU(2)_R$, which again fits that fact that the $SO(2)$ rotational symmetry in the (8,9)-plane is broken.
 
\paragraph{The dictionary of the probe brane}

 Let us now focus on the precise way that the AdS/CFT dictionary is implemented. The dynamics of the D7--brane probe is described by the Dirac--Born--Infeld action including the Chern-Simons term~\cite{Johnson:2003gi}:
\begin{equation}
\frac{S}{N_f}=-\mu_7\int\limits_{{\cal M}_8}e^{-\Phi}d^8\xi\sqrt{-det(P[G_{ab}]+(2\pi\alpha')^2{\cal F}_{ab})}+\frac{(2\pi\alpha')^2}{2}\mu_7\int\limits_{{\cal M}_8}P[C_{(4)}]\wedge{\cal F}\wedge{\cal F}\ ,
\end{equation}
where $(2\pi\alpha'){\cal F}_{ab}=P[B_{ab}]+(2\pi\alpha')F_{ab}$ and $\mu_7=[(2\pi)^7\alpha'^4]^{-1}$.
It is convenient to introduce the following coordinates:
\begin{equation}
\rho=u\cos\theta;~~~L=u\sin\theta;
\end{equation}
and consider the ansatz: 
\begin{equation}
L=L(\rho);~~~\phi=const \ .
\end{equation}
Then the lagrangian describing the D7--brane embedding is:
\begin{equation}
{\cal L}\propto \rho^3\sqrt{1+L'(\rho)^2}
\end{equation}
leading to the equation of motion:
\begin{equation}
L'(\rho)=-\frac{2c}{\sqrt{\rho^6-4c^2}}\ .
\label{adssolint}
\end{equation}
At large $\rho$ the solution has the behavior:
\begin{equation}
L(\rho)=m+\frac{c}{\rho^2}+\dots\ .
\end{equation}
Now if we introduce the field: 
\begin{equation}
\chi(u)=\frac{L(\rho)}{\rho^2+L(\rho)^2}=\frac{1}{u}m+\frac{1}{u^3}c+\dots;~~~u^2=\rho^2+L(\rho)^2 \ ,
\end{equation}
we can see that $\chi(u)$ has the same behavior as the field $\phi(u,x)$ from equation (\ref{asdict}). This is quite suggesting. The asymptotic value of $L(\infty)=m$ is exactly the separation of the D3-- and D7--branes and is thus related to the mass of the hypermultiplet via $m_q=m/(2\pi\alpha')$. Since the hypermultiplet chiral fields are our quarks we will call $m_q$ the bare quark mass. Therefore, the coefficient $c$ should be proportional to the vev of the operator that couples to the bare quark mass but this is the quark condensate! This is an example of the how the generalized AdS/CFT dictionary works at the level of a D7--brane probing. Let us provide the exact relation between the quark condensate $\langle{\cal O}_q\rangle$ and the coefficient $c$:
\begin{equation}
\langle{\cal O}_q\rangle=-\frac{N_f}{(2\pi\alpha')^3g_{YM}^2}c\ .
\label{cond-intro}
\end{equation}
We refer the reader to the appendix of Chapter~2 for more details on the last calculation and to ref.~\cite{Karch:2005ms} for an elegant presentation of the holographic renormalization of probe D--branes in AdS/CFT.

Now let us go back to equation (\ref{adssolint}) and note that in order for the D7--brane to close smoothly in the bulk of the geometry, we need to impose $L'(0)=0$. This is possible only for $c=0$ and thus we conclude that the condensate of the theory vanish. But the dual gauge theory is supersymmetric this is why it is not surprising that the quark condensate is zero. Furthermore, since there is no potential between the D3-- and D7--branes (because of the unbroken supersymmetry), the D7--brane should not bend at infinity and this is why the solution for the D7--brane embedding should be simply $L\equiv m$, as it is.

Note that the analysis that led to equation (\ref{cond-intro}) requires that the gravitational background be only asymptotically AdS$_5\times S^5$. In fact, in all cases that we are going to consider in this work, there will be some sort of the deformation of the bulk physics, coming either from the gravitational background or from the introduction of external fields. This will break the supersymmetry and will capacitate the dual gauge theory to develop a quark condensate. In Section~4, we will use this approach to provide a holographic description of magnetic catalysis of chiral symmetry breaking. Through the rest of this review, the study of the quark condensate as a function of the bare quark mass will enable us to explore the phase structure of the dual gauge theory and uncover a first order phase transition associated to the melting of the light mesons of the theory. 
\section{Magnetic Catalysis of Mass Generation in Holographic Gauge Theories.}

The phenomenon of dynamical flavor symmetry breaking catalyzed by an
arbitrarily weak magnetic field is known from
refs.~\cite{Gusynin:1994re, Gusynin:1994xp} and refs.~\cite{Klimenko:1990rh, Klimenko:1991he, Klimenko:1992ch}. This effect was shown to
be model independent and therefore insensitive to the microscopic
physics underlying the low energy effective theory. In particular the
infra-red (IR) description of the Goldstone modes associated with the
dynamically broken symmetry should be universal. One therefore expects
to be able to study this phenomenon using the holographic
formalism. 
 
\subsection{Mass Generation in the D3/D7 setup}
There are various ways in which one can study the breaking of the
chiral symmetry holographically. This has been studied in the past by
the deformation of AdS$_5\times S^5$ by a field corresponding to a
marginally irrelevant operator on the gauge theory side refs.~\cite{Babington:2003vm, Kruczenski:2003uq, Evans:2004ia}. In the
present case however we will stimulate the formation of a condensate
by turning on the magnetic components of the $U(1)$ gauge field of the
D7--branes $F_{\alpha\beta}$ (equivalent to exciting a pure gauge
$B-$field in the supergravity background). This $U(1)$ gauge field
corresponds to the diagonal $U(1)$ of the full $U(N_f)$ gauge symmetry
of the stack of D7--branes. Since the D7--branes wrap an infinite
internal volume, the dynamics of the $U(N_f)$ gauge field is frozen in
the four dimensional theory and the $U(N_f)$ gauge symmetry becomes a
global flavor symmetry $U(N_f)=U(1)_B\times SU(N_f)$. Therefore the
$U(1)$ gauge field that we consider corresponds to the gauged $U(1)_B$
baryon symmetry and the magnetic field that we introduce couples to
the baryon charge of the fundamental fields \cite{Myers:2007we}.
\subsubsection{Generalities}
The problem thus boils down to studying embeddings of probe D7--branes
in the AdS$_5\times S^5$ background parameterized as follows:
\begin{eqnarray}
ds^2&=&\frac{\rho^2+L^2}{R^2}[ - dx_0^2 + dx_1^2 +dx_2^2 + dx_3^2 ]+\frac{R^2}{\rho^2+L^2}[d\rho^2+\rho^2d\Omega_{3}^2+dL^2+L^2d\phi^2]\ ,\nonumber\\
d\Omega_{3}^2&=&d\psi^2+\cos^2\psi d\beta^2+\sin^2\psi d\gamma^2, \label{geometry1}\\
g_{s}C_{(4)}&=&\frac{u^4}{R^4}dx^0\wedge dx^1\wedge dx^2 \wedge dx^3;~~~e^\Phi=g_s;~~~R^4=4\pi g_{s}N_{c}\alpha'^2\ ,\nonumber
\end{eqnarray}
where $\rho, \psi, \beta,\gamma$ and $L,\phi$ are polar coordinates in
the transverse $\mathbb{R}^4$ and $\mathbb{R}^2$ planes respectively.

Here $x_{a=1..3},\rho,\psi,\beta,\gamma$ parameterize the world volume
of the D7--brane and the following ansatz is considered for its
embedding:
\begin{eqnarray}
\phi\equiv {\rm const}\ ,\quad L\equiv L(\rho)\nonumber \label{anzatsEmb}\ ,
\end{eqnarray}
leading to the following induced metric on its worldvolume:
\begin{equation}
d\tilde s=\frac{\rho^2+L(\rho)^2}{R^2}[ - dx_0^2 + dx_1^2 +dx_2^2
+dx_3^2]+\frac{R^2}{\rho^2+L(\rho)^2}[(1+L'(\rho)^2)d\rho^2+\rho^2d\Omega_{3}^2] \  .
\label{inducedMetric}
\end{equation}
The D7--brane probe is described by the DBI action:
\begin{eqnarray}
S_{\rm{DBI}}=-N_f\mu_{7}\int\limits_{{\cal M}_{8}}d^{8}\xi e^{-\Phi}[-{\rm det}(G_{ab}+B_{ab}+2\pi\alpha' F_{ab})]^{1/2}\  . \label{DBI}
\end{eqnarray}

Here $\mu_{7}=[(2\pi)^7\alpha'^4]^{-1}$ is the D7--brane tension,
$G_{ab}$ and $B_{ab}$ are the induced metric and $B$-field on the
D7--brane's world volume, while $F_{ab}$ is its world--volume gauge
field. A simple way to introduce a magnetic field is to consider
a pure gauge $B$--field along the $x_2,x_3$ directions:
\begin{equation}
B^{(2)}= Hdx_{2}\wedge dx_{3} \label{anzats}\ .
\end{equation}
Since $B_{ab}$ and $F_{ab}$ appear on equal footing in the DBI action,
the introduction of such a $B$-field is equivalent to introducing an
external magnetic field of magnitude $H/(2\pi\alpha')$ to the dual
gauge theory.

Though the full solution of the embedding can only be calculated
numerically, the large $\rho$ behaviour (equivalently the ultraviolet
(UV) regime in the gauge theory language) can be extracted
analytically:
\begin{equation}
L(\rho)=m+\frac{c}{\rho^2}+\cdots \ .
\end{equation}
As discussed in ref.~\cite{Kruczenski:2003uq}, the parameters $m$ (the
asymptotic separation of the D7- and D3- branes) and $c$ (the degree
of bending of the D7--brane in the large $\rho$ region) are related to
the bare quark mass $m_{q}=m/2\pi\alpha'$ and the fermionic condensate
$\langle\bar\psi\psi\rangle\propto -c$ respectively. It should be
noted that the boundary behavior of $L(r)$ really plays the role of
source and vacuum expectation value (vev) for the full ${\cal N}=2$
hypermultiplet of operators. In the present case, where supersymmetry
is broken by the gauge field configuration, we are only interested in
the fermionic bilinears and this will refer only to quarks, and not
their supersymmetric counterparts.

At this point it is convenient to introduce dimensionless parameters
$\tilde c=c/R^3H^{3/2}$ and $\tilde m=m/R\sqrt{H}$. By performing a
numerical shooting method from the infrared while varying the small
$\rho$ boundary value, $L(\rho\rightarrow 0)=L_{IR}$, we recover the
parametric plot presented in figure~\ref{fig:fig1}, the main result
explored in ref.~\cite{Filev:2007gb}.

\begin{figure}[h] 
   \centering
   \includegraphics[width=10cm]{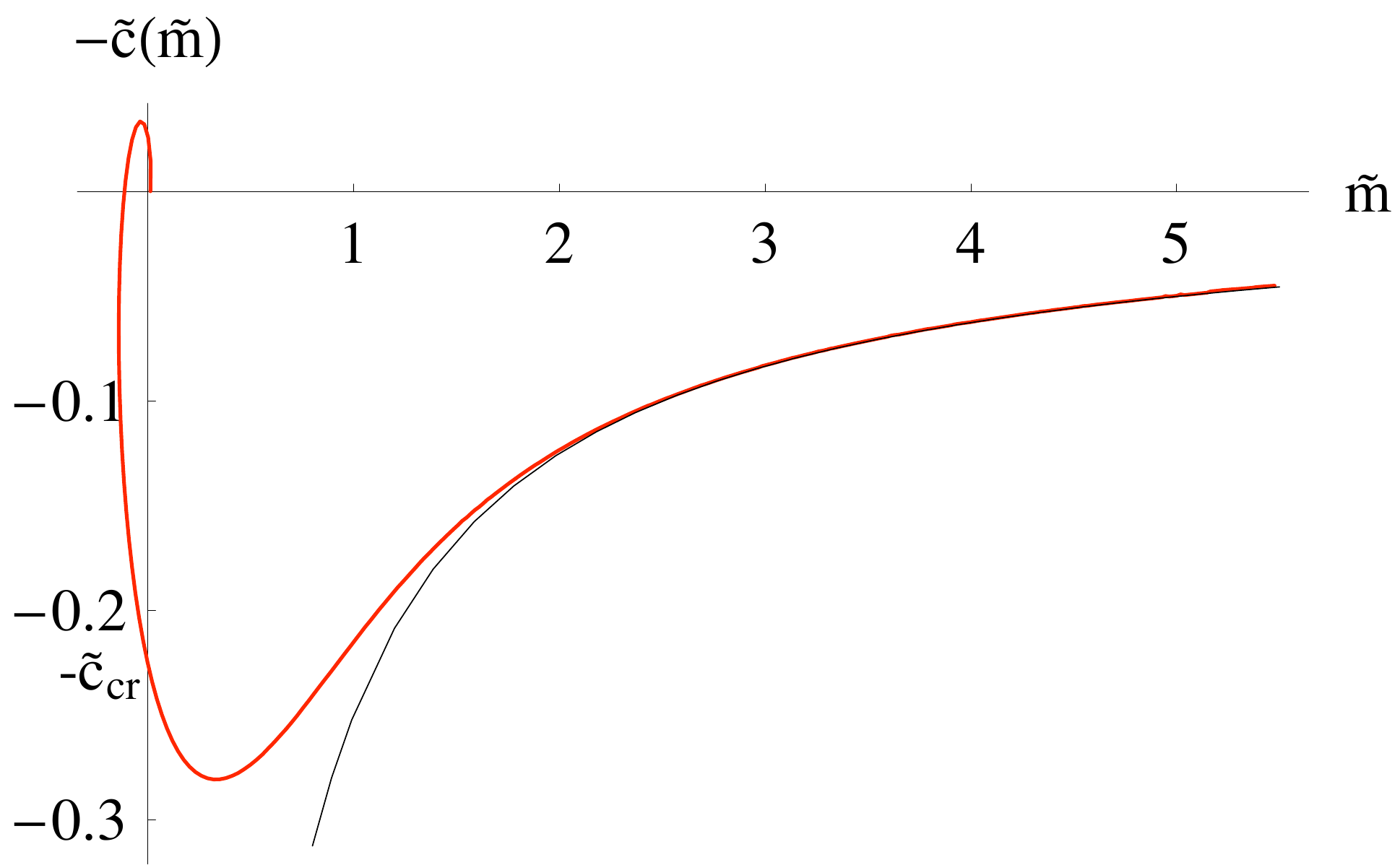}
   \caption{\small Parametric plot of $\tilde{c}$ against $\tilde{m}$
     for fundamental matter in the presence of an external magnetic
     field. The lower (black) line represents the curve $1/\tilde m$,
     fitting the large $\tilde{m}$ behavior. It is also evident that
     for the outer branch of the spiral, for $\tilde m=0$ the
     condensate, $\langle\bar\psi\psi\rangle$ is non-zero. The
     corresponding value of the condensate is $\tilde c_{\rm
       cr}=0.226$.}  \label{fig:fig1}
\end{figure}

The lower (black) curve corresponds to the analytic behavior of
$\tilde c(\tilde m)=1/\tilde m$ for large $\tilde m$. The most
important observation is that at $\tilde m=0$ there is a non-zero
fermionic condensate:
\begin{equation}
\langle\bar\psi\psi\rangle=-\frac{N_fN_c}{(2\pi\alpha')^3\lambda}c=-\frac{N_fN_c\tilde c_{\rm{cr}}}{(2\pi^2)^{3/4}\lambda^{1/4}}\left(\frac{H}{2\pi\alpha'}\right)^{3/2}\ .\label{cond}
\end{equation}

Where $\lambda=g_{YM}^2N_c$ is the 't Hooft coupling and $\tilde
c_{\rm{cr}}\approx 0.226$ is a numerical constant corresponding to the
$y$-intercept of the outer spiral from figure~\ref{fig:fig1}.
Equation (\ref{cond}) is telling us that the theory has developed a
negative condensate that scales as
$\left(\frac{H}{2\pi\alpha'}\right)^{3/2}$. This is not surprising,
since the theory is conformal in the absence of the scale introduced
by the external magnetic field.  The energy scale controlled by the
magnetic field, $\left(\frac{H}{2\pi\alpha'}\right)^{1/2}$, leads to an
energy density proportional to
$\left(\frac{H}{2\pi\alpha'}\right)^{2}$. In order to lower the
energy, the theory responds to the magnetic field by developing a
negative fermionic condensate.
 
Another interesting feature of the theory is the discrete--self--similar
structure of the equation of state ($\tilde c$ {\it vs.} $\tilde m$) in the
vicinity of the trivial $\tilde m=0$ embedding, namely the origin of
the plot from figure~\ref{fig:fig1} presented in figure~\ref{fig:fig2}.
\begin{figure}[h] 
   \centering
   \includegraphics[width=12cm]{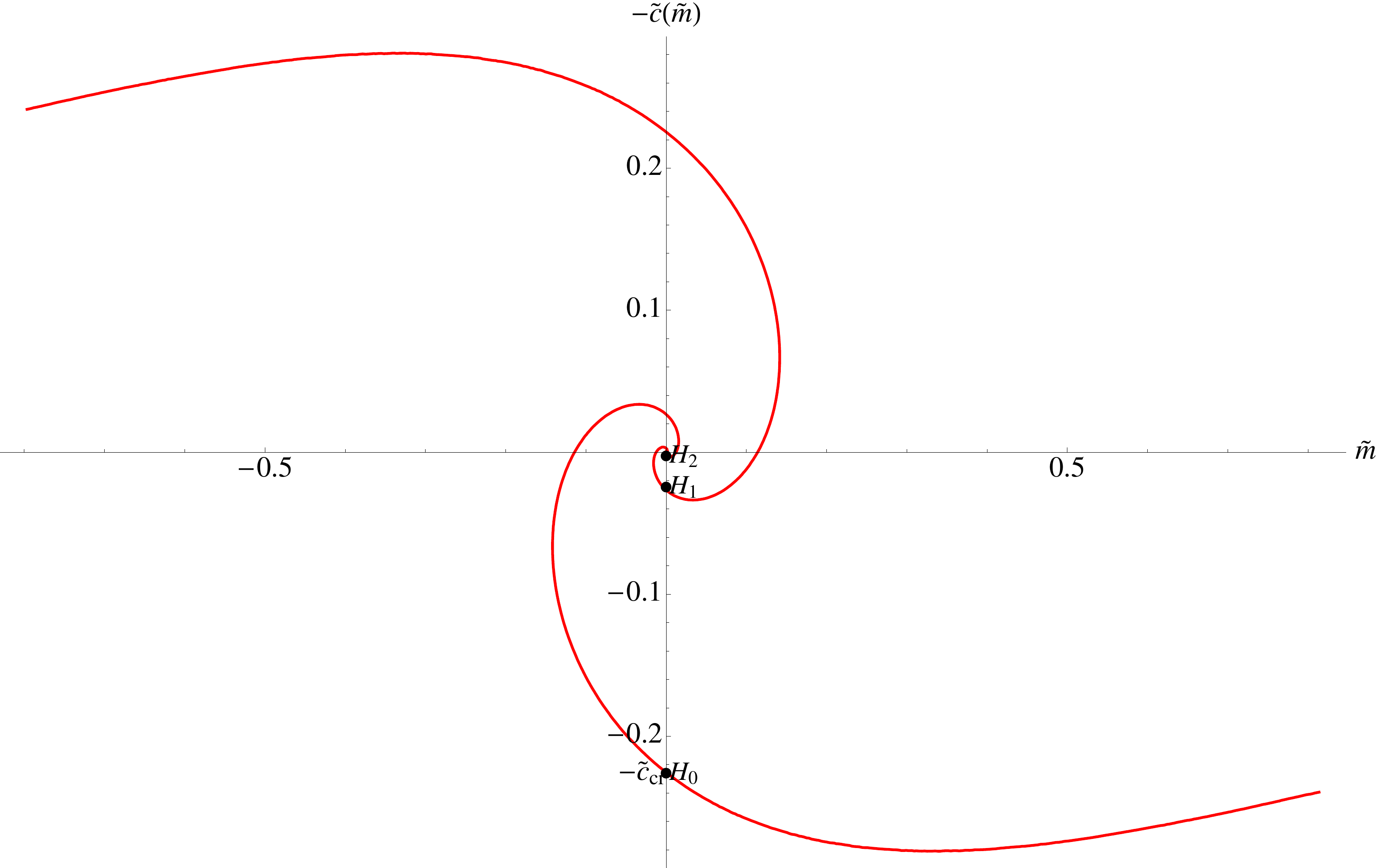}
   \caption{\small A magnification of figure~\ref{fig:fig1} shows the spiral behavior near the origin of the $(-\tilde c,\tilde m)$-plane. The second (left) spiral arm represents the $(\tilde m, -\tilde c)\rightarrow (-\tilde m,\tilde c)$ symmetry of the theory.}
   \label{fig:fig2}
\end{figure}

This double logarithmic structure has been analyzed in
ref.~\cite{Filev:2007qu}, where a study of the meson spectrum revealed
that only the outer branch of the spiral is tachyon free and
corresponds to a stable phase having spontaneously broken chiral
symmetry. In ref.~\cite{Filev:2009xp} it has been shown that an identical
structure is also present for the D3/D5 system and it has been demonstrated
 that this structure is a universal feature of the
magnetic catalysis of mass generation for gauge theories
holographically dual to Dp/Dq intersections.

A further result of
refs.~\cite{Filev:2007gb,Filev:2007qu,Erdmenger:2007bn} was the
detailed analysis of the light meson spectrum of the theory. In
ref.~\cite{Filev:2007gb} it was shown that the introduction of an
external magnetic field breaks the degeneracy of the spectrum studied
in ref.~\cite{Kruczenski:2003be}. This manifests itself as Zeeman
splitting of the energy levels. In the limit of zero quark mass, the
study also revealed the existence of a massless ``$\eta'$ meson"
corresponding to the spontaneously broken $U(1)_R$ symmetry. In the
next subsection we will review the study of the meson spectrum of the
theory. 
 
\subsubsection{ Meson spectrum}
\paragraph{General properties.}
To study the scalar meson spectrum one considers quadratic fluctuations \cite{Kruczenski:2003be} of the embedding of the D7--brane in the
transverse $(L,\phi)$-plane. It can be shown that because of the diagonal form of the metric the fluctuation modes along the $\phi$ coordinate decouple
from the one along $L$. However, because of the non--commutativity introduced by the $B$--field we may expect the scalar fluctuations to couple to the vector
fluctuations. This has been observed in ref.~\cite{Arean:2005ar}, where the authors considered the geometric dual to non--commutative super Yang Mills as well as in the studies performed in refs.~\cite{Filev:2007gb,Filev:2007qu,Erdmenger:2007bn}.

Let us proceed with obtaining the action for the fluctuations. To obtain the contribution from the DBI part of the action we consider the expansion:
\begin{eqnarray}
L=L_0(\rho)+2\pi\alpha'\chi,\label{fluct}\quad\phi=0+2\pi\alpha'\ ,
\end{eqnarray}
where $L_0(\rho)$ is the classical embedding of the D7--brane solution to the equation of motion derived from the action (\ref{DBI}). To second order in $\alpha'$ we have the following expression:
\begin{equation}
E_{ab}=E^{0}_{ab}+2\pi\alpha'E^{1}_{ab}+(2\pi\alpha')^2E^{2}_{ab}\ ,
\label{33}
\end{equation}
where $E^0,E^1,E^2$ are given by:
\begin{eqnarray}
&&\hspace{-0.6cm}E^{0}_{ab}=G_{ab}(\rho,L_0(\rho),\psi)+B_{ab},\\
&&\hspace{-0.6cm}E^{1}_{ab}=\frac{R^2{L_0}'}{\rho^2+L_0^2}\left(\partial_a\chi\delta_{b}^{\rho}+\partial_b\chi\delta_{a}^{\rho}\right)+\partial_{L_0}G_{ab}\chi+F_{ab}\label{ExpnMetrc}\nonumber\\
&&\hspace{-0.6cm}E^{2}_{ab}=\frac{R^2}{\rho^2+L_0^2}\left(\partial_{a}\chi\partial_{b}\chi+L_0^2\partial_{a}\Phi\partial_{b}\Phi\right)-\frac{2R^2L_0L_0'}{(\rho^2+L_0^2)^2}\left(\partial_{a}\chi\delta_{b}^{\rho}+\partial_{b}\chi\delta_{a}^{\rho}\right)\chi+\frac{1}{2}\partial_{L_0}^{2}G_{ab}\chi^2\ . \nonumber
\end{eqnarray}
Here $G_{ab}$ and $B_{ab}$ are the induced metric and B field on the D7--brane's world volume. Now we can substitute equation (\ref{ExpnMetrc}) into equation (\ref{DBI}) and expand to second order in $\alpha'$. It is convenient \cite{Arean:2005ar} to introduce
the following matrices:

\begin{equation}
||{E_{ab}^0}||^{-1}=S+J,
\end{equation}
where $S$ is diagonal and $J$ is antisymmetric:
\begin{eqnarray}
&||S^{ab}||&={\rm diag}\{-G_{11}^{-1},G_{11}^{-1},\frac{G_{11}}{G_{11}^{2}+H^2},\frac{G_{11}}{G_{11}^{2}+H^2},G_{\rho\rho}^{-1},G_{\psi\psi}^{-1},G_{\alpha\alpha}^{-1},G_{\beta\beta}^{-1}\}\ ,\label{S}\\
&J^{ab}&=\frac{H}{G_{11}^{2}+H^2}(\delta_{3}^{a}\delta_{2}^{b}-\delta_{3}^{b}\delta_{2}^{a})\ ,\label{J}\\
&G_{11}&=\frac{\rho^2+{L_0}^2}{R^2}\ ;~~~G_{\rho\rho}=R^2\frac{(1+{L'_0}^2)}{\rho^2+L_0^2}\ ;~~~G_{\psi\psi}=\frac{R^2\rho^2}{\rho^2+L_0^2};\nonumber\\
&G_{\alpha\alpha}&=\cos^2\psi G_{\psi\psi}\ ;~~~G_{\beta\beta}=\sin^2\psi G_{\psi\psi}\ .
\end{eqnarray}

Now it is straightforward to get the effective action. At first order in $\alpha'$ the action for the scalar fluctuations is the first variation of
the classical action (\ref{DBI}) and is satisfied by the classical equations of motion. Therefore we focus on the second order contribution from the DBI action.

After integrating by parts and taking advantage of the Bianchi identities for the gauge field, we end up with the following terms \cite{Filev:2007gb}.
For $\chi$:
\begin{eqnarray}
&{\cal L_{\chi}} \propto \frac{1}{2}\sqrt{-E^0}\frac{R^2}{\rho^2+{L_0}^2}\frac{S^{ab}}{1+{L'_0}^2}\partial_{a}\chi\partial_{b}\chi+\left[\partial_{L_0}^2\sqrt{-E^0}-\partial_\rho\left(\partial_{L_0}\sqrt{-E^0}\frac{L'_0}{1+{L'_0}^2}\right)\right]\frac{1}{2}\chi^2\ ,\nonumber\\ 
&\label{Schi}
\end{eqnarray}
and for $F$:
\begin{eqnarray}
{\cal L}_{F} \propto\frac{1}{4}\sqrt{-E^0}S^{aa'}S^{bb'}F_{ab}F_{a'b'}\ ,
\label{SF}
\end{eqnarray}
and the mixed $\chi$--$F$ terms:
\begin{eqnarray}
&{\cal L}_{F\chi}&\propto \frac{\sin2\psi}{2}f\chi F_{23}\ ,
\label{SFchi}
\end{eqnarray}
and for $\Phi$:
\begin{equation}
{\cal L}_{\Phi}\propto\frac{1}{2}\sqrt{-E^0}\frac{R^2{L_0}^2}{\rho^2+L_0^2} S^{ab}\partial_{a}\Phi\partial_{b}\Phi\ ,
\label{Sphi}
\end{equation}
where the function $f$ in (\ref{SFchi}) is given by:
\begin{eqnarray}
&f(\rho)&=\partial_{\rho}\left(g(\rho)\frac{L'_0}{1+{L_0}'^{2}}J^{23}\right)+J^{32}\partial_{L_0}g(\rho)+2g(\rho)J^{23}S^{22}\partial_{L_0}G_{11}\ ,\\
{\rm with}\quad&g(\rho)&=\frac{\sqrt{-E^0}}{\sin\psi\cos\psi}=\rho^3\sqrt{1+{L_0}'^2}\sqrt{1+\frac{R^4H^2}{(\rho^2+L_0^2)^2}}\ .\nonumber
\label{Fchi}
\end{eqnarray}

As can be seen from equation (\ref{SFchi}) the $A_2,A_3$ components of the gauge field couple to the scalar field $\chi$ via the function $f$. Note that since
for $\rho \rightarrow \infty$ and $L\rightarrow\infty$, we see that $J^{23}\rightarrow 0$, the mixing of the scalar and vector field decouples asymptotically. In order
to proceed with the analysis we need to take into account the contribution from the Wess-Zumino part of the action. The relevant terms to second order
in $\alpha'$ are \cite{Arean:2005ar}:
\begin{equation}
S_{WZ}=\frac{(2\pi\alpha')^2}{2}\mu_{7}\int{F_{(2)}\wedge F_{(2)}\wedge C_{(4)}}+(2\pi\alpha')\mu_{7}\int F_{(2)}\wedge B_{(2)}\wedge \tilde P[C_{(4)}]\ ,
\label{WZ}
\end{equation}
where $C_{(4)}$ is the background R-R potential given in equation~(\ref{geometry1}) and $\tilde C_{(4)}$ is the pull back of its magnetic dual. One can show
that:
\begin{equation}
\tilde C_{4}=-\frac{1}{g_{s}}\frac{R^4\rho^4}{(\rho^2+L^2)^2} \sin\psi\cos\psi d\psi\wedge d\alpha\wedge d\beta\wedge d\phi\ .
\end{equation}
Writing $\phi=2\pi\alpha'\Phi$  we write for the pull back $P[\tilde C_{(4)}]$:
\begin{equation}
P[\tilde C_{(4)}]=-\frac{2\pi\alpha'}{g_s}\frac{\sin2\psi}{2}K(\rho)\partial_{a}\Phi d\psi\wedge d\alpha\wedge d\beta\wedge dx^a,
\label{pulC}
\end{equation}
where we have defined:
\begin{equation}
K(\rho)=\frac{R^4\rho^4}{(\rho^2+L_0^2)^2}
\end{equation}
Now note that the $B$--field has components only along $x^2$ and $x^3$, therefore $dx^a$ in equation (\ref{pulC}) can be only $d\rho,dx^0$ or $dx^1$.
This will determine the components of the gauge field which can mix with $\Phi$. However, after integrating by parts and using the Bianchi identities
one can get the following simple expression for the mixing term:
\begin{equation}
-(2\pi\alpha')^2\frac{\mu_7}{g_s}\int d^8\xi\frac{\sin2\psi}{2}H\partial_{\rho}K\Phi F_{01}\  ,
\label{SPhiF}
\end{equation}
resulting in the following contribution to the complete lagrangian:\\
\begin{equation}
{\cal L}_{F\Phi}\propto \frac{\sin2\psi}{2}H\partial_{\rho}K\Phi F_{01}\  .
\label{mixing}
\end{equation}
Note that this means that only the $A_0$ and $A_1$ components of the gauge field couple to the scalar field $\Phi$. Next the contribution from the first term in (\ref{WZ}) is given by:
\begin{equation}
(2\pi\alpha')^2\frac{\mu_{7}}{g_s}\int d^8\xi\frac{(\rho^2+L_0^2)^2}{8R^4}F_{ab}F_{cd}\epsilon^{a b c d}\ ,
\end{equation}
where the indices take values along the $\rho,\psi,\alpha,\beta$ directions of the world volume. This will contribute to the equation of motion for
$A_{\rho},A_{\psi},A_{\alpha}$ and $A_{\beta}$ which do not couple to the scalar fluctuations. In this section we will be interested in analyzing the
spectrum of the scalar modes, therefore we will not be interested in the components of the gauge field transverse to the D3--branes world volume.
However, although there are no sources for these components from the scalar fluctuations, they still couple to the components along the D3--branes as
a result setting them to zero will impose constraints on the $A_{0}\dots A_3$. Indeed, from the equation of motion for the gauge field along the
transverse direction one gets:
\begin{equation}
\sum\limits_{a=0}^{3}S^{aa}\partial_b\partial_a{A_a}=0,~~b=\rho,\psi,\alpha,\beta\ ,
\label{Lorenzbr}
\end{equation}
(Here, no  summation on repeated indices is intended.)
However, the non--zero $B$--field explicitly breaks the Lorentz symmetry along the D3--branes' world volume. In particular we have:
\begin{eqnarray}
S^{00}=-S^{11}\ ,\quad S^{22}=S^{33}\neq S^{11}\ ,
\end{eqnarray}
which suggests that we should impose:
\begin{eqnarray}
-\partial_0{A_0}+\partial_1{A_1}=0\label{constrA}\ ,\quad\partial_2{A_2}+\partial_{3}{A_{3}}=0\ .
\end{eqnarray}
We will see that these constraints are consistent with the equations of motion for $A_{0}\dots A_3$. Indeed, with this constraint the equations of
 motion for $\chi$, $\Phi$ and $A_{\mu},\mu=0\dots 3$ are,
 for $\chi$:
\begin{eqnarray}
&&\frac{1+{L'_0}^2}{g}\partial_\rho\left(\frac{g\partial_\rho\chi}{(1+{L'_0}^2)^2}\right)+\frac{\Delta_{\Omega_3}\chi}{\rho^2}
+\frac{R^4}{(\rho^2+L_0^2)^2}\widetilde{\Box}\chi+\label{EMCHI}\\
&&+\frac{1+{L'_0}^2}{g}\left(-\partial_{\rho}\left(\frac{\partial{g}}{\partial L_0}\frac{L'_0}{1+{L'_0}^2}\right)+\frac{\partial^2{g}}{\partial L_0^2}\right)\chi+\frac{1+{L'_0}^2}{g}f
F_{23}=0\ ,\nonumber
\end{eqnarray}
and for $\Phi$:
\begin{eqnarray}
\frac{1}{g}\partial_\rho\left(\frac{{g}L_0^2\partial_\rho\Phi}{1+{L'_0}^2}\right)+\frac{L_0^2\Delta_{\Omega_3}\Phi}{\rho^2}+
\frac{R^4L_0^2}{(\rho^2+L_0^2)^2}\widetilde{\Box}\Phi-\frac{H\partial_\rho
K}{g}F_{01}=0\ ,\label{eqnPHI}
\label{eqnPhi}
\end{eqnarray}
and  finally for $A_a$:
\begin{eqnarray}
\frac{1}{g}\partial_\rho\left(\frac{{g}\partial_\rho{A_0}}{1+{L'_0}^2}\right)+\frac{\Delta_{\Omega_3}{A_0}}{\rho^2}+
\frac{R^4}{(\rho^2+L_0^2)^2}\widetilde{\Box}{A_0}+\frac{H\partial_\rho
K}{g}\partial_1\Phi&=&0\ ,\label{EqGauge}\\
\frac{1}{g}\partial_\rho\left(\frac{{g}\partial_\rho{A_1}}{1+{L'_0}^2}\right)+\frac{\Delta_{\Omega_3}{A_1}}{\rho^2}+
\frac{R^4}{(\rho^2+L_0^2)^2}\widetilde{\Box}A_1+\frac{H\partial_\rho
K}{g}\partial_0\Phi&=&0\ ,\nonumber\\
\frac{1}{g}\partial_\rho\left(\frac{{g}\partial_\rho{A_2}}{(1+{L'_0}^2)(1+\frac{R^4H^2}{(\rho^2+L_0^2)^2})}\right)+
\frac{R^4}{(\rho^2+L_0^2)^2+R^4H^2}\widetilde{\Box}{A_2}&+&\frac{\Delta_{\Omega_3}{A_2}}{\rho^2(1+\frac{R^4H^2}{(\rho^2+L_0^2)^2})}\nonumber\\  -\frac{f}{g}\partial_3\chi&=&0\ ,\nonumber\\
\frac{1}{g}\partial_\rho\left(\frac{{g}\partial_\rho{A_3}}{(1+{L'_0}^2)(1+\frac{R^4H^2}{(\rho^2+L_0^2)^2})}\right)+
\frac{R^4}{(\rho^2+L_0^2)^2+R^4H^2}\widetilde{\Box}{A_3}&+&\frac{\Delta_{\Omega_3}{A_3}}{\rho^2(1+\frac{R^4H^2}{(\rho^2+L_0^2)^2})}\nonumber\\ +\frac{f}{g}\partial_2\chi&=&0\  .\nonumber
\end{eqnarray}
We have defined:
\begin{equation}
\widetilde\Box=-\partial_0^2+\partial_1^2+\frac{\partial_2^2+\partial_3^2}{1+\frac{R^4H^2}{(\rho^2+L_0^2)^2}}\ .
\end{equation}
As one can see the spectrum splits into two independent components, namely the vector modes $A_0,A_1$ couple to the scalar fluctuations along $\Phi$, while the vector modes $A_2,A_3$ couple to the scalar modes along $\chi$. However, it is possible to further simplify the equations of motion for the gauge field. Focusing on the equations of motion for $A_0$ and $A_1$ in equation~(\ref{EqGauge}), it is possible to rewrite them as:
\begin{eqnarray}
&&\frac{1}{g}\partial_\rho\left(\frac{{g}\partial_\rho{F_{01}}}{1+{L'_0}^2}\right)+\frac{\Delta_{\Omega_3}{F_{01}}}{\rho^2}+
\frac{R^4}{(\rho^2+L_0^2)^2}\widetilde{\Box}{F_{01}}-\frac{H\partial_\rho
K}{g}(-\partial_0^2+\partial_1^2)\Phi=0\label{Elec}\\
&&\frac{1}{g}\partial_\rho\left(\frac{{g}\partial_\rho{(-\partial_0{A_0}+\partial_1{A_1})}}{1+{L'_0}^2}\right)+\frac{\Delta_{\Omega_3}{(-\partial_0{A_0}+\partial_1{A_1})}}{\rho^2}\\ \nonumber&&\hspace{6.3cm}+
\frac{R^4}{(\rho^2+L_0^2)^2}\widetilde{\Box}{(-\partial_0{A_0}+\partial_1{A_1})}=0\ .\nonumber
\end{eqnarray}
Note that the first constraint in (\ref{constrA}) trivially satisfies the second equation in (\ref{Elec}). In this way we are left with the first equation in (\ref{Elec}). Similarly one can show that using the second constraint in (\ref{constrA}) the equations of motion in (\ref{EqGauge}) for $A_2$ and $A_3$ boil down to a single equation for $F_{23}$:
\begin{eqnarray}
\frac{1}{g}\partial_\rho\left(\frac{{g}\partial_\rho{F_{23}}}{(1+{L'_0}^2)(1+\frac{R^4H^2}{(\rho^2+L_0^2)^2})}\right)&+&
\frac{R^4}{(\rho^2+L_0^2)^2+R^4H^2}\widetilde{\Box}{F_{23}}\label{teschi}\nonumber\\
&+&\frac{\Delta_{\Omega_3}{F_{23}}}{\rho^2(1+\frac{R^4H^2}{(\rho^2+L_0^2)^2})}+\frac{f}{g}(\partial_2^2+\partial_3^2)\chi=0\ .
\end{eqnarray}
Now let us proceed with a study of the fluctuations along $\Phi$.
\
\paragraph{Fluctuations along $\Phi$ for a weak magnetic field.}

To proceed, we have to take into account the $F_{01}$ component of the gauge field strength and solve the coupled equations of motion. Since the classical solution for the embedding of the D7--brane is known only numerically we have to rely again on numerics to study the meson spectrum. However, if we look at equation of motion derived from  (\ref{DBI}) we can see that the terms responsible for the non--trivial embedding of the D7--branes are of order $H^2$ \cite{Filev:2007gb}. On the other hand, the mixing of the scalar and vector modes due to the term (\ref{mixing}) appear at first order in $H$. Therefore it is possible to extract some non--trivial properties of the meson spectrum even at linear order in $H$ and as it turns out \cite{Filev:2007gb}, we can observe a Zeeman--like  effect: A splitting of states that is proportional to the magnitude of the magnetic field. Let us review the study performed in ref.~\cite{Filev:2007gb}.

To first order in $H$ the classical solution for the D7--brane profile is given by:
\begin{equation}
L_{0}=m+O(H^2),
\end{equation}
where $m$ is the asymptotic separation of the D3-- and D7--branes and corresponds to the bare quark mass. In this approximation the expressions for $g(\rho)$ and $\partial_{\rho}K(\rho)$, become:
\begin{eqnarray}
g(\rho)=\rho^3\ ,\quad\partial_{\rho}K(\rho)=\frac{4m^2R^4\rho^3}{(\rho^2+m^2)^3}\ ,\nonumber
\end{eqnarray}
and the equations of motion for $\Phi$ and $F_{01}$, equations (\ref{eqnPhi}) and (\ref{Elec}), simplify to:
\begin{eqnarray}
&&\frac{1}{\rho^3}\left(\rho^3m^2\partial_{\rho}\Phi\right)+\frac{m^2\Delta_{\Omega_{3}}}{\rho^2}\Phi+\frac{m^2R^4}{(\rho^2+m^2)^2}\Box\Phi-4H\frac{m^2R^4}{(\rho^2+m^2)^3}F_{01}=0\ ,\\
&&\frac{1}{\rho^3}\partial_\rho\left(\rho^3\partial_\rho F_{01}\right)+\frac{\Delta_{\Omega_3}{F_{01}}}{\rho^2}+\frac{R^4}{(\rho^2+m^2)^2}\Box{F_{01}}-4H\frac{m^2R^4}{(\rho^2+m^2)^3}{\cal P}^2\Phi=0\nonumber\ ,\\
&&{\rm where}\quad\Box=-\partial_{0}^2+\partial_{1}^2+\partial_{2}^2+\partial_{3}^2,~~~{\cal P}^2=-\partial_{0}^2+\partial_{1}^2\nonumber\ .
\label{simplified}
\end{eqnarray}
This system has become similar to the system studied in ref.~\cite{Arean:2005ar} and in order to decouple it we can define the fields:
\begin{equation}
\phi_{\pm}=F_{01}\pm m{\cal P}\Phi\ ,
\end{equation}
where ${\cal P}=\sqrt{-\partial_{0}^2+\partial_{1}^2}$. The resulting equations of motion are:
 \begin{equation}
\frac{1}{\rho^3}\partial_{\rho}(\rho^3\partial_{\rho}\phi_{\pm})+\frac{\Delta_{\Omega_{3}}}{\rho^2}\phi_{\pm}+\frac{R^4}{(\rho^2+m^2)^2}\Box\phi_{\pm}\mp H\frac{4R^4m}{(\rho^2+m^2)^3}{\cal P}\phi_{\pm}=0\ .
\label{eqnMotSimpl}
\end{equation}
Note that ${\cal P}^2$ is the Casimir operator in the $(x_{0},x_{1})$ plane only, while $\Box$ is the Casimir operator along the D3--branes' world volume. If we consider a plane wave $e^{ix.k}$ then we can define:
\begin{equation}
\Box e^{ix.k}=M^2 e^{ix.k},~~~{\cal P}^2 e^{ix.k}=M_{01}^2e^{ix.k}\nonumber\ ,
\end{equation}
and we have the relation:
\begin{equation}
M^2=M_{01}^2-k_{2}^2-k_{3}^2\ .
\end{equation}
The corresponding spectrum of $M^2$ is continuous in $k_{2}, k_{3}$. However, if we restrict ourselves to motion in the $(x_{0}, x_{1})$-plane the spectrum is discrete. Indeed, let us consider the ansatz:
\begin{equation}
\phi_{\pm}=\eta_{\pm}(\rho)e^{-ix_{0}k_{0}+ik_{1}x_{1}}\ .
\end{equation}
Then we can write:
 \begin{eqnarray}
&&\frac{1}{\rho^3}\partial_{\rho}(\rho^3\partial_{\rho}\eta_{\pm})+\frac{R^4}{(\rho^2+m^2)^2}M_{\pm}^2\eta_{\pm}\mp H\frac{4R^4m}{(\rho^2+m^2)^3}{M_{\pm}}\eta_{\pm}=0\ ,\label{eqn01Sm}\\
&&M_{\pm}\equiv{M_{01}}_{\pm}\nonumber\  .
\end{eqnarray}
Let us analyze equation (\ref{eqn01Sm}). It is convenient to introduce:
\begin{eqnarray}
&& y=-\frac{\rho^2}{m^2};~~~\bar M_{\pm}=\frac{R^2}{m}M_{\pm};~~~P_{\pm}(y)=(1-y)^{\alpha_{\pm}}\eta_{\pm};\label{varchan}\\
&& 2\alpha_{\pm}=1+\sqrt{1+\bar M_{\pm}^2};~~~\epsilon=H\frac{R^2}{m^2}\ .\nonumber
\end{eqnarray}
With this change of variables equation (\ref{eqn01Sm}) is equivalent to:
\begin{equation}
y(1-y)P_{\pm}''+2(1-(1-\alpha_{\pm})y)P'-\alpha_{\pm}(\alpha_{\pm-1})P_{\pm}\pm\epsilon\frac{\bar M_{\pm}}{(1-y)^2}P_{\pm}=0\ .
\label{hyper}
\end{equation}
Next we can expand:
\begin{eqnarray}
&&P_{\pm}=P_{0}\pm\epsilon P_{1}+O(\epsilon^2)\ ;~~~\alpha_{\pm}=\alpha_{0}\pm\epsilon\alpha_1+O(\epsilon^2)\ ;\label{expansions}\\
&&\bar M_{\pm}=\bar M_{0}\pm\epsilon\alpha_{1}\frac{(4\alpha_{0}+2)}{\bar M_{0}}+O(\epsilon^2)\ ;~~~\bar M_{0}=2\sqrt{\alpha_{0}(\alpha_{0}+1)} \ .\nonumber
\end{eqnarray}
leading to the following equations for $P_{0}$ and $P_{1}$:
\begin{eqnarray}
y(1-y)P_{0}''+2(1-(1-\alpha_{0})y)P_{0}'-\alpha_{0}(\alpha_{0}-1)P_{0}&=&0\ ,\label{EqPert}\\
y(1-y)P_{1}''+2(1-(1-\alpha_{0})y)P_{1}'-\alpha_{0}(\alpha_{0}-1)P_{1}&=&(\alpha_{1}(2\alpha_{0}-1)\nonumber \\ \quad &-&\frac{\bar M_{0}}{(1-y)^2})P_{0}-2\alpha_{1}y P_{0}'\ .\nonumber
\end{eqnarray}
The first equation in (\ref{EqPert}) is the hypergeometric equation and corresponds to the fluctuations in pure $AdS_{5}\times S^5$. It has the regular solution \cite{Kruczenski:2003be}:
\begin{equation}
P_{0}(y)=F(-\alpha_{0},1-\alpha_{0},2,y)\ .
\end{equation}
Furthermore, regularity of the solution for $\eta(\rho)$ at infinity requires \cite{Kruczenski:2003be} that $\alpha_{0}$  be discrete, and hence the spectrum of $\bar M_{0}$:
\begin{eqnarray}
 &&1-\alpha_{0}=-n,~~~n=0,1,\dots\label{degen}\\
 &&\bar M_{0}=2\sqrt{(n+1)(n+2)}\nonumber\ .
\end{eqnarray}
The second equation in (\ref{EqPert}) is an inhomogeneous hypergeometric equation. However, for the ground state, namely $n=0$, $P_{0}=F(-1,0,2,y)=1$ and one can easily get the solution:
\begin{equation}
P_{1}(y)=\frac{\bar M_{0}}{6}\ln(1-y)+(6\alpha_{1}-\bar M_{0})(\ln(-y)+\frac{1}{y})-\frac{\bar M_{0}}{4(1-y)}\ .
\end{equation}
On the other hand, using the definition of $P_{\pm}(y)$ in (\ref{varchan})  to first order in $\epsilon$ we can write:
\begin{equation}
\eta_{\pm}=\frac{1}{(1-y)^{\alpha_{0}}}\left(1\mp\epsilon\frac{\alpha_{1}}{\alpha_0}\ln(1-y)\right)\left(1\pm\epsilon P_{1}(y)\right)\ ,
\end{equation}
for the ground state $\alpha_{0}=1$ and we end up with the following expression for $\eta_{\pm}$:
\begin{equation}
\eta_{\pm}=\frac{1}{1-y}\pm\epsilon\frac{\bar M_{0}}{4(1-y)^2}\pm\frac{\epsilon}{1-y}(6\alpha_1-\bar M_0)\left(\ln(-y)+\frac{1}{y}-\frac{\ln(1-y)}{6}\right)\ .
\label{groundstate}
\end{equation}
Now if we require that our solution is regular at $y=0$ and goes as $1/\rho^2\propto1/y $ at infinity, the last term in (\ref{groundstate}) must vanish. Therefore we have:
\begin{equation}
\alpha_{1}=\frac{\bar M_0}{6}\ .
\label{correction}
\end{equation}
\begin{figure}[h] 
   \centering
   \includegraphics[ width=11cm]{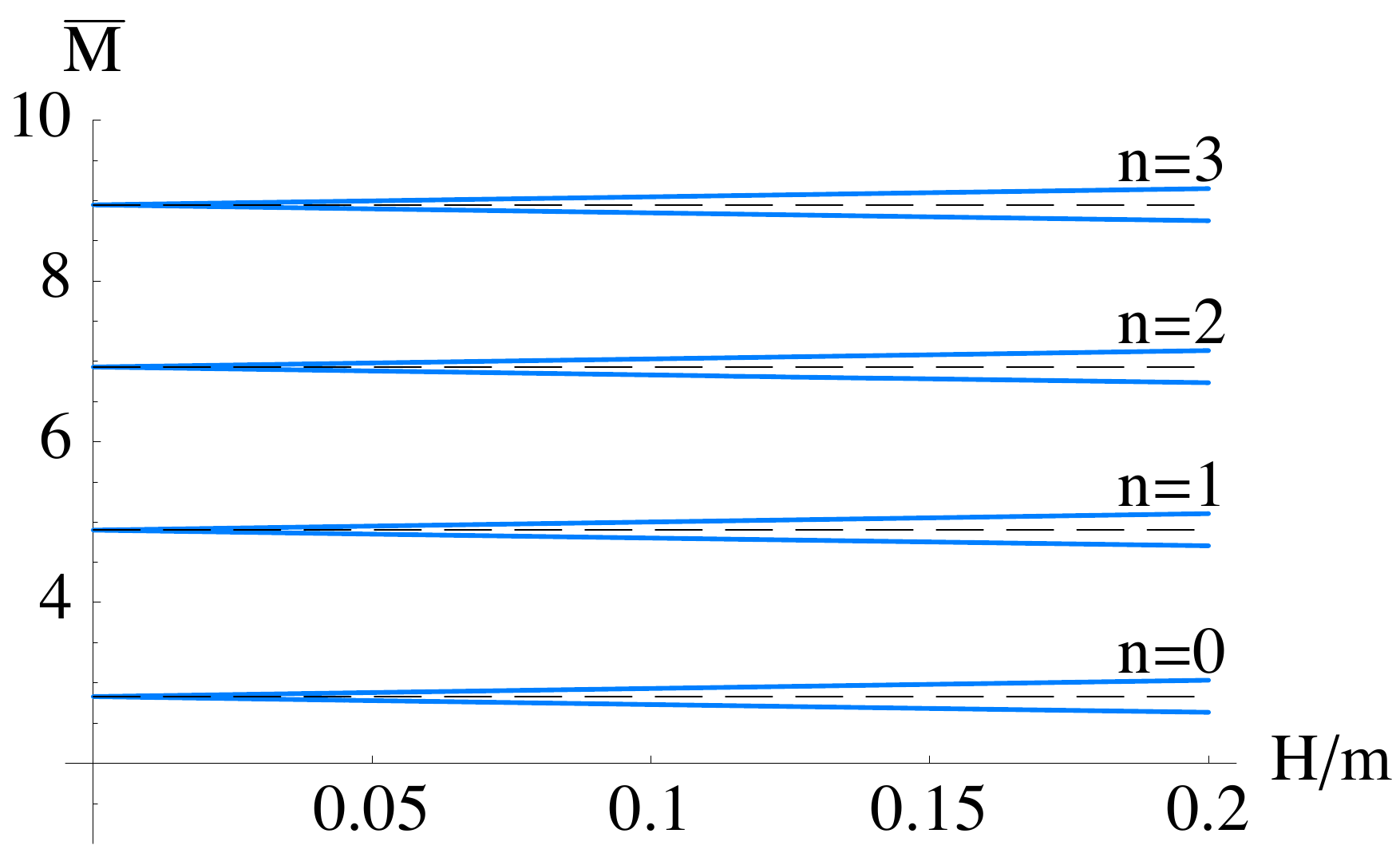}
   \caption{Plot of $\bar M=M{R^2}/{m}$ vs. $H/m$ for the first three states. The dashed black lines correspond to the spectrum given by equation~(\ref{degen}). }
   \label{fig: Zeeman}
\end{figure}
After substituting in (\ref{expansions}) and (\ref{varchan}) we end up with the following correction to the ground sate \cite{Filev:2007gb}:
\begin{equation}
M_{\pm}=M_{0}\pm\frac{H}{m}\ .
\label{Zeeman}
\end{equation}
We observe how the introduction of an external magnetic field breaks the degeneracy of the spectrum given by equation (\ref{degen}) and results in Zeeman splitting of the energy states, proportional to the magnitude of $H$. Although equation (\ref{Zeeman}) was derived using the ground state it is natural to expect that the same effect takes place for higher excited states. To demonstrate this it is more convenient to employ numerical techniques for solving equation (\ref{eqn01Sm}) and use the methods described in ref.~\cite{Babington:2003vm} to extract the spectrum. The resulting plot is presented in Figure~\ref{fig: Zeeman}. As expected we observe Zeeman splitting of the higher excited states. It is interesting that equation (\ref{Zeeman}) describes well not only the ground state, but also the first several excited states.

It turns out that one can easily generalize equation (\ref{Zeeman}) to the case of non--zero momentum in the $(x_{2},x_{3})$-plane. Indeed, if we start from equation (\ref{eqnMotSimpl}) and proceed with the ansatz:
\begin{equation}
\phi_{\pm}=\tilde {\eta}_{\pm}(\rho)e^{-ix.k}\ ,
\end{equation}
we end up with:
\begin{eqnarray}
&&\frac{1}{\rho^3}\partial_{\rho}(\rho^3\partial_{\rho}\tilde\eta_{\pm})+\frac{R^4}{(\rho^2+m^2)^2}M_{\pm}^2\tilde\eta_{\pm}\mp H\frac{4R^4m}{(\rho^2+m^2)^3}{{M_{01}}_{\pm}}\tilde\eta_{\pm}=0\ ,\label{eqn01Sm+mom}\\
&&{M_{01}}_{\pm}=\sqrt{M_{\pm}^2+k_{23}^2};~~~k_{23}\equiv\sqrt{k_2^2+k_3^2}\ .\nonumber
\end{eqnarray}
After going through the steps described in equations (\ref{varchan})-(\ref{groundstate}), equation (\ref{correction}) gets modified to:
\begin{equation}
\alpha_{1}=\frac{\bar M_0}{6}\sqrt{1+\frac{k_{23}^2}{M_0^2}}\ .
\end{equation}
Note that validity of the perturbative analysis suggests that $\alpha_1$ is of the order of $\alpha_0$ and therefore we can trust the above expression as long as $k_{23}$ is of the order of $M_0$. Now it is straightforward to obtain the correction to the spectrum \cite{Filev:2007gb}:
\begin{equation}
M_{\pm}=M_{0}\pm\frac{H}{m}\sqrt{1+\frac{k_{23}^2}{M_0^2}}\ .
\label{ZeemanGen}
\end{equation}
We see that the addition of momentum along the $(x_{2}-x_{3})$-plane enhances the splitting of the states. Furthermore, the spectrum depends continuously on $k_{23}$.

\paragraph{Fluctuations along $\Phi$ for a strong magnetic field.}

For strong magnetic field we have to take into account terms of order $H^2$, which means that we no longer have an expression for $L_{0}(\rho)$ in a closed form and we have to rely on numerical calculations. We consider the ans\"atz:
\begin{equation}
\Phi=e^{i(k_0x^0+k_1x^1)}h(\rho);~~~F_{01}=e^{i(k_0x^0+k_1x^1)}f(\rho)\label{anz1}\, ,
\end{equation}
and define:
\begin{equation}
M^2=k_0^2-k_1^2\ .
\end{equation}
The equations (\ref{eqnPHI}) and (\ref{Elec}) simplify to:
\begin{eqnarray}
&&\frac{1}{g}\partial_{\rho}\left(\frac{gL_0^2}{1+L'^2_0}\partial_{\rho}h\right)+\frac{R^4L_0^2}{(\rho^2+L_0^2)^2}M^2h-\frac{H\partial_{\rho}K}{g}f=0\ ,\label{EOMsmpl} \\
&&\frac{1}{g}\partial_{\rho}\left(\frac{g}{1+L'^2_0}\partial_{\rho}f\right)+\frac{R^4}{(\rho^2+L_0^2)^2}M^2f-\frac{M^2H\partial_{\rho}K}{g}h=0\ .\nonumber
\end{eqnarray}
Note that for large bare masses $m$ (and correspondingly large values
of $L$) the term proportional to the magnetic field is suppressed and
the meson spectrum should approximate to the result for the pure
AdS$_5\times S^5$ space-time case studied in
ref.~\cite{Kruczenski:2003be}, where the authors obtained the
following relation:
\begin{equation}
M_n=\frac{2m}{R^2}\sqrt{(n+1)(n+2)}\ ,\label{purespect}
\end{equation}
between the eigenvalue of the $n^{th}$ excited state $\omega_n$ and the bare mass $m$. If one imposes the boundary conditions:
\begin{equation}
h(\epsilon)=1;~~~h'(\epsilon)=0;~~~f(\epsilon)=1;~~~f'(\epsilon)=0\ , 
\end{equation}
the coupled system of differential equations can be solved
numerically \cite{Filev:2009xp}.  Then by requiring the functions $h(\rho)$ and $f(\rho)$
to be regular at infinity one can quantize the spectrum of the
fluctuations. It is also convenient to define the following
dimensionless parameter $\tilde M=MR/\sqrt{H}$. The resulting plot for
the first three excited states is presented in figure
\ref{fig:mesonD7} \cite{Filev:2009xp}. There is Zeeman splitting of the states due to the
magnetic field. (In the absence of the field there are three straight
lines emanating from the origin; these are split to form six curves.)
Also, at zero bare quark mass there is indeed a massless Goldstone
mode, appearing at the end of the lowest curve. Furthermore the plot
in figure~\ref{fig:zoomedD7} shows that for small bare quark mass one
can observe a characteristic $\tilde M\propto \sqrt{\tilde m}$
dependence. In the next section we will review the analysis of the Goldstone mode performed in ref.~\cite{Filev:2009xp}, where  an analytic proof of
the Gell-Mann--Oakes--Renner relation \cite{GellMann:1968rz}:
\begin{equation} 
M_{\pi}^2=-\frac{2\langle\bar\psi\psi\rangle}{f_{\pi}^2}m_q\ ,\label{GellMann1}
 \end{equation} 
 in the spirit of ref.~\cite{Kruczenski:2003uq}. has been obtained.

\begin{figure}[h] 
   \centering
   \includegraphics[width=9cm]{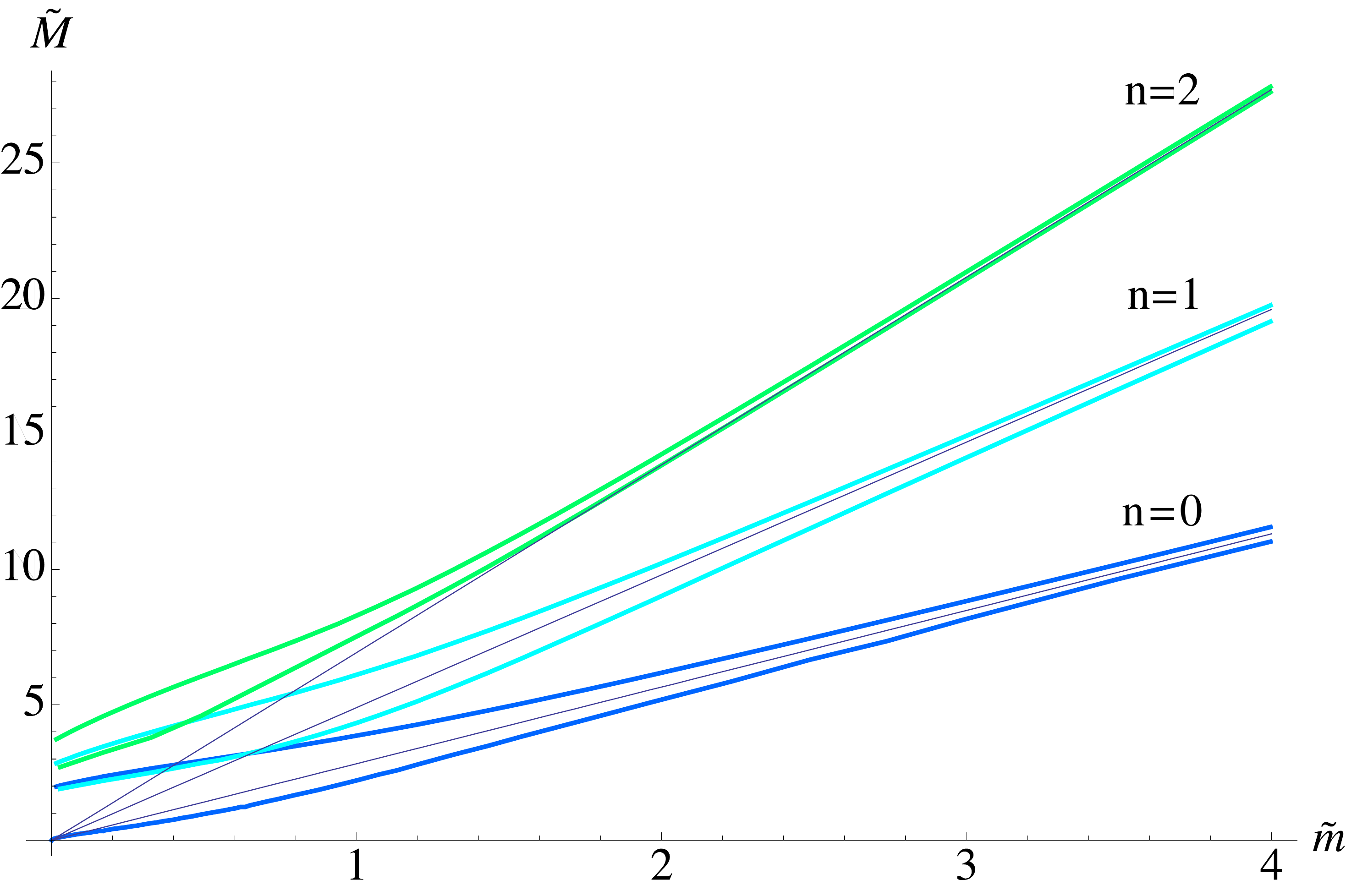}
   \caption{\small There is  Zeeman splitting of the states
     due to the magnetic field. In the absence of the field there are
     three straight lines emanating from the origin; these are split
     to form six curves. At zero bare quark
     mass (the end of the lowest curve) there is indeed a massless
     Goldstone mode. The straight lines correspond to the asymptotic
     AdS results.}
   \label{fig:mesonD7}
\end{figure}

\begin{figure}[h] 
   \centering
   \includegraphics[width=9cm]{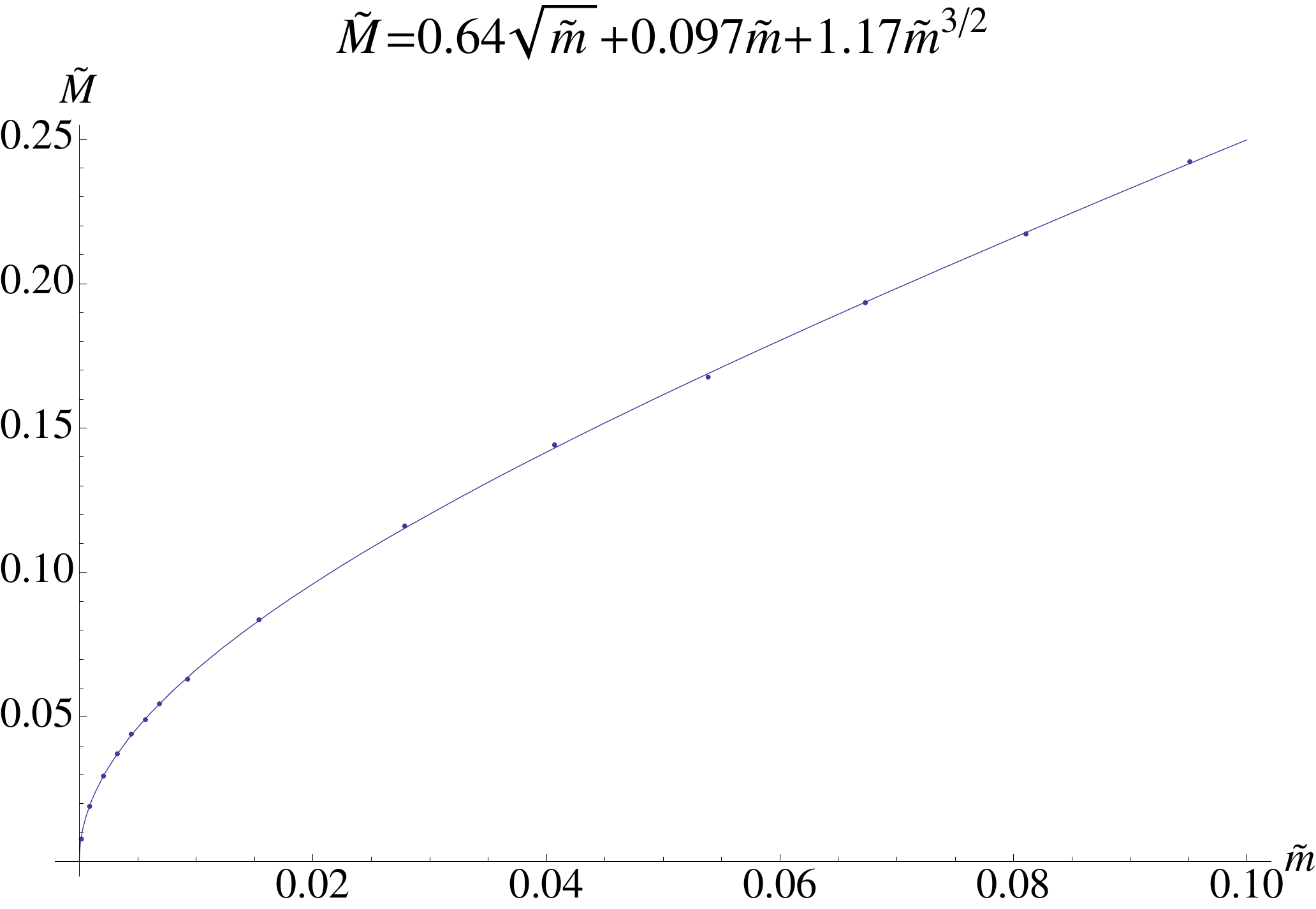}
   \caption{\small There is a characteristic $\tilde M\propto
     \sqrt{\tilde m}$ behavior at small bare quark mass.}
   \label{fig:zoomedD7}
\end{figure}


Furthermore we will generalize the ans\"atz (\ref{anz1})
to consider fluctuations depending on both the momentum along the
magnetic field $\vec k_{||}=(k_1,0,0)$ and the transverse momentum
$\vec k_{\perp}=(0,k_2,k_3)$ \cite{Filev:2009xp}:
\begin{equation}
\Phi=e^{i(\omega t+\vec k . \vec x)}h(\rho)\ ;~~~F_{01}=e^{i(\omega t+\vec k.\vec x)}f(\rho)\label{anz2}\ .
\end{equation}
 We also review the result of ref.~\cite{Filev:2009xp} that for small $\omega=k_0$ and $|{\vec k}|$ the following dispersion relation holds:
\begin{equation}
\omega(\vec k)^2=M^2+\vec k_{||}^2+\gamma\vec k_{\perp}^2\ ;~~~\omega=k_0\ ;~~~\vec k_{||}=(k_1,0,0)\ ;~~~\vec k_{\perp}=(0,k_2,k_3)\ ,\label{disp-rel}
\end{equation}
where $\gamma$ is a constant that has been determined.

\paragraph{The $M^2\propto m$ dependence.}

Using an approach similar to the one employed in ref.~\cite{Kruczenski:2003uq} the authors of ref.~\cite{Filev:2009xp} define:
\begin{eqnarray}
&&\Psi^2=\frac{gL_0^2}{1+L'^2_0}\ ;~~~\nu=R^4\frac{1+L'^2_0}{(\rho^2+L_0^2)^2}\ ;~~~\tilde\nu=R^4\frac{1+L'^2_0}{(\rho^2+L_0^2)^2}\frac{1}{1+\frac{R^4H^2}{(\rho^2+L_0^2)^2}}\ ,\label{PSIS}\\ 
&&\Psi_1=\Psi/L_0\ ;~~~\psi=h\Psi\ ;~~~\psi_1=f\Psi_1\ .\nonumber
\end{eqnarray}
The equations of motions (\ref{eqnPHI}) and (\ref{Elec}) can then be written in the compact form:
\begin{eqnarray}
&&\ddot\psi-\frac{\ddot\Psi}{\Psi}\psi=-(\omega^2-\vec k_{||}^2)\nu\psi+\vec k_{\perp}^2\tilde\nu\psi+\frac{H\partial_{\rho}K}{\Psi\Psi_1}\psi_1\ ,\label{EOMc}\\
&&\ddot\psi_1-\frac{\ddot\Psi_1}{\Psi_1}\psi_1=-(\omega^2-\vec k_{||}^2)\nu\psi_1+\vec k_{\perp}^2\tilde\nu\psi_1+\frac{H\partial_{\rho}K}{\Psi\Psi_1}(\omega^2-\vec k_{||}^2)\psi\nonumber \ .
\end{eqnarray}
Let us remind the reader that for large $\rho$, $L_0(\rho)$ has the behavior:
\begin{equation}
L_0\propto m+\frac{c}{\rho^2}+\cdots\ ,\label{exp2}
\end{equation}
Let us denote by $\bar L_0$ the classical embedding corresponding to
$(m=0,c=c_{\rm{cr}})$. It is relatively easy to verify that at
$m=0,\vec k_{\perp}=\vec 0$ and correspondingly $M^2=\omega^2-\vec
k_{||}^2=0$ the choice:
\begin{equation}
\psi=\bar\Psi\equiv\Psi|_{\bar L_0}\ ; ~~~\psi_1=0\ ,
\end{equation}
is a solution to the system (\ref{EOMc}).  Next we consider embeddings
corresponding to a small bare quark mass $\delta m$. This will
correspond to small nonzero values of $M^2$ and $\vec k_{\perp}^2$. It
is then natural to consider the following variations:
\begin{eqnarray}
&&\psi=\bar\Psi+\delta\psi \label{variation}\ ,\\
&&\psi_1=0+\delta\psi_1\nonumber\ ,
\end{eqnarray}
where $\delta\psi$ and $\delta\psi_1$ are of order $M^2$. Note that
$M$ corresponds to the mass of the ground state at $m_q=\delta
m/2\pi\alpha'$ and we are assuming that the variations of the wave
functions $\delta\psi$ and $\delta\psi_1$ are infinitesimal for
infinitesimal $m_q$. After expanding in equation~(\ref{EOMc}) we get
the following equations of motion:
\begin{eqnarray}
  &&\delta\ddot\psi-\frac{\ddot{\bar\Psi}}{\bar\Psi}\delta\psi-\delta\left(\frac{\ddot{\Psi}}{\Psi}\right)\bar\Psi=-(\omega^2-\vec k_{||}^2)\bar\nu\bar\Psi+\vec k_{\perp}^2\bar{\tilde v}\bar\Psi+\frac{H\partial_{\rho}K}{\bar\Psi_1\bar\Psi}\delta\psi_1\ ,\label{EOMVAR}\\
  &&\bar\Psi_1\delta\ddot{\psi_1}-\ddot{\bar{\Psi}}_1\delta\psi_1=H\partial_{\rho} K(\omega^2-k_{||}^2)\nonumber\ ,
\end{eqnarray}
where $\bar{\nu}=\nu|_{\bar L_0}$. The second equation in (\ref{EOMVAR}) can be integrated to give:
\begin{equation}
{\bar\Psi}_1\delta\dot\psi_1-\dot{\bar\Psi}_1\delta\psi_1=HK(\omega^2-k_{||}^2)+\rm{constant}\ .
\end{equation}
From the boundary conditions that $K|_{\rho=0}=0$ and
$\bar\Psi_1(0)=0,\dot{\bar\Psi}_1(0)=0$ we see that the constant of
integration is zero and arrive at:
\begin{equation}
\partial_{\rho}\left(\frac{\delta\psi_1}{\bar\Psi_1}\right)=\frac{HK(\omega^2-k_{||}^2)}{\bar\Psi_1^2}\ .\label{der}
\end{equation}
Next we multiply the first equation in (\ref{EOMVAR}) by $\bar\Psi$ and integrate along $\rho$ to obtain:
\begin{eqnarray}
&&(\omega^2-\vec k_{||}^2)\int\limits_0^{\infty}d\rho\bar\nu\bar\Psi^2-\vec k_{\perp}^2\int\limits_0^{\infty}d\rho\bar{\tilde\nu}\bar\Psi^2=-\int\limits_0^{\infty}(\bar\Psi\delta\ddot\psi-\ddot{\bar\Psi}\delta\psi)d\rho+\int\limits_0^{\infty}{\bar\Psi}^2\delta\left(\frac{\ddot\Psi}{\Psi}\right)d\rho+\label{integr}\\ 
&&+H\int\limits_0^{\infty}\frac{\partial_{\rho}K\delta\psi_1}{\bar\Psi_1}d\rho=-(\bar\Psi\delta\dot\psi-\dot{\bar\Psi}\delta\psi)\Big |_0^{\infty}+(\bar\Psi\delta\dot\Psi-\dot{\bar\Psi}\delta\Psi)\Big |_0^{\infty}-H\int\limits_0^{\infty}K\partial_\rho\left(\frac{\delta\psi_1}{{\bar\Psi}_1}\right)d\rho\nonumber\ ,
\end{eqnarray}
where the last term on the right-hand side of equation~(\ref{integr})
has been integrated by parts using the fact that $\delta\psi_1$ should
be regular at infinity. From the definition of $\bar\Psi$ it follows
that $\bar\Psi\propto \rho^{3/2}L_0(0)$ as $\rho\rightarrow 0$ and
$\bar\Psi\propto c/\rho^{1/2}$ as $\rho\rightarrow\infty$. This together with
the requirement that $\psi_1$ is regular at $\rho=0$ and vanishes at
infinity, suggests that the first term on the right-hand side of
equation~(\ref{integr}) vanishes. For the next term, we use the fact
that:
\begin{equation}
\delta\Psi=\rho^{3/2}\delta\left({\frac{1+\frac{H^2R^4}{(\rho^2+L^2_0)^2}  }{1+L'^2_0}  }\right)^{1/4}L_0+\rho^{3/2}\left(\frac{1+\frac{H^2R^4}{(\rho^2+L^2_0)^2}  }{1+L'^2_0} \right)^{1/4}\delta L_0\ ,
\end{equation}
and therefore obtain:
\begin{eqnarray}
&&\delta\Psi|_0=0;\quad \delta\dot\Psi |_0=0\ ,\\ \nonumber
&&\delta\Psi |_{\infty}\propto\rho^{3/2}\delta m\ ;\quad\delta\dot\Psi|_{\infty}\propto \frac{3}{2}\sqrt{\rho}\delta m\ .
\end{eqnarray}
The second term in equation~(\ref{integr}) then becomes:
\begin{equation}
(\bar\Psi\delta\dot\Psi-\dot{\bar\Psi}\delta\Psi)\Big |_0^{\infty}=2c\delta m\ .
\end{equation}
Finally using the equality in equation~(\ref{der}) we arrive at the result:
\begin{equation}
(\omega^2-\vec k_{||}^2)\int\limits_0^\infty d\rho\left\{\bar\nu{\bar\Psi}^2+\frac{H^2{\bar K}^2}{{{\bar\Psi}^2_1}}\right\}-\vec k_{\perp}^2\int\limits_0^{\infty}d\rho\bar{\tilde\nu}\bar\Psi^2=2c\delta m\ .
\end{equation}
Now we define \cite{Filev:2009xp}:
\begin{equation}
\gamma=\left(\int\limits_0^{\infty}d\rho\bar{\tilde\nu}\bar\Psi^2\right)/\left(\int\limits_0^\infty d\rho\left\{\bar\nu{\bar\Psi}^2+\frac{H^2{\bar K}^2}{{{\bar\Psi}^2_1}}\right\}\right)\ ,\label{GAMMA}
\end{equation}
and solve for $M^2$ from equation (\ref{disp-rel}) to obtain:
\begin{equation}
M^2\int\limits_0^\infty d\rho\left\{\bar\nu{\bar\Psi}^2+\frac{H^2{\bar K}^2}{{{\bar\Psi}^2_1}}\right\}=2c\delta m\ .\label{GMOR1}
\end{equation}
Equation (\ref{GMOR1}) suggests that the mass of the ``pion"
associated to the softly broken global $U(1)$ symmetry satisfies the
Gell-Mann--Oakes--Renner relation \cite{GellMann:1968rz}:
 \begin{equation} 
M_{\pi}^2=-\frac{2\langle\bar\psi\psi\rangle}{f_{\pi}^2}m_q\ .\label{GellMann2}
 \end{equation}
 In order to prove equation~(\ref{GellMann2}) we need to evaluate the
 effective coupling of the ``pion" $f_{\pi}^2$. Noting that $\delta
 m\propto m_q$ and $c\propto -\langle\bar\psi\psi\rangle$, we conclude
 that:
 \begin{equation}
f_{\pi}^2\propto\int\limits_0^{\infty}d\rho\left\{\bar\nu{\bar\Psi}^2+\frac{H^2{\bar K}^2}{{{\bar\Psi}^2_1}}\right\}\ .
\end{equation}
To verify the consistency of their analysis the authors of ref.~\cite{Filev:2009xp} compare the coefficient in equation (\ref{GMOR1}) to the numerically
determined coefficient $0.64$ from the plot in
figure~\ref{fig:zoomedD7}. Indeed from equation~(\ref{GMOR1}) it follows that:
\begin{equation}
\tilde M/\sqrt{\tilde m}=\left[\frac{1}{2\tilde c_{\rm{cr}}}\int\limits_0^{\infty}d\tilde\rho\left\{\bar{\hat\nu}{\bar{\hat\Psi}}^2+\frac{{\bar {\hat K}}^2}{{{\bar{\hat\Psi}}^2_1}}\right\}\right]^{-1/2}\approx 0.655\ ,
\end{equation}
where the dimensionless quantities:
\begin{equation}
\hat\nu=H^2\nu;~~\hat\Psi^2=\Psi^2/R^5H^{5/2};~~\hat\Psi_1^2=\Psi_1^2/R^3H^{3/2};~~\hat K=K/R^4
\end{equation}
have been defined. There is excellent agreement with the fit from
figure~\ref{fig:zoomedD7}.

Next we will obtain an effective four dimensional action for
the ``pion" and from this derive an exact expression for $f_\pi^2$.

\paragraph{Effective chiral action and $f_{\pi}^2$.}

In this section we will reduce the eight dimensional world-volume
action for the quadratic fluctuations of the D7--brane to an effective
action for the massless ``pion" associated to the spontaneously broken
global $U(1)$ symmetry. As rigid rotations along $\phi$ correspond to chiral rotations, (the
asymptotic value of $\phi$ at infinity corresponds to the phase of the
condensate in the dual gauge theory) the spectrum of $\Phi$ at zero
quark mass contains the Goldstone mode that we are interested in. This
is why we first integrate out the gauge field components $A_0$ and
$A_1$ and then dimensionally reduce to four dimensions \cite{Filev:2009xp}.

Furthermore as mentioned earlier, because of the magnetic field the
$SO(1,3)$ Lorentz symmetry is broken down to $SO(1,1)\times SO(2)$
symmetry. This is why in order to extract the value of $f_{\pi}^2$ we
consider excitations of $\Phi$ depending only on the $x_0,x_1$
directions and read off the coefficient in front of the kinetic term. The resulting on-shell effective action for $\Phi$ is \cite{Filev:2009xp}:
\begin{equation}
S^{\rm{eff}}=-{\cal{N}}\int d^4x\left[-(\partial_0\Phi)^2+(\partial_1\Phi)^2\right]\ ,
\end{equation}
where ${\cal N}$ is given by:
\begin{equation}
{\cal N}=(2\pi\alpha')^2\frac{\mu_7}{g_s}N_f\pi^2\int\limits_0^{\infty}d\rho\left\{\bar\nu{\bar\Psi}^2+\frac{H^2{\bar K}^2}{{{\bar\Psi}^2_1}}\right\}\ .
\end{equation}
We refer the reader to the Appendix of ref.~\cite{Filev:2009xp} for a detailed derivation of the 4D
effective action $S^{\rm{eff}}$.

We have defined $\Phi$ {\it via} $\phi=(2\pi\alpha')\Phi$, where $\phi$ corresponds to rotations in the transverse $\IR^2$ plane and is the angle of chiral rotation in the dual gauge theory. The chiral Lagrangian is then given by: 
\begin{equation}
S^{\rm{eff}}=-(2\pi\alpha')^2\frac{f_{\pi}^2}{4}\int d^4x \partial_{\mu}\Phi\partial^{\mu}\Phi\ ;~~~\mu=0\,\, {\rm or} \,\, 1\, ,
\end{equation}  
and therefore:
\begin{equation}
f_{\pi}^2=N_f 4\pi^2\frac{\mu_7}{g_s}\int\limits_0^\infty d\rho\left(\bar\nu\bar\Psi^2+\frac{H^2K^2}{{\bar\Psi}_1^2}\right)\label{Fpi}\ .
\end{equation}

The D7--brane charge in equation~(\ref{Fpi}) is given by
$\mu_7=[(2\pi)^7\alpha'^4]^{-1}$ and the overall prefactor in equation
(\ref{Fpi}) can be written as $N_fN_c/2(2\pi\alpha')^4\lambda$. Now,
recalling the expressions for the fermionic condensate,
equation~(\ref{cond}), and the bare quark mass, $m_q=m/2\pi\alpha'$,
one can easily verify that equation (\ref{GMOR1}) is indeed the
Gell-Mann--Oakes--Renner relation:
 \begin{equation} 
M_{\pi}^2=-\frac{2\langle\bar\psi\psi\rangle}{f_{\pi}^2}m_q\ .\label{GellMann3}
 \end{equation}

 It turns out that for small momenta $\vec k_{||}, \vec k_{\perp}$ and
 small mass $M_{\pi}^2$ one can obtain the following more general
 effective 4D action (see Appendix  A of ref.~\cite{Filev:2009xp} for a detailed derivation):
\begin{equation}
S_{\rm{eff}}=-{\cal N}\int d^4x\left\{[-(\partial_0\tilde\Phi)^2+(\partial_1\tilde\Phi)^2]+\gamma[(\partial_2\tilde\Phi)^2+(\partial_3\tilde\Phi)^2]-\frac{2\langle\bar\psi\psi\rangle}{f_{\pi}^2}m_q\tilde\Phi^2\right\}+\cdots\ ,\label{effacttext}
\end{equation}
where $\gamma$ is defined in equation (\ref{GAMMA}). As one can
see, the action (\ref{effacttext}) is the most general quadratic
action consistent with the $SO(1,1)\times SO(2)$ symmetry and suggests
that pseudo Goldstone bosons satisfy the dispersion relation
(\ref{disp-rel}).


\subsection{Mass Generation in the D3/D5 setup}
In this section we review the results of ref.~\cite{Filev:2009xp} providing a holographic description of the magnetic
catalysis of chiral symmetry breaking in $1+3$ dimensional $SU(N_c)$
$\N=4$ supersymmetric Yang-Mills theory coupled to~$N_f$ $\N=2$
fundamental hypermultiplets confined to a $1+2$ dimensional
defect.  In ref.~\cite{Filev:2009xp} it has been shown that the system develops a dynamically
generated mass and negative fermionic condensate leading to a
spontaneous breaking of a global $SO(3)$ symmetry down to a $U(1)$
symmetry. On the gravity side this symmetry corresponds to the
rotational symmetry in the transverse $\IR^3$. The authors of \cite{Filev:2009xp} had shown that the 1+2 dimensional nature of
the defect theory leads to a coupling of the transverse scalars
corresponding to the coset generators and as a result there is only a
single Goldstone mode. It has also been shown that the characteristic
$M_{\pi}\propto\sqrt{m}$ Gell-Mann--Oakes--Renner relation is modified to a
linear $M_{\pi}\propto m$ behavior. These features
has been understood from a low energy effective theory point of
view. Indeed in $1+2$ dimensions the effect of the magnetic field is
to break the $SO(1,2)$ Lorentz symmetry down to $SO(2)$ rotational
symmetry and as a result the theory is non-relativistic. A single time
derivative chemical potential term is allowed. It is this
term that is responsible for the modified counting rule of the number
of Goldstone bosons \cite{Nielsen:1975hm} and leads to a quadratic
dispersion relation as well as to the modified linear
Gell-Mann--Oakes--Renner relation. 


\subsubsection{Generalities}

Let us consider the AdS$_5\times S^5$ supergravity background
(\ref{geometry1}) and introduce the following parameterization:
\begin{eqnarray}
ds^2&=&\frac{u^2}{R^2}[ - dx_0^2 + dx_1^2 +dx_2^2 + dx_3^2 ]+\frac{R^2}{u^2}[dr^2+r^2d\Omega_{2}^2+dl^2+l^2d\tilde\Omega_2^2]\ ,\\
u^2&=&r^2+l^2\ ;~~d\Omega_2^2=d\alpha^2+\cos^2\alpha d\beta^2\ ;~~d\tilde\Omega_2^2=d\psi^2+\cos^2\psi d\phi^2\ .\nonumber
\end{eqnarray}
We have split the transverse $\IR^6$ to $\IR^3\times\IR^3$ and
introduced spherical coordinates $r,\Omega_2$ and $l,\tilde\Omega_2$
in the first and second $\IR^3$ planes respectively. Next we introduce
a stack of probe $N_f$ D5--branes extended along the $x_0,x_1,x_2$
directions, and filling the $\IR^3$ part of the geometry parameterized
by $r,\Omega_2$. As mentioned above on the gauge theory side this
corresponds to introducing $N_f$ fundamental $\N=2$ hypermultiplets
confined on a $1+2$ dimensional defect. The asymptotic separation of
the D3 and D5 --branes in the transverse $\IR^3$ space parameterized
by $l$ corresponds to the mass of the hypermultiplet. In the following
we will consider the following ans\"atz for a single D5--brane:
\begin{equation}
l=l(r)\ ;~~~\psi=0\ ;~~~\phi=0\ .\label{anz2-a}
\end{equation}
The asymptotic separation $m=l(\infty)$ is related to the bare mass of
the fundamental fields {\it via} $m_q=m/2\pi\alpha'$. If the D3 and D5
branes overlap, the fundamental fields in the gauge theory are
massless and the theory has a global $SO(3)\times SO(3)$
symmetry. Clearly a non-trivial profile of the D5--brane $l(r)$ in the
transverse $\IR^3$ would break the global symmetry down to
$SO(3)\times U(1)$, where $U(1)$ is the little group in the transverse
$\IR^3$. If the asymptotic position of the D5--brane vanishes ($m=0$)
this would correspond to a spontaneous symmetry breaking, the non-zero
separation $l(0)$ on the other hand would naturally be interpreted as
the dynamically generated mass of the theory.

Note that the D3/D5 intersection is T--dual to the D3/D7 intersection
from the previous section and thus the system is supersymmetric. The
D3 and D5 --branes are BPS objects and there is no attractive potential
for the D5--brane, hence the D5--brane has a trivial profile $l\equiv
const$. However a non-zero magnetic field will break the supersymmetry
and as we are going to demonstrate, the D5--brane will feel an
effective repulsive potential that will lead to dynamical mass
generation. In order to introduce a magnetic field perpendicular to
the plane of the defect, we consider a pure gauge $B$-field in the
$x_1,x_2$ plane given by:
\begin{equation}
B=Hdx_1\wedge dx_2\ .
\end{equation}
This is equivalent to turning on a non-zero value for the $0,1$
component of the gauge field on the D5--brane. The magnetic field
introduced into the dual gauge theory in this way has a magnitude
$H/2\pi\alpha'$. The D5--brane embedding is determined by the DBI
action:
\begin{eqnarray}
S_{\rm{DBI}}=-N_f\mu_{5}\int\limits_{{\cal M}_{6}}d^{6}\xi e^{-\Phi}[-{\rm det}(G_{ab}+B_{ab}+2\pi\alpha' F_{ab})]^{1/2}\  . \label{DBI2}
\end{eqnarray}
Where $G_{ab}$ and $B_{ab}$ are the pull-back of the metric and the $B$-field respectively and $F_{ab}$ is the gauge field on the D5--brane.

With the ans\"atz (\ref{anz2-a}) the lagrangian is given by:
\begin{equation}
{\cal L}\propto r^2\sqrt{1+l'^2}\sqrt{1+\frac{R^4H^2}{(r^2+l^2)^2}}\ .\label{lagr2}
\end{equation}
From this it is trivial to solve the equation of motion for $l(r)$
numerically, imposing $l(0)=l_{in}$ and $l'(0)$ as initial
conditions. Clearly, at large $r$ the lagrangian (\ref{lagr2})
asymptotes to that at zero magnetic field and hence we have the
asymptotic solution \cite{Mateos:2007vn}:
\begin{equation}
l(r)=m+\frac{c}{r}+\cdots\ ,\label{exp3}
\end{equation}
where $c\propto\langle\bar\psi\psi\rangle$ the condensate of the fundamental fields. 

\paragraph{Spontaneous symmetry breaking}

Before solving the equation of motion it is convenient to introduce dimensionless variables:
\begin{equation}
\tilde r=r/R\sqrt{H}\ ;~~~\tilde l=l/R\sqrt{H}\ ;~~~\tilde m=m/R\sqrt{H}\ ;~~~\tilde c=c/R^2H\ .
\end{equation}
The lagrangian (\ref{lagr2}) can then be written as:
\begin{equation}
{\cal L}\propto {\tilde r}^2\sqrt{1+{\tilde l}'^2}\sqrt{1+\frac{1}{(\tilde r^2+\tilde l^2)^2}}\ .
\end{equation}
The corresponding equation of motion is:
\begin{equation}
\partial_{\tilde r}\left(\frac{\tilde r^2l'}{\sqrt{1+\tilde l'^2}}\frac{\sqrt{1+(\tilde r^2+\tilde l^2)^2}}{
(\tilde r^2+\tilde l^2)}\right)=-2\frac{\tilde r^2\tilde l\sqrt{1+\tilde l'^2}}{(\tilde r^2+\tilde l^2)^2\sqrt{1+(\tilde r^2+\tilde l^2)^2}}\ .\label{EOM2}
\end{equation}
Before solving equation~(\ref{EOM2}) it will be useful to extract the
asymptotic behavior of $\tilde c(\tilde m)$ at large $\tilde m$. To
this end we use that at large $\tilde m$ the separation $\tilde
l(\tilde r)\approx \tilde m=\rm{const}$. The equation of motion then
simplifies to:
\begin{equation}
\partial_{\tilde r}(\tilde r^2\tilde l')=-\frac{2\tilde r^2\tilde m}{(\tilde r^2+\tilde m^2)^3}\ ,
\end{equation}
and hence:
\begin{equation}
\tilde r^2\tilde l'=-2\tilde m\int\limits_0^{\tilde r}d\tilde r\frac{\tilde r^2}{(\tilde r^2+\tilde m^2 )^2}\ .
\end{equation}
Using the expansion (\ref{exp3}) one can verify that:
\begin{equation}
\lim_{\tilde r \rightarrow +\infty}\tilde r^2\tilde l'=\tilde c=2\tilde m\int\limits_0^{\infty}d\tilde r\frac{\tilde r^2}{(\tilde r^2+\tilde m^2)^3}=\frac{\pi}{8\tilde m^2}\label{largm}\ .
\end{equation}
Equation (\ref{largm}) can thus be used as a check of the accuracy of
our numerical results. Indeed the numerically generated plot of
$-\tilde c$ {\it vs.} $\tilde m$ is presented in figure
\ref{fig:fig3}. The most important observation is that at zero bare
mass $\tilde m$ the theory has developed a negative condensate
$\langle\bar\psi\psi\rangle\propto -\tilde c_{\rm{cr}}\approx
-0.59$. It can also be seen that for large $\tilde m$ the numerically
generated plot is in good agreement with equation (\ref{largm})
represented by the lower (black) curve. Another interesting feature of the
equation of state is the spiral structure near the origin of the
parameter space analogous to the one presented in figure
\ref{fig:fig2} for the case of the D3/D7 system. 
\begin{figure}[h] 
   \centering
   \includegraphics[width=9cm]{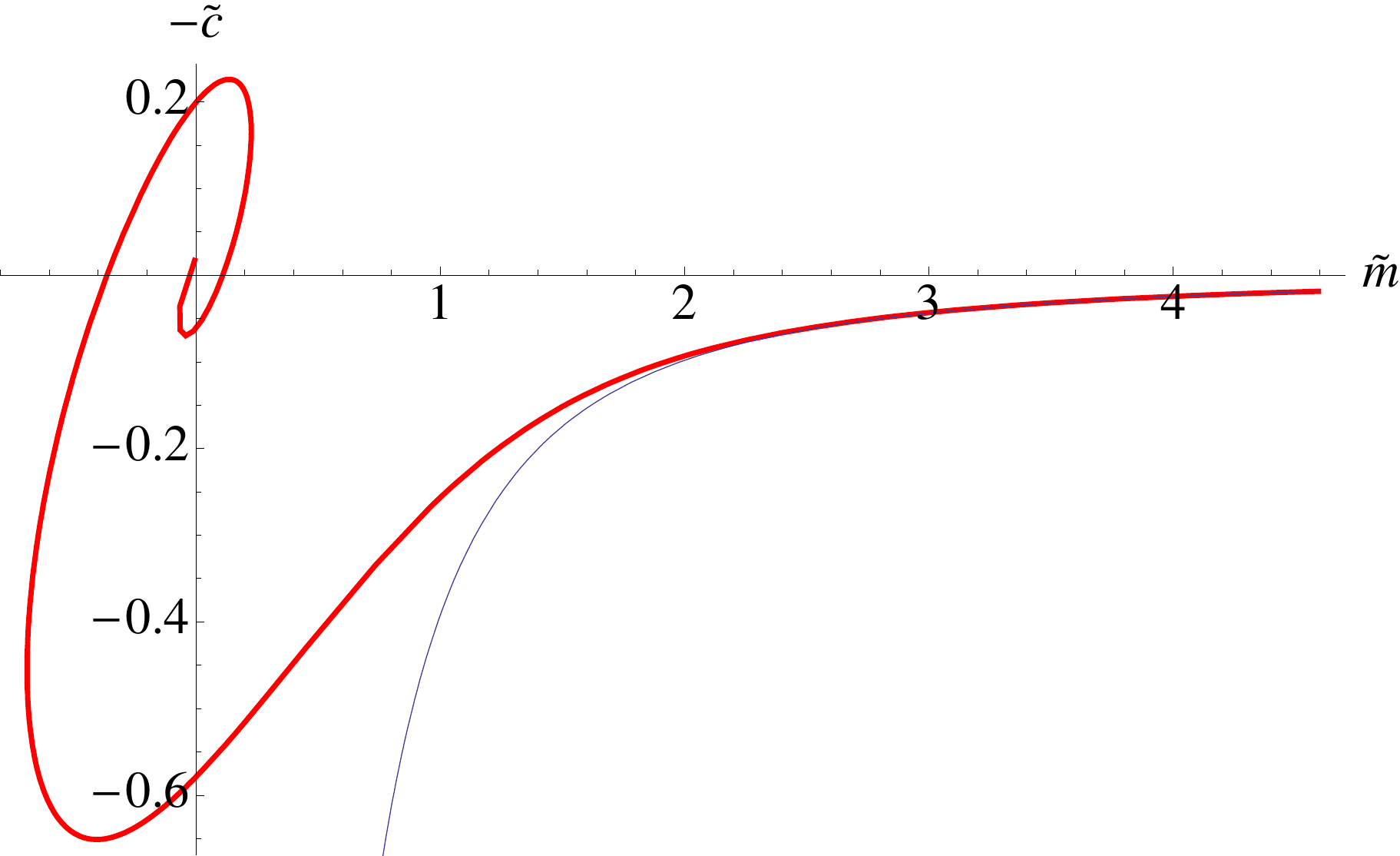}
   \caption{\small A plot of $-\tilde c$ {\it vs.} $\tilde m$. At zero
     bare mass $\tilde m=0$ the theory has developed a negative
     condensate $\langle\bar\psi\psi\rangle\propto -\tilde
     c_{\rm{cr}}\approx -0.59$. For large $\tilde m$ there is 
     excellent  agreement with equation~(\ref{largm}), as  represented by the
     lower (blue) curve. }
   \label{fig:fig3}
\end{figure}

In order to show that the global $SO(3)$ symmetry is indeed
spontaneously broken we need to study the free energy of the
theory. Indeed the existence of the spiral structure suggests that
there is more than one phase at zero bare mass, corresponding to the
different $y$-intercepts of the $-\tilde c$ {\it vs.} $\tilde m$
plot. We will demonstrate below that the lowest positive branch of the
curve presented in figure~\ref{fig:fig3} is the stable one.

Following ref.~\cite{Mateos:2007vn} we will identify the regularized
wick rotated on-shell action of the D5--brane with the free energy of
the theory. Let us introduce a cut-off at infinity, $r_{\max}$, The
wick rotated on-shell action is given by:
\begin{equation}
S=N_f\frac{\mu_5}{g_s}4\pi V_3R^3H^{3/2}\int\limits_0^{\tilde r_{\rm{max}}} d\tilde r\tilde r^2\sqrt{1+\tilde l'^2}\sqrt{1+\frac{1}{(\tilde r^2+\tilde l^2)^2}}\ ,\label{actionwick}
\end{equation}
where $V_3=\int d^3x$ and $\tilde l(\tilde r)$ is the solution of
equation (\ref{EOM2}). It is easy to verify, using the expansion from
equation (\ref{exp3}),  that the integral in equation~(\ref{actionwick})
has the following behavior at large $\tilde r_{max}$:
\begin{equation} 
\int\limits_0^{\tilde r_{\rm{max}}} d\tilde r\tilde r^2\sqrt{1+\tilde l'^2}\sqrt{1+\frac{1}{(\tilde r^2+\tilde l^2)^2}}=\frac{1}{3}{r_{\rm{max}}^3}+O\left(\frac{1}{r_{\rm{max}}}\right)\ .
\end{equation}
It is important that in these coordinates the divergent term is
independent of the field $\tilde l$, it is therefore possible to
regularize the on-shell action by subtracting the free energy of the
$\tilde l\equiv 0$ embedding. The resulting regularized expression for
the free energy is:
\begin{equation}
F=S_{\rm{reg}}=N_f\frac{\mu_5}{g_s}4\pi V_3R^3H^{3/2}\tilde I_{\rm{D5}}\ ,
\end{equation}
where
\begin{equation}
\tilde I_{\rm{D5}}=\int\limits_0^{\infty}d\tilde r\left[\tilde r^2\sqrt{1+\tilde l'^2}\sqrt{1+\frac{1}{(\tilde r^2+\tilde l^2)^2}} -\sqrt{1+\tilde r^4}\right]\, .
\end{equation}

A plot of $\tilde I_{\rm{D5}}$ {\it vs.} $|\tilde m|$ is presented in
figure~\ref{fig:fig4}. The states from the lowest
positive branch in figure~\ref{fig:fig4} have the lowest free energy
and correspond to the stable phase of the theory. Therefore there is a
spontaneous breaking of the global $SO(3)$ symmetry and the theory at
$\tilde m=0$ develops a negative condensate proportional to $-\tilde
c_{\rm{cr}}\approx -0.59$. Note that only the absolute value of
$\tilde m$ corresponds to the bare mass of the fundamental fields. The
states with negative $\tilde m$ correspond to D5--brane embeddings that
intercept the $\tilde l=0$ line in the $\tilde l$ {\it vs.} $\tilde r$
plane and as seen from figure~\ref{fig:fig4} are unstable. It is to be 
expected that the meson spectrum of the theory in such a phase would
contain tachyons based on an analogy with the meson spectrum of the
D3/D7 system studied in ref.~\cite{Filev:2007qu}.
\begin{figure}[h] 
   \centering
   \includegraphics[width=9cm]{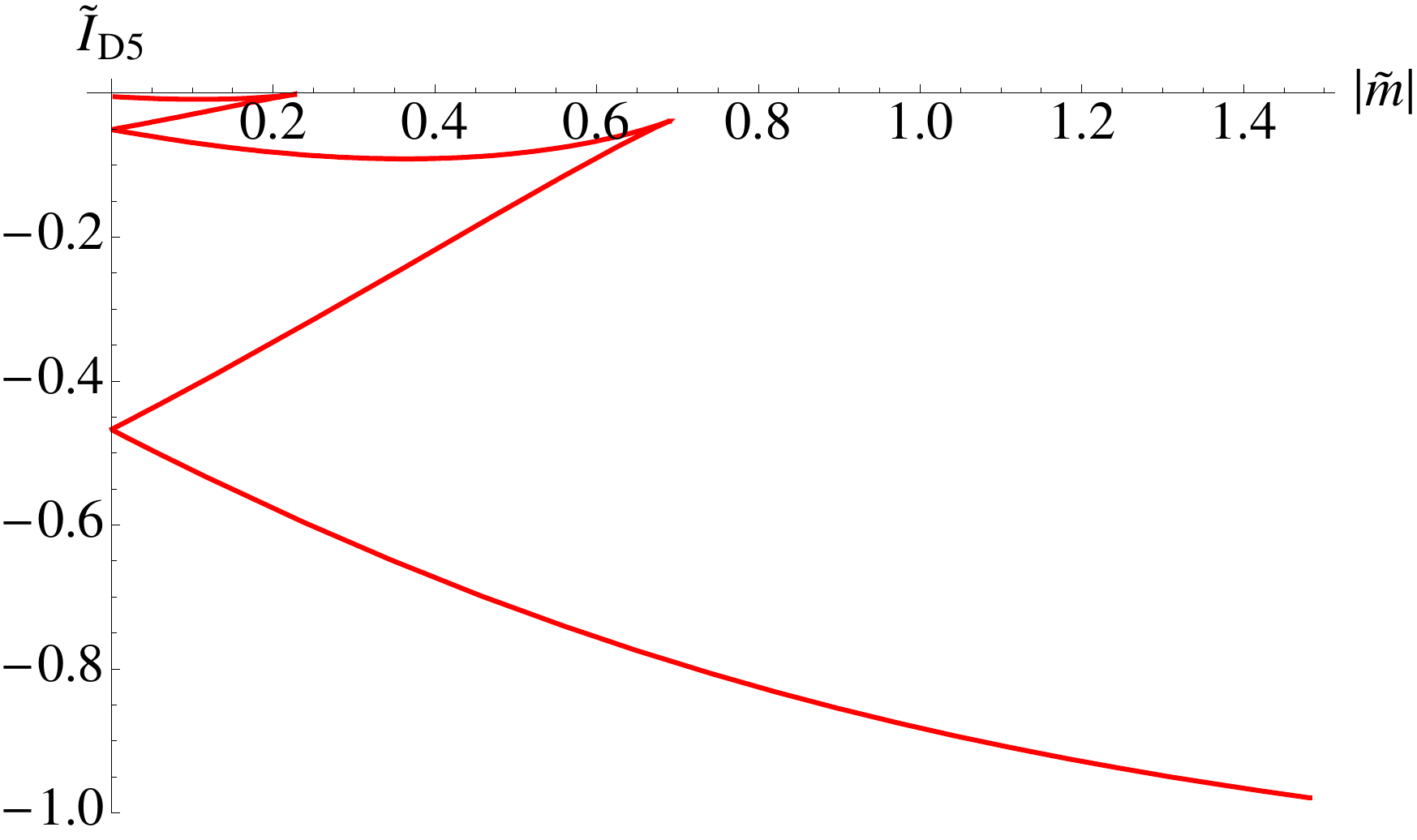}
   \caption{\small The states corresponding to the
     lowest positive branch of the plot in figure~\ref{fig:fig3} have
     the lowest free energy and thus correspond to the stable phase of
     the theory.  }
   \label{fig:fig4}
\end{figure}
Before we proceed with the analysis of the meson spectrum of the
theory let us write an expression for the condensate of the theory
$\langle\bar\psi\psi\rangle\propto -c_{\rm{cr}}=R^2H\tilde
c_{\rm{cr}}$ at zero bare quark mass. The coefficient of
proportionality is given by \cite{Mateos:2006nu}:
\begin{equation}
\langle\bar\psi\psi\rangle=-8\pi^2\alpha'\frac{\mu_5}{g_s}c_{\rm{cr}}=-16\pi^3\alpha'^2\frac{\mu_5}{g_s}\tilde c_{\rm{cr}}R^2(H/2\pi\alpha')\, .
\end{equation}
Note that the condensate is proportional to the magnitude of the
magnetic field $H/2\pi\alpha'$.


\subsubsection{Meson spectrum and pseudo-Goldstone bosons}

In this section we review the analysis of ref.~\cite{Filev:2009xp} on the normal modes of the
D5--brane. The authors of ref.~\cite{Filev:2009xp} had focused on the normal modes corresponding to the Goldstone bosons of the spontaneously
broken $SO(3)$ symmetry and studied their spectrum as a function of the
bare quark mass $m_q$. The study shows that the external magnetic
field splits the degeneracy of the meson spectrum and gives mass to
one of the pions of the theory. It also modifies the standard
$M_{\pi}^2\propto m$ GMOR relation for the remaining Goldstone mode to
a linear relation $M_{\pi}\propto m$. It has been shown that these results
are in accord with the behavior expected from the effective chiral
lagrangian of the theory.

In order to study the light meson spectrum of the theory one looks for
the quadratic fluctuations of the D5--brane embedding along the
transverse directions parametrized by $l,\psi,\phi$. To this end one
expands:
\begin{equation}
l=\bar l+2\pi\alpha'\delta l;~~~\psi=2\pi\alpha'\delta\psi;~~~\phi=2\pi\alpha'\delta\phi\, ,
\end{equation}
in the action (\ref{DBI2}) and leave only terms of order
$(2\pi\alpha')^2$. Note that fluctuations of the $U(1)$ gauge field
$F_{\alpha\beta}$ of the D5--brane will also contribute to the
expansion. There is also an additional contribution from the
Wess-Zumino term of the D5--brane's action:
\begin{equation}
S_{\rm{WZ}}=N_f\mu_5\int\limits_{{\cal M}_6}\sum_p[C_{p}\wedge e^{\cal F}];~~~{\cal F}=B+2\pi\alpha'F\, .
\end{equation}
The relevant term is:
\begin{equation}
S_{\rm{WZ}}=N_f\mu_5\int\limits_{{\cal M}_6}B\wedge P[\tilde C_{4}]\ ,
\end{equation}
where $P[\tilde C_4]$ is the pull-back of the magnetic dual, $\tilde
C_{4}$, to the background $C_{4}$ R-R form. It is given by:
\begin{equation}
\tilde C_4=\frac{1}{g_s}\frac{4r^2l^2}{(r^2+l^2)^3}R^4\sin\psi(l dr- rdl)\wedge d\Omega_2\wedge d\phi\ .
\end{equation}
The action for the quadratic fluctuations along $l$ is given by:
\begin{eqnarray}
  &&{\cal L}_{ll}^{(2)}\propto\frac{1}{2}\sqrt{-E}G_{ll}\frac{S^{\alpha\beta}}{1+l'^2}\partial_{\alpha}\delta l\partial_{\beta}\delta l+\frac{1}{2}\left[\partial_{l}^2\sqrt{-E}-\frac{d}{dr}\left(\frac{l'}{1+l'^2}\partial_l\sqrt{-E}\right)\right]\delta l^2\ ,\\ 
  &&{\cal L}_{lF}^{(2)}\propto\frac{\sqrt{-E}}{1+l'^2}(\partial_l J^{12}-\partial_{r}J^{12}l')F_{21}\delta l\nonumber\ ,\\
  &&{\cal L}_{FF}^{(2)}\propto\frac{1}{4}\sqrt{-E}S^{\alpha\beta}S^{\gamma\lambda}F_{\beta\gamma}F_{\alpha\lambda}\nonumber\ ,
\end{eqnarray}
and along $\phi$ and $\psi$:
\begin{eqnarray}
&&{\cal L}_{\psi\psi,\phi\phi}^{(2)}\propto\frac{1}{2}\sqrt{-E}S^{\alpha\beta}(G_{\psi\psi}\partial_{\alpha}\delta\psi\partial_{\beta}\delta\psi+G_{\phi\phi}\partial_{\alpha}\delta\phi\partial_{\beta}\delta\phi)\ \label{qvfl},\\
&&{\cal L}_{\psi\phi}^{(2)}\propto (\cos\alpha) PH\delta\psi\partial_0\delta\phi\nonumber \ .
\end{eqnarray}
Here $E_{\alpha\beta}$ is the pull-back of the generalized metric on
the classical D5--brane embedding:
\begin{equation}
E_{\alpha\beta}=\partial_{\alpha}\bar X^{\mu}\partial_{\beta}\bar X^{\nu}(G_{\mu\nu}+B_{\mu\nu})\, ,
\end{equation}
while $S^{\alpha\beta}$ and $J^{\alpha\beta}$ are the
symmetric and anti-symmetric elements of the inverse generalized
metric $E^{\alpha\beta}$:
\begin{equation}
E^{\alpha\beta}=S^{\alpha\beta}+J^{\alpha\beta}\ .
\end{equation}
The determinant $E$ and the function $K=P$ are given by:
\begin{eqnarray}
&&\sqrt{-E}=(\cos\alpha) r^2\sqrt{1+l'^2}\sqrt{1+\frac{R^4H^2}{(r^2+l^2)^2}}\equiv g(r)\cos\alpha \ ,\\
&&P=\frac{4R^4r^2l^2}{(r^2+l^2)^3}(rl'-l)\ .
\end{eqnarray}
As one can see, the fluctuations along $\psi$ and $\phi$ decouple from
the fluctuations along $l$ and the fluctuations of the gauge field
$A_{\alpha}$. To study the pseudo-Goldstone modes of
the dual theory one should focus on the fluctuations along $\psi$ and
$\phi$. The equations of motion derived from the quadratic action
(\ref{qvfl}) are the following:
\begin{eqnarray}
&&\partial_r\left(\frac{g(r)l^2}{1+l'^2}\partial_r\delta\psi\right)+\frac{g(r)R^4l^2}{(r^2+l^2)^2}\tilde\Box\delta\psi+\frac{g(r)l^2}{r^2}\Delta_{(2)}\delta\psi-PH\partial_0\delta\phi=0\ ,\\
&&\partial_r\left(\frac{g(r)l^2}{1+l'^2}\partial_r\delta\phi\right)+\frac{g(r)R^4l^2}{(r^2+l^2)^2}\tilde\Box\delta\phi+\frac{g(r)l^2}{r^2}\Delta_{(2)}\delta\phi+PH\partial_0\delta\psi=0\ ,\nonumber
\end{eqnarray}
where 
\begin{equation}
\tilde\Box=-\partial_0^2+\frac{\partial_1^2+\partial_2^2}{1+\frac{R^4H^2}{(r^2+l^2)^2}}\ .
\label{LP}
\end{equation}
Note that the background magnetic field breaks the $SO(1,2)$ Lorentz
symmetry to $SO(2)$, which manifests itself in the modified laplacian
(\ref{LP}). Next one considers a plane-wave ansatz:
\begin{equation}
\delta\phi=e^{i(\omega t-\vec k\dot \vec x)}\eta_1(r);~~~\delta\psi=e^{i(\omega t-\vec k\dot \vec x)}\eta_2(r)\label{anzatsqv}\, ,
\end{equation}
using the ans\"atz (\ref{anzatsqv}) one gets:
\begin{eqnarray}
&&\partial_r\left(\frac{g(r)l^2}{1+l'^2}\eta_1'\right)+\frac{g(r)R^4l^2}{(r^2+l^2)^2}(\omega^2-\frac{{\vec k}^2}{1+\frac{R^4H^2}{(r^2+l^2)^2}})\eta_1-i\omega PH\eta_2=0\ ,\label{coupled}\\
&&\partial_r\left(\frac{g(r)l^2}{1+l'^2}\eta_2'\right)+\frac{g(r)R^4l^2}{(r^2+l^2)^2}(\omega^2-\frac{{\vec k}^2}{1+\frac{R^4H^2}{(r^2+l^2)^2}})\eta_2+i\omega PH\eta_1=0\ .\nonumber
\end{eqnarray}
The equations of motion in (\ref{coupled}) can be decoupled by the
definition $\eta_{\pm}=\eta_1\pm i\eta_2$. The result is:
\begin{eqnarray}
&&\partial_r\left(\frac{g(r)l^2}{1+l'^2}\eta_+'\right)+\frac{g(r)R^4l^2}{(r^2+l^2)^2}(\omega^2-\frac{{\vec k}^2}{1+\frac{R^4H^2}{(r^2+l^2)^2}})\eta_+-\omega PH\eta_+=0\ ,\label{decoupled}\\
&&\partial_r\left(\frac{g(r)l^2}{1+l'^2}\eta_-'\right)+\frac{g(r)R^4l^2}{(r^2+l^2)^2}(\omega^2-\frac{{\vec k}^2}{1+\frac{R^4H^2}{(r^2+l^2)^2}})\eta_-+\omega PH\eta_-=0\ .\nonumber
\end{eqnarray}
Because of the broken Lorentz symmetry, the $1+2$ dimensional mass
$M^2=\omega^2-{\vec k}^2$ depends on the choice of frame. One can
define the spectrum of excitations as the rest energy (consider the
frame with $\vec k=0$).  The spectrum is
discrete. Furthermore just as in the D3/D7 case there is a Zeeman
splitting of the spectrum due to the external magnetic
field. Interestingly, at low energy the splitting is breaking the
degeneracy of the lowest energy state and as a result there is only
one pseudo-Goldstone boson. This is not in contradiction
with the Goldstone theorem because there is no Lorentz symmetry. This
opens the possibility of having two types of Goldstone modes: type I
and type II satisfying odd and even dispersion relations
correspondingly. In this case there is a modified counting rule
(ref.~\cite{Nielsen:1975hm}, see also ref.~\cite{Brauner:2005di})
which states that {\it the number of GBs of type I plus twice the
  number of GBs of type II is greater than or equal to the number of
  broken generators.} As we are going to show below the single
Goldstone mode that we see satisfies a quadratic dispersion relation
(hence is type II) and the modified counting rule is not
violated. Note also that for large bare masses $m$ (and
correspondingly large values of $l$) the term proportional to the
magnetic field is suppressed and the meson spectrum should approximate
to the result for the pure AdS$_5\times S^5$ space-time case studied
in refs.~\cite{{Arean:2006pk},{Myers:2006qr}}, where the authors
obtained the following relation:
\begin{equation}
\omega_n=\frac{2m}{R^2}\sqrt{(n+1/2)(n+3/2)}\ ,\label{purespect-a}
\end{equation}
between the eigenvalue of the $n^{th}$ excited state $\omega_n$ and the bare mass $m$.  

In order to obtain the meson spectrum, one solves numerically the
equations of motion (\ref{decoupled}) in the rest frame ($\vec
k=0$). The quantization condition for the spectrum comes from imposing
regularity at infinity. More precisely one requires that $\eta_{\pm}
\sim 1/r$ at infinity ($r\rightarrow\infty$). The results are summarized in
figure~\ref{fig:fig6}. It is convenient to define the
dimensionless quantities $\tilde m=m/R\sqrt{H}$ and
$\tilde\omega=\omega R/\sqrt{H}$. As one can see from figure
\ref{fig:fig6}, for large $\tilde m$ the spectrum asymptotes to that
of pure AdS$_5\times S^5$, given by equation~(\ref{purespect-a}). The
Zeeman splitting of the spectrum is also evident. It is interesting
that as a result of the splitting of the ground state there is only a
single pseudo-Goldstone mode. Furthermore, as can be seen from figure
\ref{fig:fig7}, for small bare masses instead of the usual
Gell-Mann--Oakes--Renner relation there is a linear dependence
$\tilde\omega\sim\tilde m$. As we will review in the next subsection the
slope is given by the relation:
\begin{equation}
\tilde\omega=\frac{4\tilde c_{\rm{cr}} }{\pi}\tilde m\approx 0.736\tilde m\, . \label{lingmor}
\end{equation}

\begin{figure}[h] 
   \centering
   \includegraphics[width=10cm]{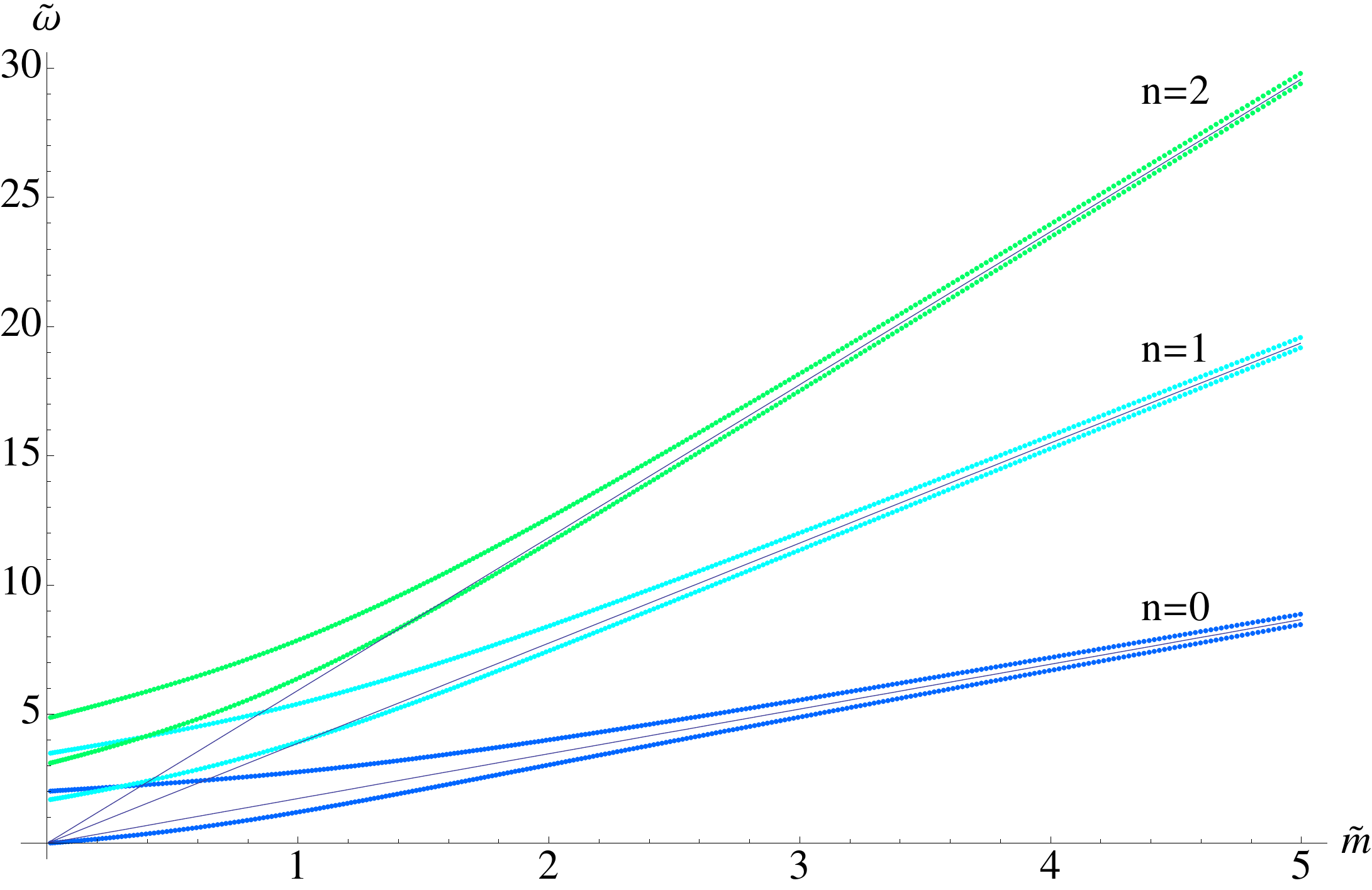}
   \caption{\small The meson spectrum of the first three excited
     states is plotted. There is Zeeman splitting of the
     spectrum and the existence of a mass gap at $\tilde m=0$ as well
     as a single Goldstone boson mode. For large $\tilde m$ the
     spectrum asymptotes to that of zero magnetic field given by
     equation (\ref{purespect}) (straight lines).}
   \label{fig:fig6}
\end{figure}

\begin{figure}[h] 
   \centering
   \includegraphics[width=10cm]{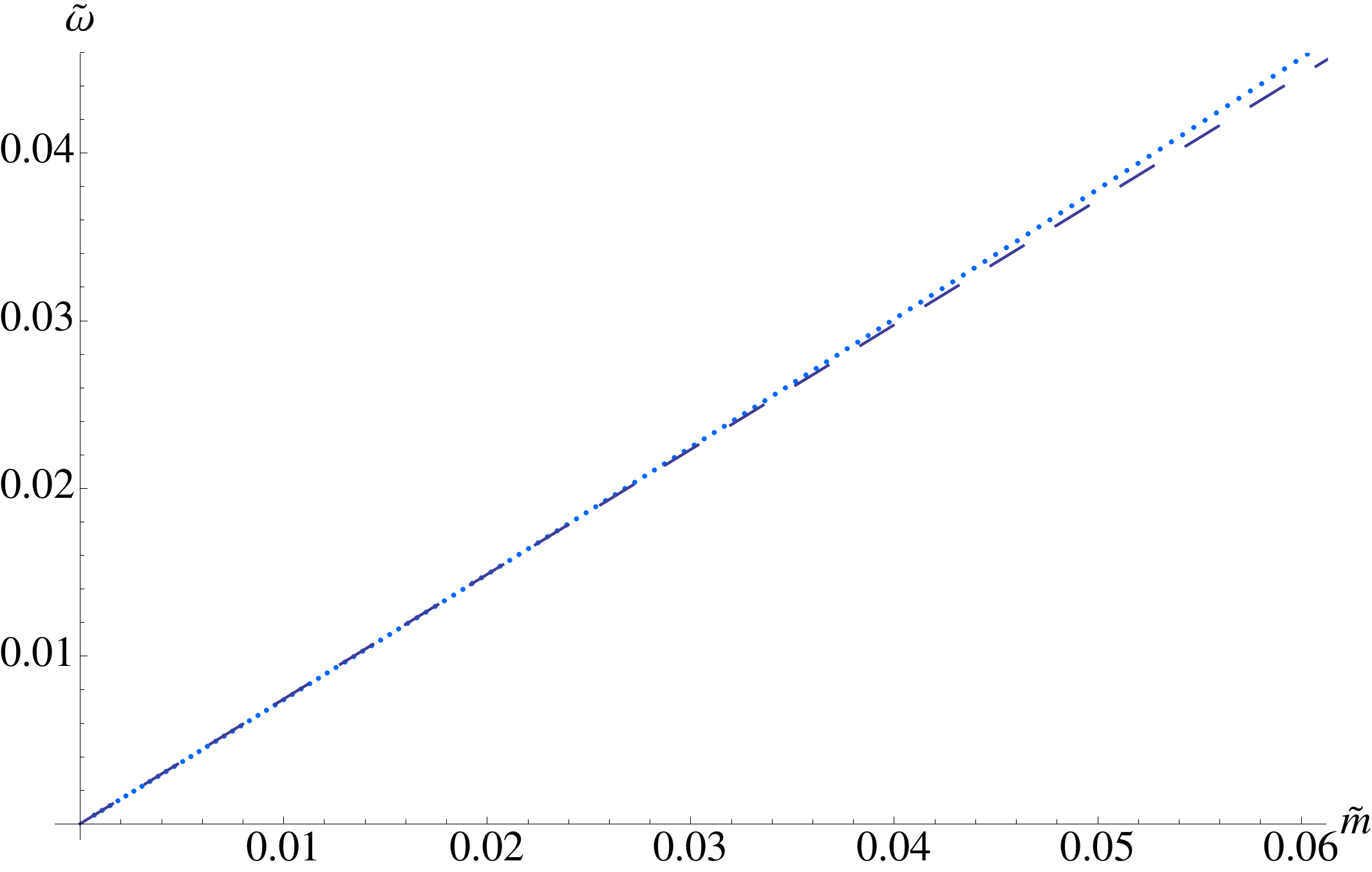}
   \caption{\small Plot of the spectrum of the ground state from
     figure~\ref{fig:fig6} for small bare masses. The dashed line
     corresponds to the linear behavior from equation
     (\ref{lingmor}).}
\label{fig:fig7}
\end{figure}

It is also interesting to study the dispersion relation of the
Goldstone mode. Since the Lorentz symmetry is broken and only
one pseudo-Goldstone mode is observed one anticipates a quadratic dispersion relation (see
refs.~\cite{Brauner:2005di} and~\cite{Brauner:2007uw} for discussion).

In order to obtain the dispersion relation of the Goldstone mode the authors of ref.~\cite{Filev:2009xp}
numerically solve equations (\ref{decoupled}) at very small bare mass
$\tilde m\approx 0.0007$ and for a range of small momenta $\tilde{\vec
  k}=\vec k R/\sqrt{H}$. The result is presented in figure
\ref{fig:fig8}. There is  indeed  a quadratic dispersion
relation. It has been shown \cite{Filev:2009xp} that the dispersion relation is given
by:
\begin{equation}
\tilde \omega=\gamma {\tilde{\vec k}}^2+\frac{4}{\pi}\tilde c_{\rm{cr}}\tilde m\label{dispersion}\ ,
\end{equation}
where:
\begin{equation}
\gamma=\frac{4}{\pi}\int\limits_{0}^{\infty}d\tilde r\frac{\tilde r^2\tilde l^2\sqrt{1+\tilde l'^2}}{(\tilde r^2+\tilde l^2)\sqrt{1+(\tilde r^2+\tilde l^2)^2}}\, .\label{gamma}
\end{equation}
For $\tilde m\approx 0.0007$ the relation (\ref{dispersion}) is given by:
\begin{equation}
\tilde \omega\approx 0.232 {\tilde{\vec k}}^2+0.000515\, ,\label{dispfit}
\end{equation}
and is represented by the fitted curve in figure~\ref{fig:fig8}.

\begin{figure}[h] 
   \centering
   \includegraphics[width=10cm]{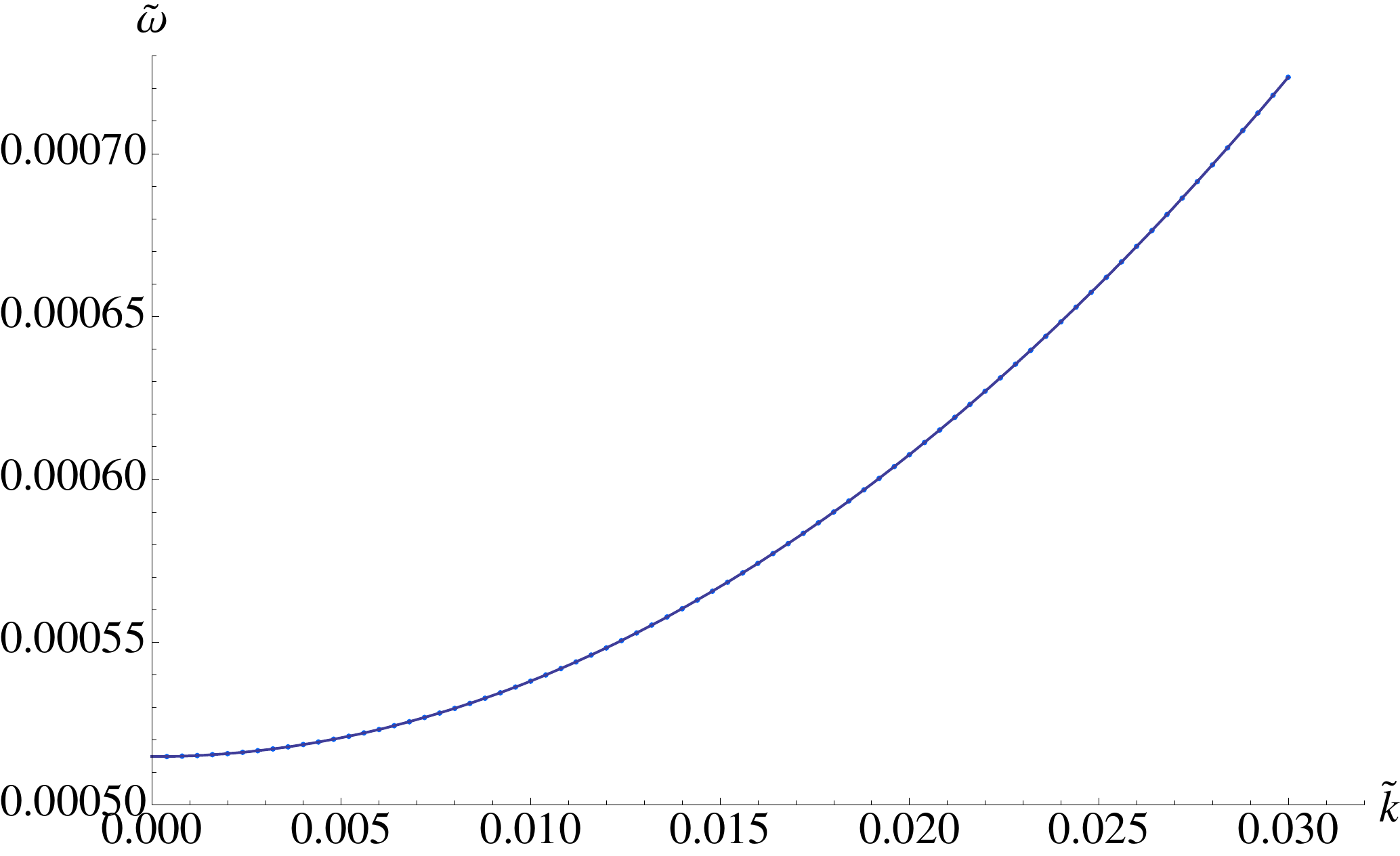}
   \caption{\small Plot of the  dispersion relation of the pseudo-Goldstone mode for $\tilde m\approx 0.0007$. The parabolic fit corresponds to equation (\ref{dispfit}).}
\label{fig:fig8}
\end{figure}

\paragraph{Low energy dispersion relation.}  
In order to obtain the dispersion relation for the pseudo-Goldstone
mode one considers ref.~\cite{Filev:2009xp} the limit of small $\omega$ in the first equation in (\ref{decoupled}) thus leaving only the linear potential term in $\tilde \omega$. In view of
the observed quadratic dispersion relation (\ref{dispersion}) one also keeps the ${\vec k}^2$ term in equation~(\ref{decoupled}).
\begin{equation}
\partial_r\left(\frac{g(r)l^2}{1+l'^2}\eta_+'\right)-\left(\omega PH+\frac{g(r)R^4l^2}{(r^2+l^2)^2+R^4H^2}{\vec k}^2\right)\eta_+=0\, .\label{etapl}
\end{equation}
It is convenient to define the following variables:
\begin{equation}
\Theta^2=\frac{g(r)l^2}{1+l'^2}\ ;~~~\xi=\eta_+\Theta\ .
\end{equation}
Then equation~(\ref{etapl}) can be written as:
\begin{equation}
\ddot\xi-\frac{\ddot\Theta}{\Theta}\xi-\left(\omega PH+\frac{g(r)R^4l^2}{(r^2+l^2)^2+R^4H^2}{\vec k}^2\right)\frac{\xi}{\Theta^2}=0\ .\label{xi}
\end{equation}
Where the overdots represent derivatives with respect to $r$. Now if one takes
the limit $m\rightarrow 0$ one has that $\omega\rightarrow 0$ and $k\rightarrow 0$ and obtains
that:
\begin{equation}
\xi=\Theta|_{\omega=0}\equiv\bar\Theta\ ,
\end{equation}
is a solution to equation (\ref{xi}). The next step is to consider small $m$ and expand:
\begin{equation}
\xi=\bar\Theta+\delta\xi\ ;~~~\Theta=\bar\Theta+\delta\Theta\ ,\label{expcrit}
\end{equation}
where the variations $\delta\xi$ and $\delta\Theta$ are vanishing in
the $m\rightarrow 0$ limit. Then, to leading order in $m$ (keeping in mind
that $\omega\sim m$ and $\vec k^2\sim m$) one obtains:
\begin{equation}
\delta\ddot\xi-\frac{\ddot{\bar\Theta}}{\bar\Theta}\delta\xi-\delta\left(\frac{\ddot\Theta}{\Theta}\right)\bar\Theta-\left(\omega PH+\frac{g(r)R^4l^2}{(r^2+l^2)^2+R^4H^2}{\vec k}^2\right)\frac{1}{\bar\Theta}=0\ .
\label{prib}
\end{equation}
Now one multiplies equation (\ref{prib}) by $\bar\Theta$ and integrates
along $r$. The result is:
\begin{equation}
(\bar\Theta\delta\dot\xi-\dot{\bar\Theta}\delta\xi)\Big |_0^{\infty}-(\bar\Theta\delta{\dot\Theta}-\dot{\bar\Theta}\delta\Theta)\Big |_0^{\infty}-\omega H\int\limits_0^{\infty}{dr}P(r)-\frac{\pi}{4} R^5\sqrt{H}\gamma\vec k^2=0\ .\label{intermid}
\end{equation}
Using the definitions of $\Theta, P(r)$ and $\xi$ and requiring regularity at infinity for $\eta_+$, one can show that the first term in equation~(\ref{intermid}) vanishes and that:
\begin{equation}
 (\bar\Theta\delta{\dot\Theta}-\dot{\bar\Theta}\delta\Theta)\Big |_0^{\infty}=c\delta  m\ ;~~~\int\limits_0^{\infty}{dr}P(r)=-R^4\pi/4\ ,
\end{equation}
and hence using the previous definitions, $\tilde m =m/R\sqrt{H}$, $\tilde c=c/R^2 H$, $\tilde\omega=\omega R/\sqrt{H}$ and $\tilde{\vec k}=\vec k R/\sqrt{H}$, one obtains equation (\ref{dispersion}).

\paragraph{Effective chiral lagrangian.}

In order to obtain the $1+2$ dimensional effective action describing
the pseudo-Goldstone mode one considers \cite{Filev:2009xp} the $1+5$ dimensional action
(\ref{qvfl}) for a classical embedding in the vicinity of the critical
embedding, namely that embedding corresponding to a very small bare
mass $\tilde m$. Next one takes the ans\"atz for the
fields $\delta\phi$ and $\delta\psi$:
 \begin{equation} 
\delta\phi=\frac{\xi_1(r)}{\Theta(r)}\chi_1(x)\ ;~~~\delta\psi=\frac{\xi_2(r)}{\Theta(r)}\chi_2(x)\ .
\end{equation}
Since one is close to the critical embedding one considers the same expansion as in equation~(\ref{expcrit}):
\begin{equation}
\xi_i=\bar\Theta+\delta\xi_i\ ,~~~i=1\, {\rm or}\, 2\ ;~~~\Theta=\bar\Theta+\delta\Theta\ .
\end{equation}
By definition it follows that as $\tilde m\rightarrow 0$, $\delta \xi_i$ and $\delta \Theta$ vanish. Then to leading order one has that:
\begin{eqnarray}
&&\partial_r\delta\phi=\frac{1}{\bar\Theta^2}[(\bar\Theta\delta\dot\xi_1-\dot{\bar\Theta}\delta\xi_1)+(\dot{\bar\Theta}\delta\Theta-\bar\Theta\delta\dot\Theta)]\chi_1(x)\ ;~~~\partial_{\mu}\delta\phi=\partial_{\mu}\chi_1(x)\ ;~~\mu=0,1,2\, ,\\
&&\partial_r\delta\psi=\frac{1}{\bar\Theta^2}[(\bar\Theta\delta\dot\xi_2-\dot{\bar\Theta}\delta\xi_2)+(\dot{\bar\Theta}\delta\Theta-\bar\Theta\delta\dot\Theta)]\chi_2t(x)\ ;~~~\partial_{\mu}\delta\psi=\partial_{\mu}\chi_2(x)\ ;~~\mu=0,1,2\, .\nonumber
\end{eqnarray}
Next one integrates equation~(\ref{qvfl}) along $r$ from $0,\infty$ and
along the internal unit sphere $\tilde\Omega_2$. The interesting term
is the part of the kinetic term involving derivatives along $r$. After
integration by parts it boils down to a mass term for the $1+2$
dimensional fields $\chi_1,\chi_2$. Explicitly:
\begin{eqnarray}
&&\int drd\tilde\Omega_2\frac{1}{2}\frac{g(r)l^2}{1+l'^2}\partial_r\delta\phi\partial_r\delta\phi=-\int drd\tilde\Omega_2\frac{1}{2}\partial_r\left(\frac{g(r)l^2}{1+l'^2}\partial_r\delta\phi\right)\delta\phi=\label{kinrph}\\
&&\hskip2cm =-4\pi[(\bar\Theta\delta\dot\xi_1-\dot{\bar\Theta}\delta\xi_1)+(\dot{\bar\Theta}\delta\Theta-\bar\Theta\delta\dot\Theta)]\Big|_0^{\infty}\frac{1}{2}\chi_1^2=4\pi m c\frac{1}{2}\chi_1^2\nonumber\ .
\end{eqnarray}
Here the same arguments as in equation (\ref{intermid}) have been used. It is clear that one can perform an analogous
calculation for the term involving $\partial_r\delta\psi$. The rest of
the terms are dealt with straightforwardly by integrating along
$r$. The resulting action is:
\begin{equation}
\frac{S_{\rm{eff}}}{(2\pi\alpha')^2}=\int d^3x\left\{\frac{f_{\pi||}^2}{4}\partial_0\chi^{*}\partial_0\chi-\frac{f_{\pi\bot}^2}{4}\partial_i\chi^{*}\partial_i\chi-\mu\frac{i}{2}(\chi\partial_0\chi^{*}-\chi^{*}\partial_0\chi)+\frac{m_q}{2}\langle\bar\Psi\Psi{\rangle}_0\chi^*\chi\right\}\ ,\label{effectivef}
\end{equation}
where the complex scalar field $\chi=\chi_1+i\chi_2$ has been defined. The constants in the effective action are given
by:
\begin{eqnarray}
&&\frac{f_{\pi||}^2}{4}=\frac{{\cal N}}{2}\int\limits_0^{\infty} dr\frac{g(r)R^4l^2}{(r^2+l^2)^2}\ ;~~~~\frac{f_{\pi\bot}^2}{4}=\frac{{\cal N}}{2}\int\limits_0^{\infty} dr\frac{g(r)R^4l^2}{(r^2+l^2)^2+R^4H^2}\ ,\\
&&\mu=\frac{\cal N}{8}\pi R^4H;~~\langle\bar\psi\psi\rangle=-(2\pi\alpha'){\cal N}c_{\rm{cr}}\ ;~~{\cal N}=4\pi N_f\frac{\mu_5}{g_s};~~m_q=\frac{m}{2\pi\alpha'}\ .\nonumber
\end{eqnarray}
\subsection{Summary}
In this section we reviewed the studies of the influence of external magnetic field on holographic gauge theories dual to the D3/D7-- and the D3/D5-- intersections. 

In the case of the D3/D7--system we described the effect of mass generation realized as a separation of the probe D7--branes and the background D3--branes in the interior of the geometry. The review of the study of the meson spectrum revealed Zeeman splitting and the existence of Goldstone mode associated to the spontaneously broken Chiral Symmetry. In the limit of small bare masses we reviewed the analytic derivation of the effective chiral action obtained in ref.~\cite{Filev:2009xp} after dimensional reduction of the eight dimensional effective action of the probe D7--brane. As expected the mass of the pseudo-Goldstone modes satisfies the Gell-Mann-Oaks--Renner relation. The effective action also suggests an anisotropic relativistic dispersion relation consistent with the residual $SO(1,1)\times SO(2)$ symmetry. An integral expression for the parameter of anisotropy is also obtained. It would be particularly interesting if one could obtain this parameter via alternative non-prturbative approach, such as lattice simulation. 

The D3/D5--system studied in subsection 4.2 exhibit properties similar to the D3/D7 system. Again there is mass generation realized as separation of the color and flavor branes. The meson spectrum also has Zeeman splitting and a Goldstone mode. There is however a crucial difference. Due to the completely broken Lorentz symmetry the goldstone modes satisfy non-relativistic dispersion relation and modified counting rule. An integral expression for the galilean mass in terms of the parameter $\gamma$ given in equation (\ref{gamma}) has also been obtained. It would be nice if one can compare this result with a result obtained via lattice simulation.

Overall the holographic studies that we presented in this section confirm the universal nature of the phenomenon of magnetic catalysis of chiral symmetry breaking reviewed in section 2.2. It is also somewhat satisfying that some of the results (such as the Gell-Mann--Oaks-Renner relation) presented here  have been obtained in a closed form and are consistent with the results derived via standard chiral dynamics. 
\section{Conclusion}
In this review we outlined a recent application of the AdS/CFT correspondence to study the effect of magnetic catalysis of mass generation in holographic gauge theories dual to the D3/D5-- and the D3/D7--intersections. Our goal was to illustrate the potential of the correspondence to capture essential properties of the strongly coupled regime of non--abelian gauge theories especially when the investigated phenomenon is of a universal nature. We attempted to give the review a self-contained and somewhat pedagogical form:

The purpose of Section 2 was to remind the reader about some basic properties of flavored Yang-Mills theories as well as to provide the physical motivation for the holographic studies subject to our investigation. We provided a brief description of the global symmetries of the theory with emphasis on the anomalous and non-singlet Chiral Symmetries. We also outlined the effective chiral lagrangian description of the effect of Chiral Symmetry breaking and provided a short derivation of the Gell--Mann-Oaks--Renner relation. The second part of Section 2 describes the standard field theory approach to studying the phenomenon of magnetic catalysis of mass generation.

In Section 3 of the review we outlined the basics of the AdS/CFT correspondence. We started with a review of the general ideas that lead to its formulation and descried in details the physical justification of the correspondence in the framework of string theory. We also provided a brief description of the way the standard AdS/CFT dictionary operates. In the second part of Section 3 we focused on the addition of flavor degrees to the correspondence. We reviewed the standard way \cite{Karch:2002sh} to introduce quenched fundamental matter by adding probe flavor branes to the dual supergravity background. We provided some basics extract from the generalized AdS/CFT dictionary which are implemented in the holographic set up reviewed in Section 4.
  
Section 4 is the main part of the review. We reviewed the holographic study of the influence of external magnetic field on holographic gauge theories dual to the D3/D5-- and the D3/D7--intersections. We reviewed the general properties of the holographic set up. We described how spontaneous symmetry breaking is realized as a separation between the flavor and color branes in the bulk of the geometry. Investigations of the meson spectrum revealed Zeeman splitting of the energy level as well as the existence of Goldstone modes. In the case of the D3/D7--set up we reviewed the analytic derivation of the Gell-Mann--Oaks--Renner relation performed in ref.~\cite{Filev:2009xp}. We also investigated the dispersion relations of the pseudo-Goldosten modes and verified that they are consistent with the residual space time symmetry: $SO(1,1)\times SO(2)$ in the D3/D7 set up and $SO(2)$ in the D3/D5 set up. It is intriguing that the D3/D5 system exhibits a non-relativistic dispersion relation.   
  
The overall goal of the review was to review in a self-consistent way one of the many successful applications of the AdS/CFT duality to the investigation of universal properties of strongly coupled Yang-Mills theories. The results presented in Section 4 seems to be in perfect agreement with the proposed universal nature of the effect of chiral symmetry breaking in an external magnetic field. We hope that future investigations using alternative non--perturbative techniques such as lattice simulations could confirm qualitatively the results presented in Section 4. Such studies could provide a non-trivial check of the AdS/CFT correspondence in this essentially non--supersymmetric set up.

\section{Acknowledgments}
V.F. would like to thank Clifford V. Johnson and Jonathan P. Shock for fruitful collaboration in refs.~\cite{{Filev:2007gb},{Filev:2009xp}}. The work of V.F. was supported by an IRCSET 
postdoctoral fellowship. The work of R.R. was supported in part by the Austrian Research Fund FWF \# P22000, NSFB VU-F-201/06 and DO 02-257.

\end{document}